\documentclass[preprint,12pt]{aastex}
\usepackage{epsfig}
\usepackage{lscape}
\usepackage{ifthen}
\usepackage{color}

\newcommand{\nhitotal}{{18}}
\newcommand{\MHI}{M_{\rm H\sc I}}

\newcommand\kms{{km~s$^{-1}$}}
\newcommand\Ha{{H$\alpha$}}
\newcommand{\hi}{{H{\sc i}}}
\newcommand\Msun{$M_\odot$}

\slugcomment{accepted to ApJ }

\shorttitle{Arecibo Deep Survey}
\shortauthors{Freudling et al.}

\begin{document}

\title{Deep 21-cm \hi\ Observations at z$ \approx $0.1: 
The {Precursor to} the Arecibo Ultra Deep Survey \footnote{ The Arecibo Observatory is part of the National
Astronomy and Ionosphere Center which is operated by Cornell University under a Cooperative Agreement with the
National Science Foundation.
  }}

\author{Wolfram Freudling}
\affil{ European Southern Observatory, Karl-Schwarzschild-Str. 2, 85748 Garching, Germany }
\author{Lister Staveley-Smith}
\affil{International Centre for Radio Astronomy Research M468, The University of Western Australia, Crawley, 6009, Australia}
\author{Barbara Catinella}
\affil{Max-Planck-Institut f\"ur Astrophysik, Karl-Schwarzschild-Str. 1, 85748 Garching, Germany }
\author{Robert Minchin}
\affil{Arecibo Observatory, HC03 Box 53995, Arecibo, PR 00612, USA}
\author{Mark Calabretta}
\affil{ Australia Telescope National Facility, PO Box 76
Epping NSW 1710, Australia}
\author{Emmanuel Momjian}
\affil{National Radio Astronomy Observatory, P. O. Box O, Socorro, NM,
87801, USA}
\author{Martin Zwaan}
\affil{ European Southern Observatory, Karl-Schwarzschild-Str. 2, 85748 Garching, Germany }
\author{Martin Meyer}
\affil{International Centre for Radio Astronomy Research
M468, The University of Western Australia, Crawley, 6009, Australia}
\author{Karen O'Neil}
\affil{National Radio Astronomy Observatory, PO Box 2, Green Bank, WV 24944, USA}

\author{}

\clearpage
\begin{abstract}   

The ``{A}LFA {U}ltra {D}eep {S}urvey'' (AUDS) is an ongoing 21-cm spectral
survey with the Arecibo 305m telescope.  AUDS will be  the most sensitive blind
survey  undertaken with Arecibo's  300 MHz Mock spectrometer. The survey
searches for 21-cm \hi\ line emission at redshifts between 0 and 0.16.  The
main goals of the survey are  to investigate the \hi\ content and probe the
evolution of \hi\ gas within that redshift region.  In this paper, we report on
a set of precursor observations with a total integration time of 53 hours. The
survey detected a total of eighteen 21-cm emission lines at redshifts between
0.07 and 0.15 in a region centered around $\alpha_{2000} \approx 0^h$, $\delta
\approx 15^\circ42'$.  The rate of detection is consistent with the one
expected from the local \hi\ mass function. The derived relative \hi\ density
at the median redshift of the survey is $\rho_{\rm HI}[z = 0.125] =
(1.0\pm0.3) \rho_0$, where $\rho_0$ is the \hi\ density at zero redshift.

\end{abstract}

\keywords{cosmology: observations -- galaxies: evolution -- radio lines: galaxies }

\singlespace

\section{Introduction}

\subsection{Neutral hydrogen surveys of the Local Universe}

Large-area blind \hi\ surveys have come of age since the first 21-cm multibeam
receiver was installed at Parkes Observatory in 1997, to be followed by
similarly powerful receiving systems at the Lovell telescope, the Arecibo
telescope and the Effelsberg telescope. These surveys have allowed extensive
areas of sky to be mapped for extragalactic neutral hydrogen, resulting in the
detection of over $10^4$ galaxies
\citep{hipass,hipassnorth,alfalfa1,effelsberg}.  {Surveys that  specifically
targeted the zone of avoidance expanded the sky coverage even further
\citep[e.g.][]{zoadwing,zoaparkes,zoaalfa}.} Prior to this, the most extensive
neutral hydrogen surveys were those carried out with the single-beam or
dual-beam systems at Arecibo: the Arecibo \hi\ Strip Survey
\citep[AHISS]{ahiss} the Arecibo Slice \citep{spitzak} and the Arecibo Dual
Beam Survey \citep[ADBS]{adbs}. These were all limited by the available
bandpass to maximum recessional velocities of around 8000 km\,s$^{-1}$
(z$\simeq0.03$) and had relatively slow mapping speeds compared to the more
recent multibeam surveys.

Multibeam surveys have adopted the strategy of going wide rather than deep,
since this generally results in the largest observed volume and the greatest
number of detected objects.  If a survey in the local Universe is carried out
with an integration time of 1 unit, then extending it to $t$ units by making it
wider will increase the volume covered by a factor of $t$, while extending it
with a longer integration in the same area will increase the volume by a factor
of $t^{3/4}$, if not limited by receiver bandwidth {and the changing beamsize
with frequency is neglected}.  Going very deep is generally only advantageous
if it becomes possible to reach a new population of galaxies, e.g.  outside the
local region, or galaxies at an earlier epoch in their evolution.  Therefore,
most survey time at Parkes and Lovell was spent on the relatively shallow \hi\
Parkes All Sky Survey \citep[HIPASS]{hipass} and the HI Jodrell All Sky Survey
\citep[HIJASS]{hijass} respectively.  These surveys were limited by sensitivity
(as well as the available bandpass) to maximum recessional velocities of less
than 14000 km\,s$^{-1}$ (z$\simeq0.04)$, even for the most massive systems.
The deepest field survey carried out with the Parkes system
\citep[HIDEEP]{hideep} reached a noise level of only 3.2 mJy (at a velocity
resolution of 18 km$\,s^{-1}$), considerably higher than AHISS (0.75 mJy at 18
km\,s$^{-1}$). {The most sensitive \hi\ surveys in the local universe are those
pointed towards the CVn group \citep{cvnsurvey} and the Virgo cluster
\citep{virgosurvey}}.

Observational programs on the newer Arecibo L-band Feed Array (ALFA) system at
Arecibo have followed a similar pattern with the largest surveys being devoted
to large areas of sky. The Arecibo Legacy Fast ALFA Survey
\citep[ALFALFA]{alfalfa1,alfalfa2} is covering 7000 deg$^2$ to a noise level of
2.2 mJy (at 10 km\,s$^{-1}$) and  the Arecibo Galaxy Environment Survey
\citep[AGES]{ages} is covering 200 deg$^2$ to a noise level of 0.75 mJy (at 10
km\,s$^{-1}$). 

\subsection{\bf \hi\ observations above z=0.1}

Currently, the Cornell \hi\ digital archive (Springob et al. 2005), a
homogeneous database of pointed \hi\ observations of about 9000
optically-selected galaxies, includes only a few objects with recessional
velocities larger than 20,000 \kms. Even the ongoing ALFALFA and AGES surveys
mentioned above only reach maximum recessional velocities of 18,000 and 20,000
\kms, respectively. However, in spite of challenges such as limited sensitivity
and radio-frequency interference (RFI), some progress in the detection of \hi\
emission from galaxies at redshifts beyond $z\sim 0.1$ has been made with deep
integrations of galaxy clusters with the Westerbork Synthesis Radio Telescope
(WSRT) \citep{zwaan,ver}, pointed Arecibo observations of H$\alpha$-emitting
galaxies \citep{sdsssample}, and `stacked' Giant Metrewave Radio Telescope
(GMRT) integrations of fields with existing redshift information
\citep{lah,lah09}. {The parameters of these surveys are summarized in
Tab.~\ref{tab:hisurveys}.} These observations have shown that it is possible to
study the gaseous properties of galaxies in environments different to the local
Universe, and at look-back times of up to 2--4 Gyr. Nevertheless, selection
effects, limited sensitivity, cosmic variance all conspire to make it
challenging to obtain the accurate measurements required to quantify systematic
evolutionary and environmental trends.

\subsection{Evolution of the Gas Content of the Universe}\label{dla_science}

The rate of conversion of gas into stars is one of the fundamental quantities
that describe the evolution of galaxies and remains a difficult measurement to
make for all but the most nearby galaxy populations
\citep[e.g.][]{madau,haar,bouwens,wong2009}. At redshifts between zero and
unity, the global co-moving star-formation rate appears to increase by a factor
of five \citep{hopkins}.  Molecular gas content possibly also increases over
this redshift range \citep{h2evolution}.

Early models of the evolution of cold gas predicted a constant conversion
factor between gas supply and star formation  and therefore led to a similar
strong evolution of the cold gas content \citep{pei}.  {Observations of local
galaxies suggest that both hot and cold gas accretion significantly contribute
to replenish the gas content of galaxies \citep[see e.g. review by ][]{review},
and the balance between the two might depend on redshift.  The modern picture
of} a self-regulated equilibrium between the net inflow of gas and star
formation leads to predictions of somewhat smaller evolution of the cold gas
content at redshifts lower than two \citep{noevol,power}. This appears to fit
observations of damped Lyman-$\alpha$ systems \citep[DLAs,
e.g.][]{rao2005,slw,pro2004,pro2005} which suggest that the redshift density of
absorbers with $N_H>2\times10^{20}$ cm$^{-2}$ increases only mildly,
$dN_{DLA}/dz=0.05(1+z)^{1.1}$, which is consistent with constant co-moving gas
density. 

Conversion from DLA measurements to a cosmic gas density $\Omega_{\rm HI}$ is
unfortunately fraught with uncertainty, largely due to the small number of
observed systems, but also due to uncertain systematic effects such as lensing
and covering factor.  Better measurements of the gas mass density from
intermediate redshift DLAs are presently not possible because of the limited
availability of space-based UV instruments.  \cite{rosenberg}, \cite{ryan}, and
\cite{zwaan_hi} compared local \hi\ observations with the DLA number and argue
that the number of galaxies found in their ``blind'' \hi\ surveys is consistent
with the number densities of DLAs at  redshifts out to $z\approx1.5$.  

Unfortunately, direct measurement of the \hi\ in emission at high redshift is
difficult. Bridging the gap between 21-cm surveys at z$\sim0$ and DLA surveys
at high z, for example, to measure the cosmic \hi\ mass density, requires
deeper and more extensive surveys in the 21-cm line than currently available.
To be useful, such measurements ideally need to (a) have better accuracy than
existing measurements; (b) preferably not be tied to `biased' optical samples
or galaxy overdensities; (c) be over a large enough cosmic volume that cosmic
variance is not a limitation; (d) be at a significant redshift or look-back
time; and (e) sample the \hi\ mass function below the `knee' where the bulk of
the gas may may locked up.  Unfortunately, achieving all of the above probably
awaits the Square Kilometre Array (SKA).  However, many of the above goals can
be partially achieved with a deep survey of a small field with the ALFA
instrument on the Arecibo telescope.  {Such a survey  is  currently in progress
at Arecibo.  This ``Arecibo Ultra Deep Survey'' (AUDS) is the deepest blind
survey ever  carried out with the Arecibo telescope, the world's largest single
dish radio telescope.  The full survey should allow the accurate measurements
of the HI mass function up to the redshift limit of ALFA of z$\approx0.16$,
corresponding to a look-back time of about 2 Gyrs or roughly 25\% of the time
between z$\approx1$ and z$\approx0$.  In this paper, we report on a precursor
project for AUDS which tested our ability to achieve the very low noise levels
required by the full survey.}

\section{Arecibo Ultra Deep Precursor Survey}

\subsection{Motivation} 

The goal of the AUDS is to carry out a blind 21-cm line survey at the highest
redshift possible using the  Arecibo L-Band Feed Array (ALFA).  A fundamental
requirement for the survey is the ability to survey a large volume to the
highest practical redshift, which is given by the bandpass of the ALFA receiver
and is about z$ \approx 0.16$.  Sampling galaxies that  contribute {most of the
gas mass to the \hi\ mass function at z$\approx0$ requires sufficient
sensitivity to detect some galaxies in the flat part of the luminosity
function, i.e.  galaxies down to at least $10^{9}$ M$_{\odot}$.}

With a linewidth of 200 \kms, a galaxy with $M_{HI} = 10^{9.5}$ M$_{\odot}$ has
an integrated flux of {0.03 Jy km s$^{-1}$ and a mean flux density of only
140$\,\mu$Jy at the far end of our redshift range. A sensitivity of 50$\,\mu$Jy
is necessary to detect such a galaxy with at least 3$\sigma$ significance per
channel.}  AUDS will use a total of 1200 hours of observing time  to survey a
total area of 1/3 square degrees with such a sensitivity. {The integration time
per pointing will be about 40 hours.} This  will be the longest integration
time per pointing of any 21-cm survey ever conducted with the Arecibo
telescope.

Before committing this large amount of observing time, we carried out precursor
observations to test the feasibility of extremely deep surveys with the Arecibo
telescope.  These precursor observations were carried out from October 2004 to
February 2005. The maximum redshift at which \hi\ line emission can be probed
by AUDS is z$ = 0.16$ and is set by the ALFA multibeam receiver, which operates
in the 1225$-$1525 MHz frequency range. However, the lower end of such interval
(i.e., frequencies below 1270 MHz, or z$>0.12$) is heavily affected by RFI.
Thus we carefully chose a suitable target field with previously measured
optical redshifts that would  guarantee a few \hi\ detections across the whole
redshift range and that would allow us to test our algorithms of signal
extraction in presence of strong RFI.  Note that for this precursor survey an
interim backend with a total bandwidth of 100 MHz was used. The  redshift range
from $z = 0$ to $0.07$, which will be sampled by the full survey, was therefore
not accessible by the precursor survey. 

\subsection{Mapping Strategy}

To reach the deepest possible observations with the multibeam array ALFA, we
used a ``drift and chase'' mode to cover the survey region in a uniform manner.
For that purpose, we rotated the feed array to an orientation so that a single
position on the sky drifts through the three central beams. The basic strategy
was to point 10 arcminutes ahead of the target field with the central beam, and
then let the sky drift past for 100 seconds of time (or 20 arcminutes).
Subsequently, the telescope drove back to the starting point and the whole
procedure was repeated.  The feed array remained stationary during each drift,
and rotated to track the sky only between drifts. This strategy is ideal for
Arecibo because of its sensitivity to ground spillover. Keeping the orientation
and therefore the spillover constant during each drift results in the best
possible baselines. At the same time, the ``on'' and ``off'' source integration
are as close as possible in time and allow removal of temporal changes in
receiver gain and interference.

\begin{figure}
\plotone{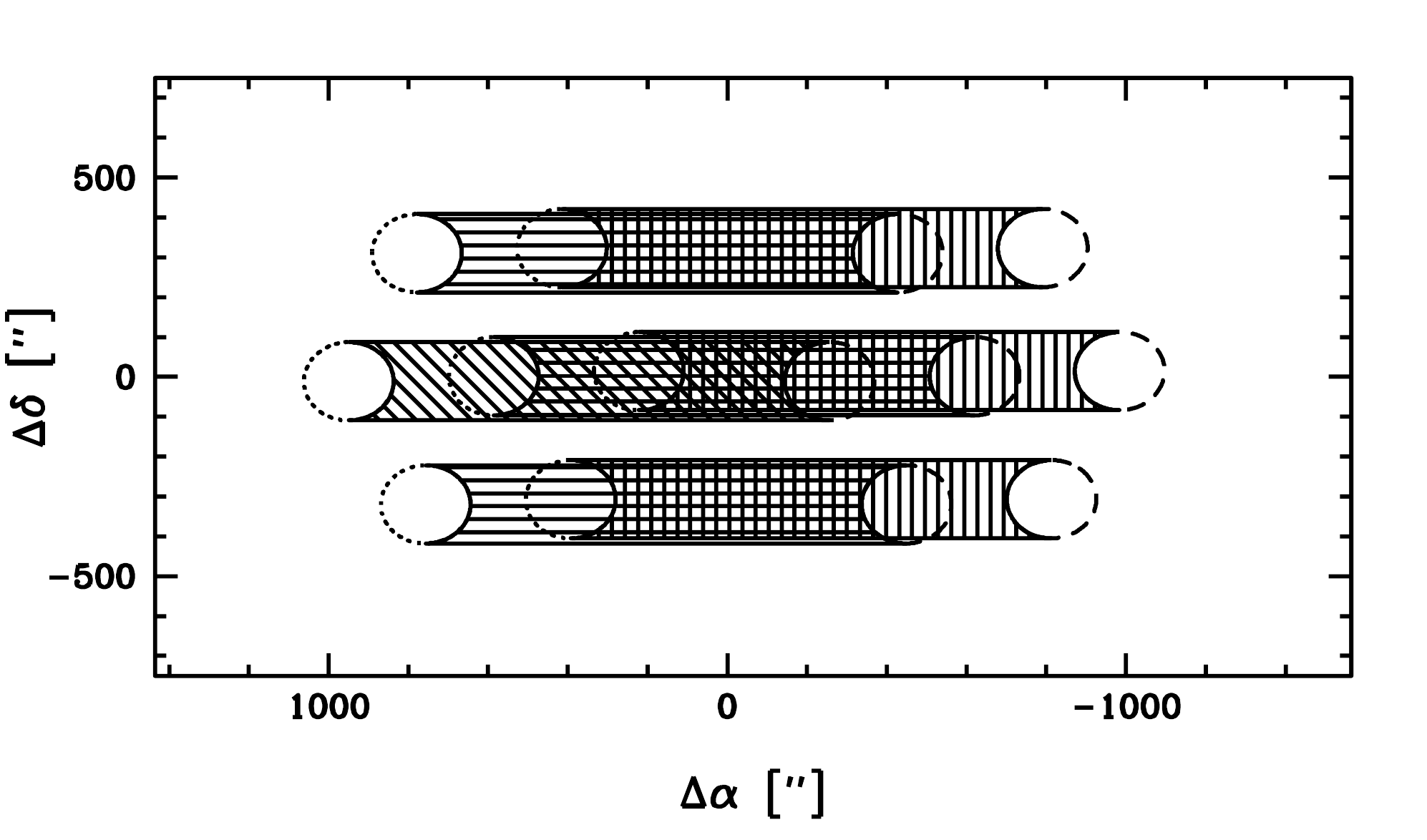}
\caption{Footprint of the AUDS precursor survey. Dashed circles are the
positions of the seven ALFA beams at the start of a scan, and dotted line
at the end of a scan. The central of the three strips is covered by
three beams, whereas the outer two strips are only covered by two
beams.
\label{fig:footprint}}
\end{figure}

An additional advantage of drift scanning is that positions and fluxes of
sources can be measured at the full telescope resolution at least in the
scanning direction. The goal of the full AUDS will be to obtain {fully sampled
2-d maps. This is achieved by introducing  offsets of about 1/10th of the beam
width between the individual drifts.}   However, to reach the highest possible
sensitivity for the precursor observations, we repeated all our scans at
exactly the same Declination. The resultant sky coverage  therefore consists of
three disjoint strips, each of length 1500 arcseconds.  This footprint is shown
in Fig.~\ref{fig:footprint}. Only a small part of the central strip is covered
by three beams in each of the scans. 

\subsection{Target Selection} 

As explained in the previous section, the deepest part of the survey was a
strip with the extension in Declination given by the size of the ALFA beams,
and about 13 arcminutes long in Right Ascension. To make the precursor
observations a useful test of our ability to detect galaxies, we have used the
SDSS DR2 \citep{sloandr2} spectroscopic data base to identify a target field
with the following characteristics: (a) observable from Arecibo during
nighttime; (b) including at least one \hi-rich galaxy in the redshift range
$0.09 \leq z \leq 0.12$, corresponding to the relatively RFI-free 1270$-$1300
MHz frequency window; (c) including at least one \hi-rich galaxy at lower
redshift; and (d) not contaminated by strong radio continuum sources.  In
practice, we generated a list of good candidates for \hi\ detection in the
highest redshift end of our survey, and visually inspected the SDSS images,
looking for presence of additional disk galaxies within a few arcminutes from
the main target.  The selection criteria for {\em good} \hi\ candidates are
similar to those successfully used by \citet{sdsssample} to identify \hi -rich
galaxies at intermediate redshifts (z$>0.16$) detectable with Arecibo.
Basically, we looked for relatively isolated, non-interacting galaxies with
disk morphology, presence of \Ha\ line emission, late-type spectra, and
redshift as close as possible to our limit, z$ = 0.16$.

\begin{landscape}
\begin{figure}
\plotone{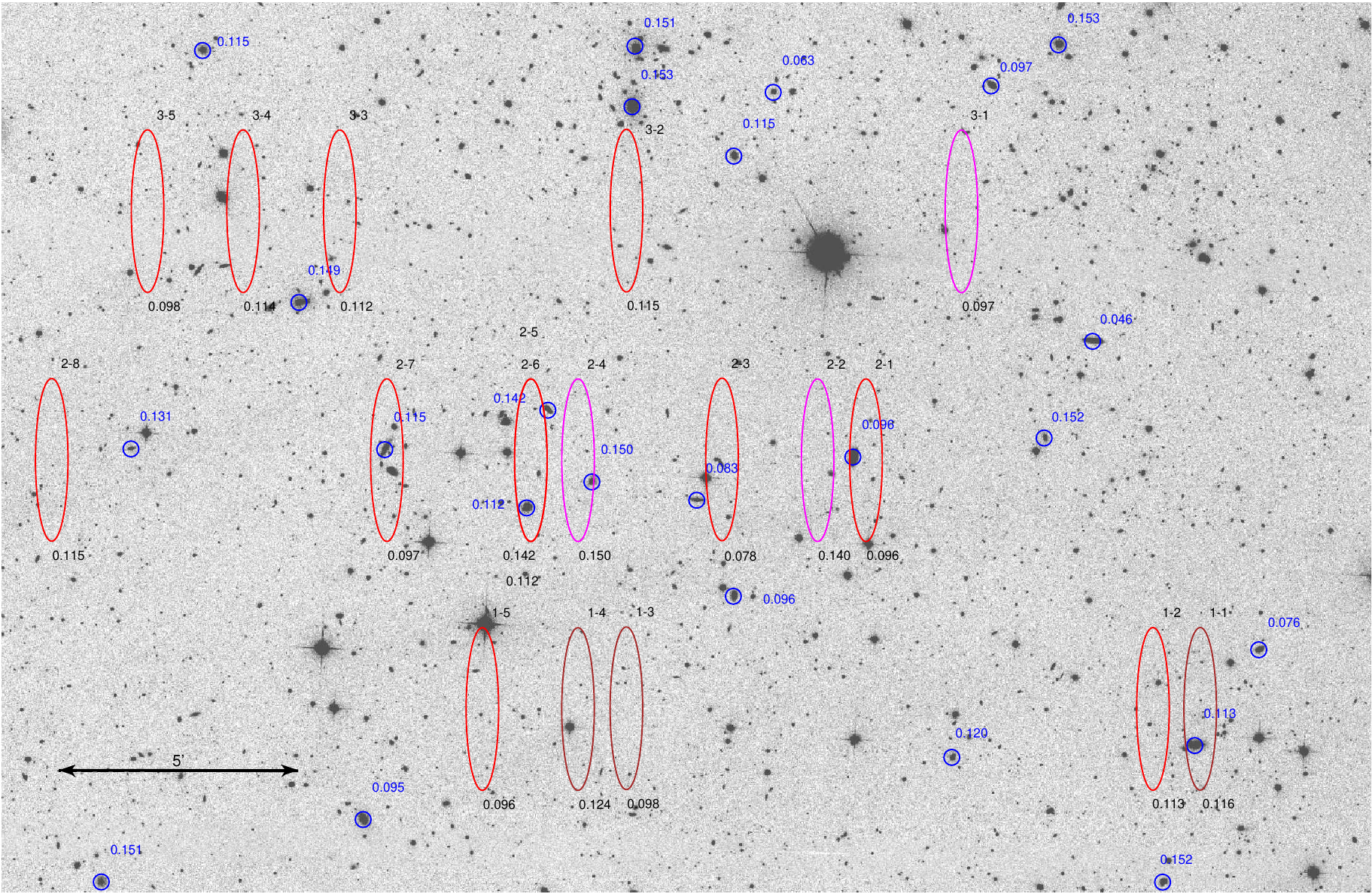}
\caption{SDSS $r$-band image of the AUDS precursor field.
   \hi\ detections are marked with the expected uncertainty in position 
   for high signal-to-noise detections. The colors of the ellipses code
   the quality of the detections, red for the best quality (q=1), brown
   for medium quality (q=2), and magenta for low quality spectra (q=3). 
  Galaxies with available SDSS
  redshifts are marked as circles with their redshift.
   Note that \hi\ can be detected at any
   position within the shown region, and that low signal-to-noise detections have much
   larger positional uncertainties (see text). 
\label{fig:field}
\label{fig:radec} } 
\end{figure}
\end{landscape}

The field selected for our precursor observations is shown in
Fig.~\ref{fig:field}. {There are a total of 31 galaxies with measured redshift
listed in the SDSS DR7 \citep{sloan}. Two of them are in the foreground of our
survey at $z < 0.07$, 17 of them within our redshift range $0.07 < z < 0.15$,
and 14 in background at $z > 0.15$.  There are more than one thousand
additional SDSS galaxies in the field without optical redshifts.  The field
includes several late-type galaxies at different redshifts. Some of them are at
or close to frequencies which are affected by RFI for a large fraction of the
time.}

\subsection{Observations and Data Reduction}

We used the {\sl Wideband Arecibo Pulsar Processors} (WAPPs)  with 100 MHz
bandwidth and 4096 channels (3-level sampling).  This resulted in a channel
separation of 24.4 kHz (5.2 km\,s$^{-1}$ at z$ = 0$).  For this experiment, the
WAPPs were tuned to a central frequency of 1275 MHz, giving coverage over the
range 1325 -- 1225 MHz (corresponding to z$  =  0.07 $--$ 0.16$).

We spent a total 53 one-hour sessions testing our ``drift-and-chase'' procedure
on this target field. In the most sensitive part of the survey, the accumulated
integration time per pointing was about 14 hours.  The total telescope time
spent to carry out these observations was about 50\% higher than the actual
time spent on the drift scans.  A significant fraction of these overheads was
due to time losses at the start of each observing session and before and after
each drift scan. {For the full AUDS survey, overheads are expected to be about
25\% since individual strips are longer. }

\begin{figure}
\plotone{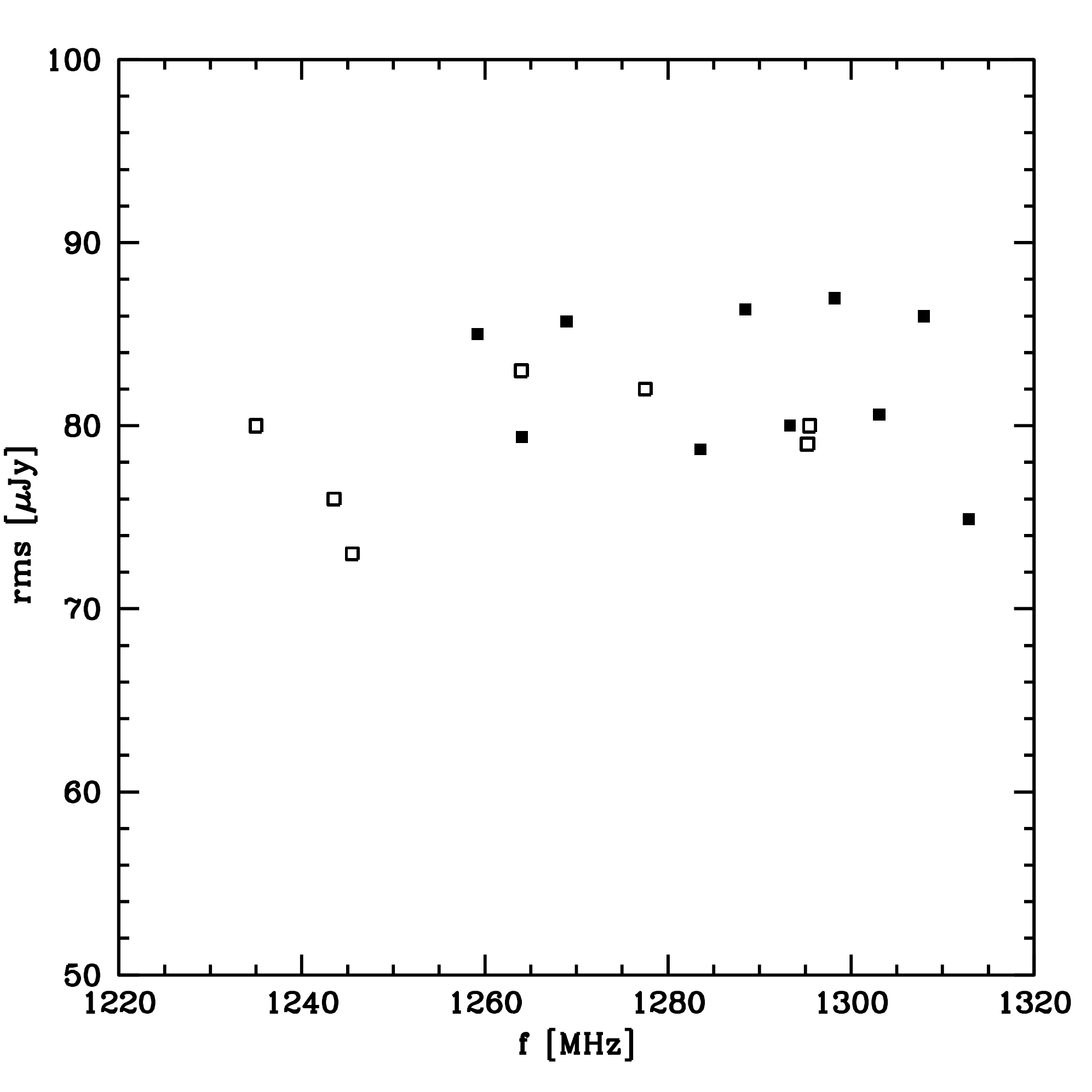}
\caption{{Measured rms in unsmoothed spectra with exposure times longer than
13.5 hours. All measurements were performed in parts of the spectra without line
emission or RFI. Solid squares are measurements on extracted galaxy spectra close to the detected \hi\ lines, 
whereas open squares are measurements at the center of the field.} \label{fig:freq_rms}}
\end{figure}

\begin{figure}
\plotone{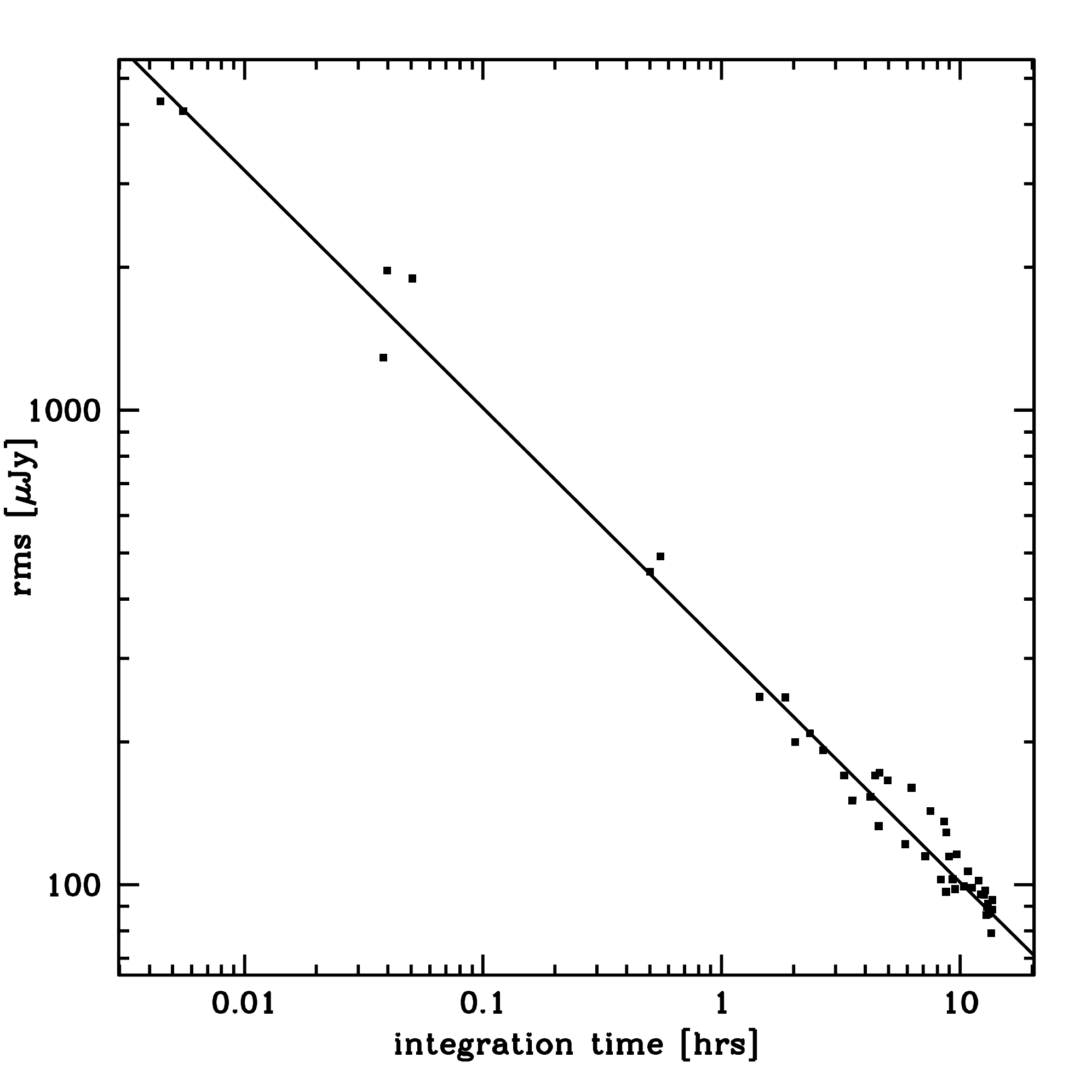}
\caption{Measured rms {at 1370 MHz} in unsmoothed spectra along the central strip of the survey
as a function of the total accumulated integration time. The line
is a fit to the data of $\sigma\propto\sqrt{1/t_{\rm int}}$}.\label{fig:rms}
\end{figure}

The measured rms in the spectra of the most sensitive part of the precursor
survey is about 80\,$\mu$Jy per 24.4 kHz channel {throughout the bandpass, see
Fig.~\ref{fig:freq_rms}. This is close to the expectations based on system
temperature of 28K, gain of 9 K/Jy, and the accumulated integration time per
pointing of 14 hours. In Fig.~\ref{fig:rms}, we show the measured rms as a
function of integration time at a frequency of 1270 MHz.}  It can be seen that
the rms noise of the averaged spectra declined as $\sigma\propto\sqrt{1/t_{\rm
int}}$ down to the longest achieved integration time.

Data reduction was performed using the {\sl Livedata} and {\sl Gridzilla}
multibeam processing packages.  {\sl Livedata} performs bandpass estimation and
removal, Doppler corrects the data, and calibrates the resulting spectrum
\citep{barnes}.  The packages were modified by one of us (MC) to read Arecibo
data, and new routines to reject RFI were implemented.  Current versions of the
{\sl Livedata/Gridzilla} packages include these routines and are available at
\url{http://www.atnf.csiro.au/computing/software/livedata.html}.  The bandpass
was calculated independently for each drift, beam, and polarization.  A clipped
median was used to compute a reliable calibration factor for each spectral
channel in the presence of strong and persistent RFI. 

After the bandpass calibration, data points with RFI were flagged, both with an
automatic procedure and manually (see below).  Finally, flagged calibrated data
were combined into a data cube using {\sl Gridzilla}.  For the precursor
experiment, the array configuration was optimized to provide depth rather than
uniform coverage, with the beams forming three independent strips separated in
Declination.  In order to grid this efficiently we chose a non-isotropic
grid-size, with pixels 1 arcminutes in Right Ascension and 5.2 arcminutes in
Declination.  This gave one row of pixels for each strip.  The cube was formed
by median-combining the data from the calibrated spectra, weighted so as to
correctly reconstruct the flux of a source at the pixel center.  The final data
cube had a size of 45 pixels in Right Ascension,  3 pixels in Declination, and
4096 spectral channels.

\subsection{RFI}\label{sec:rfi}

\begin{figure}
\plotone{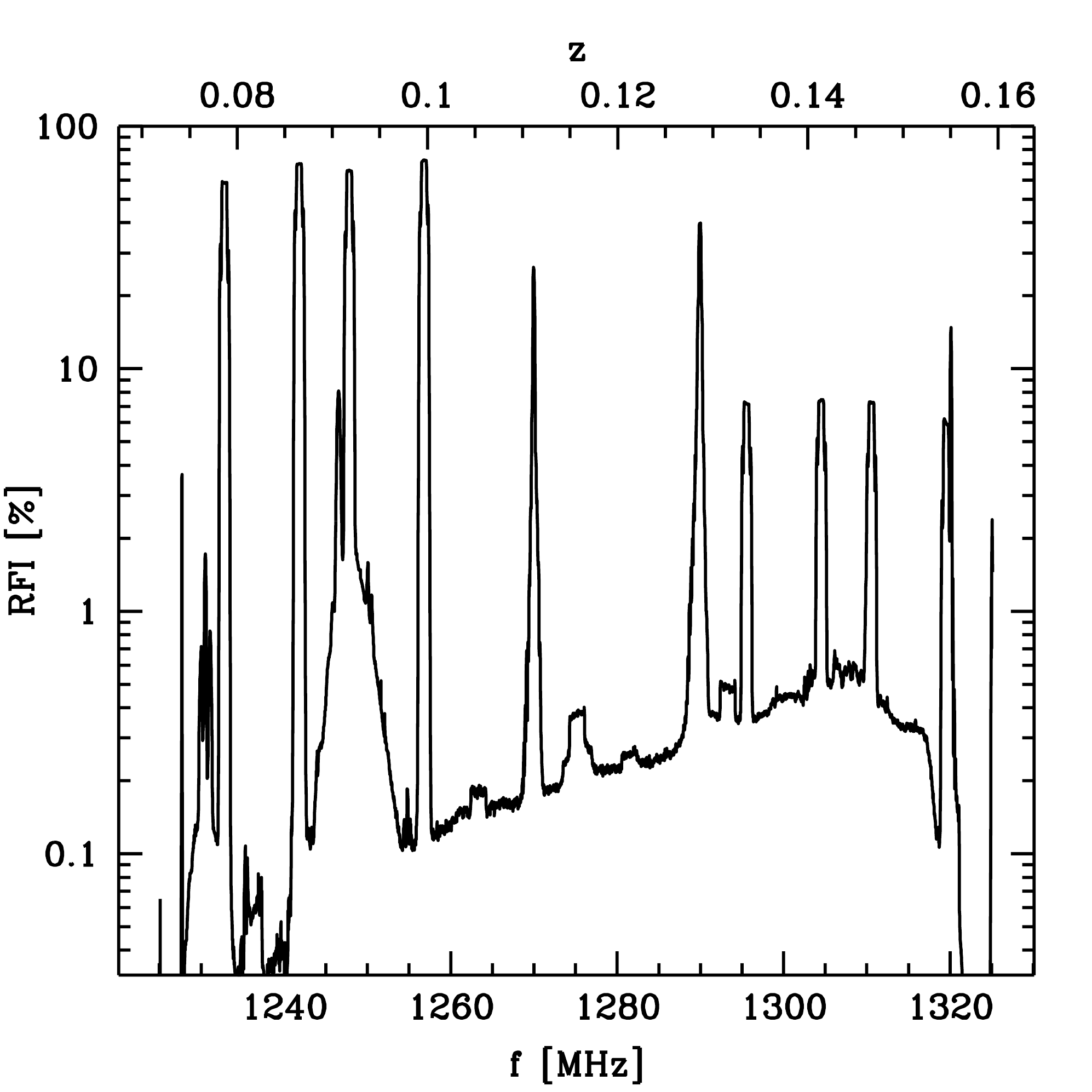}
\caption{RFI occupancy versus frequency.}\label{fig:rfispectrum}
\end{figure}

All precursor data are impacted to some extent by RFI at various frequencies.
In about 70\% of all collected data, RFI was limited to well known frequencies
as listed in the observatory RFI catalog.  In about 30\% of the data, RFI at
unusual frequencies occurred, sometimes wiping out large fractions of the
bandpass. Our strategy for removing RFI was twofold. First, we implemented an
automatic RFI removal routine based on 3.5-sigma clipping of the data smoothed
in time-frequency space. Subsequently, we carefully edited out data sets with
RFI not removed by our automatic procedure. The fraction of data rejected
because of RFI as a function of frequency is shown in
Fig.~\ref{fig:rfispectrum}. It can be seen that RFI appeared at all observed
frequencies during the observations. About 6\% of the band is severely affected
by RFI with 10\% or more of the data points being flagged. {This corresponds to
about 8\% of the sampled volume affected heavily by RFI.}

{One concern for any deep survey are possible spurious detections of line
emission caused by unrecognized  RFI.} Such a situation is most likely to
happen close to spikes in the RFI spectrum. For example, the detected sources
2-4 and 2-6 are close to RFI spikes (see section \ref{sec:detection}).  We
therefore re-examined this part of the spectrum and did not find any indication
of undetected RFI.  Later, we found that 2-6 was also detected in the SDSS at
the exact redshift we determined from the \hi\ observations. This confirms our
ability to detect lines even in spectral regions close to RFI spikes.

\section{Results}

\subsection{Detection of \hi\ Emitters}\label{sec:detection}

\begin{figure} 
\epsscale{1.1}
\plottwo{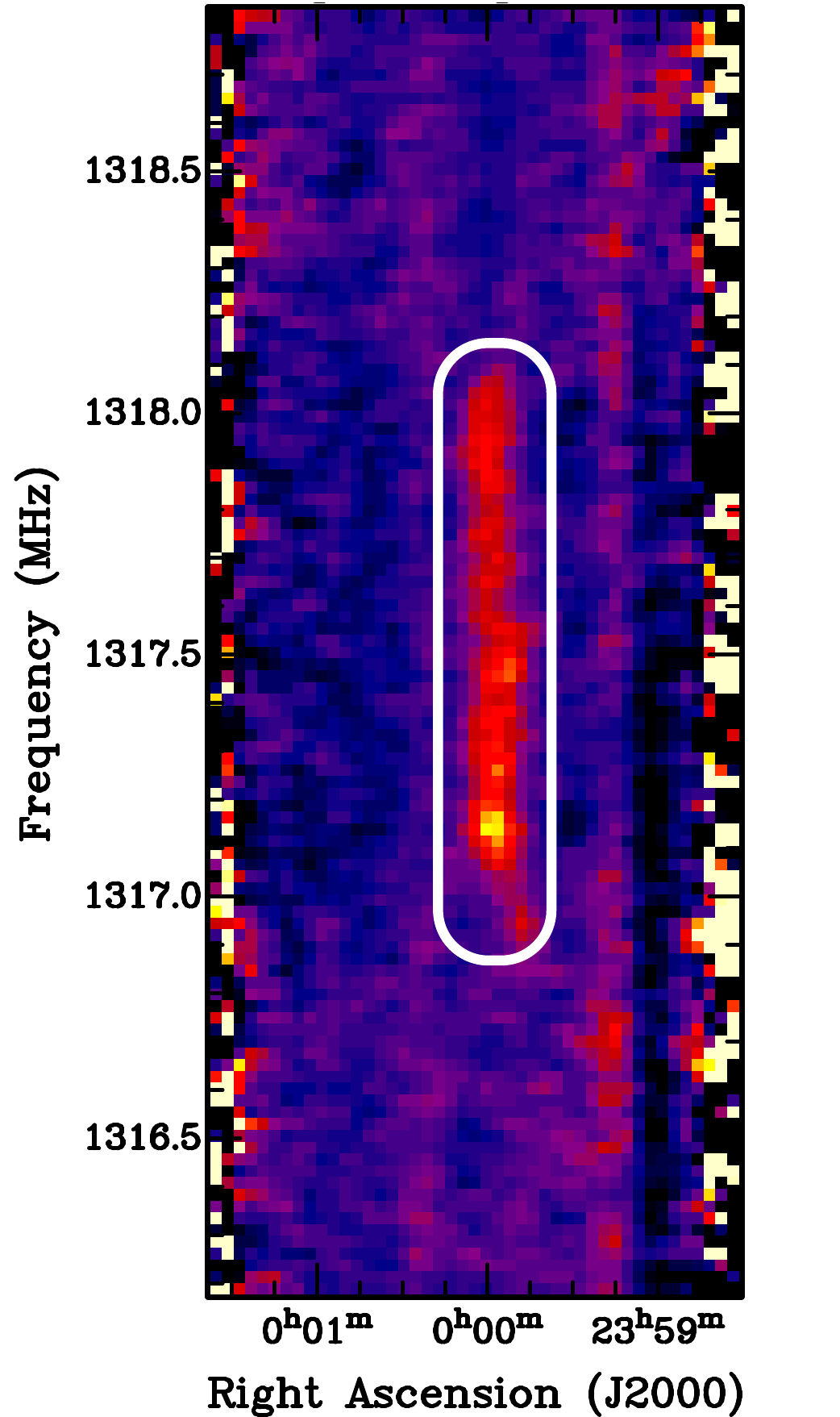}{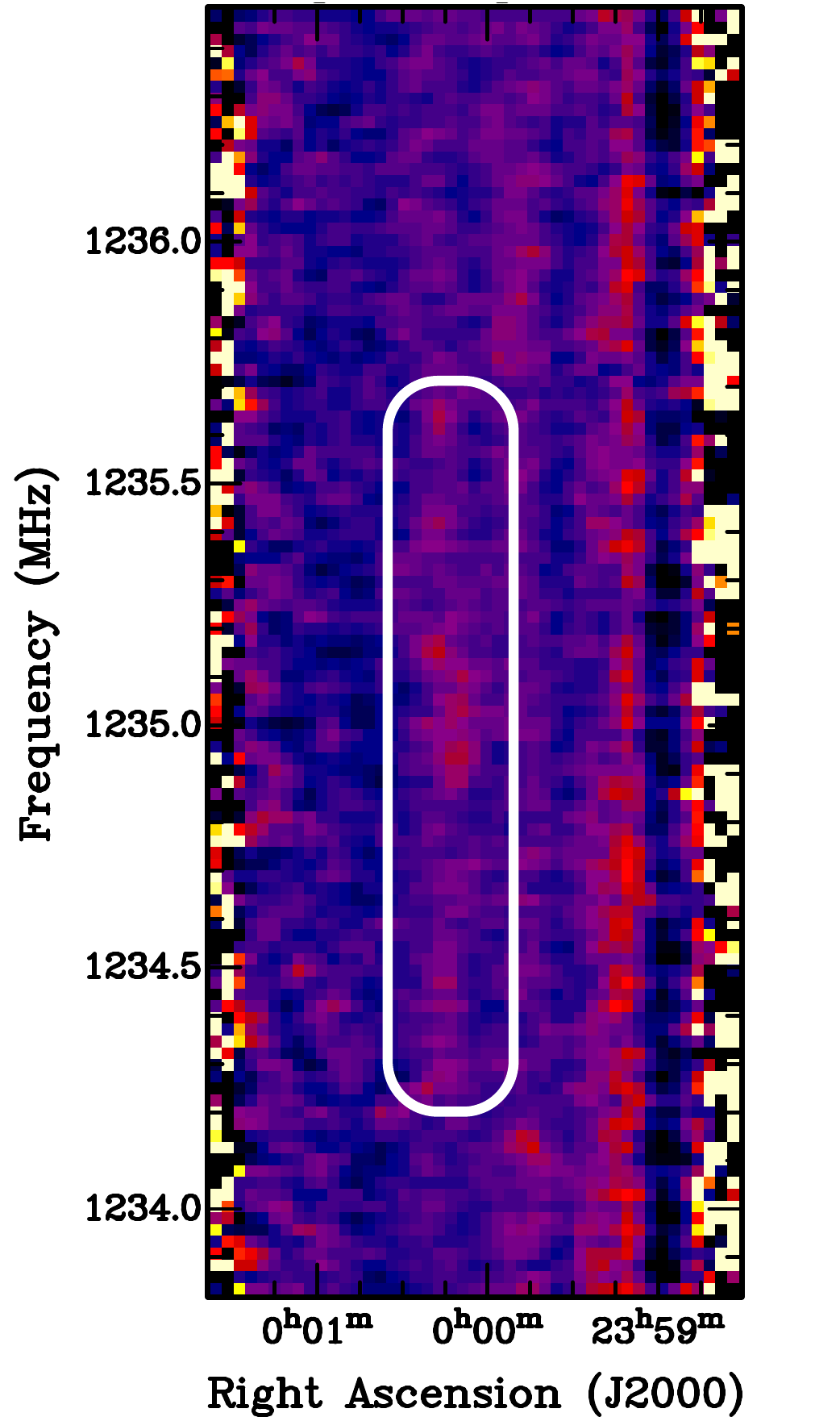} 
\caption{Examples of frequency -- Right Ascension images used to search for 
HI emission. The two images are grayscale plots around a high signal-to-noise
ratio source  (AUDS 2-3, left) and low signal-to-noise ratio source  (AUDS 2-4,  right).
The detected lines are encircled in each image.
} 
\label{fig:candidates} \end{figure}

To search for \hi\ emission lines in the final data cube, we created grayscale
plots of the flux density  as a function of {Right Ascension and frequency} for
each of the three strips. Each image was searched by eye for candidate \hi\
emitters. Examples of two such candidates are shown in
Fig.~\ref{fig:candidates}. Individual scans for each candidate were carefully
inspected to minimize the probability of including spurious RFI into the
candidate list.  We found \nhitotal\ detected \hi\ lines with a peak
signal-to-noise ratio in the unsmoothed spectrum of at least 3, {and a total
width of the line of at least 20 spectral channels. Since the channels are
mostly independent, the total signal-to-noise ratio of these lines is at least
about $2\sigma\times\sqrt{20}\approx9.$ } 

{We assigned a subjective quality code to each detected line, ranging from 1
for the best detections, to 3 for detections that can not be ruled out to be
spurious.   The criteria for the quality code were as follows.  Isolated, high
signal-to-noise ratio lines were assigned a quality code 1. Any line spectra
for which the uncertainties in the baseline were judged to be a significant
fraction of the detected line were assignd a quality code of 3. Spectral lines
in this category have low signal-to-noise ratios, large variations in the
baseline and/or are uncomfortably close to RFI spikes. Spectra for which such
an assignment was difficult to make were assigned the quality code 2. We
suspect that several of the detected lines of the lowest quality are spurious,
but note that the redshifts of two of them (AUDS 2-4 and 3-1)}  were later
confirmed by optical redshift measurements in the SDSS.

\begin{figure} 
\epsscale{1.0}
\vskip-1truecm
\plotone{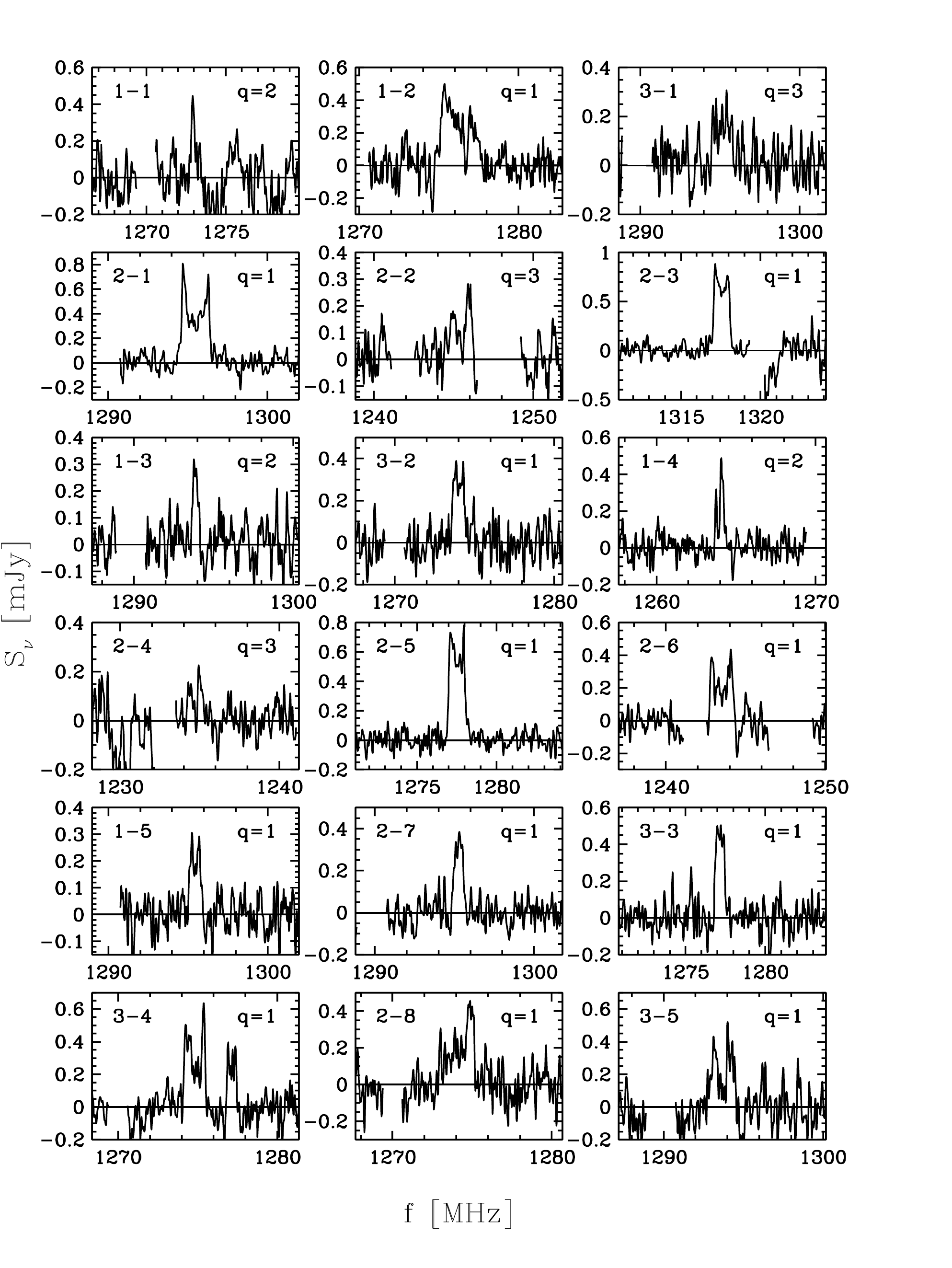} 
\vskip-1truecm
\caption{Spectra  of all eighteen \hi\ emission lines detected in the AUDS
precursor survey. A three channel Hanning smoothing and subsequent 3 channel
boxcar smoothing has been applied to each raw sepctrum.  Regions where RFI was
present more than 5\% of the time are not plotted because of the high
probability residuals from removed RFI. Each panel is labeled with the name of
the source $n$--$m$ and the quality code, which ranges from 1 for the best
spectra to 3 for the lowest quality spectra.} \label{fig:spectra} 
\end{figure}

{For each detected line source, we extracted a total spectrum by summing up all
adjacent  spectra where flux at the corresponding redshift was detected with at
least 3$\sigma$ significance.  The Right Ascension of the source was computed
as the center of flux from the positions with line emission.}  The extracted
spectra for all \nhitotal\ detections sorted by increasing Right Ascension are
shown in Fig.~\ref{fig:spectra}. Each detected source is named $s$--$n$, where
$s$ is the number of the strip 1, 2 or 3, and $n$ is a running number within
the strip in order of increasing Right Ascension.  Fig.~\ref{fig:radec} shows
the distribution of the sources on the sky. Each source is  marked by an
ellipse which indicates the uncertainty in the position for high
signal-to-noise sources. The positional uncertainty is discussed in more detail
in Appendix~A.  

\subsection{Analysis of Extracted Spectra}\label{sec:extract}

Frequencies were converted to velocities using standard relations
\citep[e.g.][]{velocities}, and \hi\ line parameters were estimated using the
MBSPECT routine in the MIRIAD package \citep{miriad} on the spectra after
applying a 15 channel Hanning smoothing.  Errors were estimated as in
\cite{miriaderr}.  The measured parameters are listed in
Tab.~\ref{tab:hiresults}.  Total \hi\ content was estimated from the integrated
line flux  using the standard relation

\begin{equation}
        M_{\rm HI}= 2.356\times10^5\  {{\rm M}_{\rm \odot}\over1+z}  \   {d_{\rm lum}^2\over{\rm Mpc}^2} \ \int  {I\over{\rm Jy}}\ {dv \over{\rm (km/sec)}}
\end{equation}
where $d_{\rm lum}$ is the luminosity distance, and $v$ is measured optical velocity.

\begin{figure}
\plotone{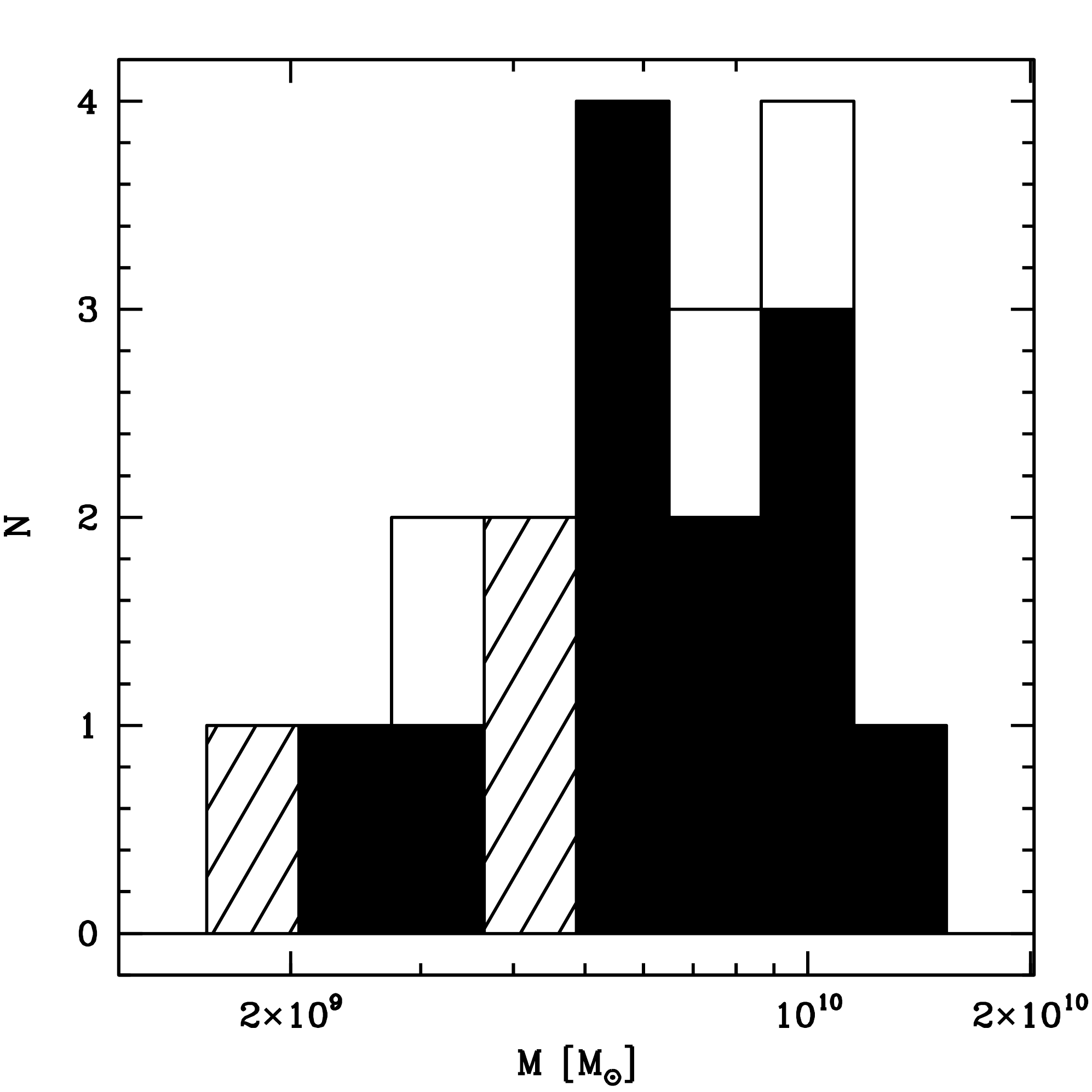}
\caption{Mass distribution of \hi\ lines detected in the AUDS precursor run.
The filled  parts of the histogram are the highest quality  spectra, the
shaded part is the medium quality, and
open parts are the lowest quality spectra.
\label{fig:mhist}}
\end{figure}

The AUDS precursor geometry of  three strips along constant Declination leads
to a substantial positional uncertainty for the detected galaxies which will
not be present in the better sampled full survey.  {The sensitivity between two
strips is only a few percent compared to the center of a strip, and the strips
are largely independent. } We estimate  the listed Right Ascension to be
accurate to about 1/5th the HPBW of the beam, i.e. about 40 arcseconds for high
signal-to-noise detections in the central part of each strip. The position in
Declination is much less accurate, galaxies  with large \hi\ masses  can be
detected even if they are located between the strips. In Appendix~A, the
positional accuracy is discussed in more detail.  Our inability to locate the
galaxies precisely causes a bias in the mass estimate which does not exist in
better sampled maps.  Since the center of the beam is more sensitive than
positions towards the edge, the positional uncertainty also leads to an
uncertainty in the measured flux. \hi\ masses computed using the gain for the
center of the beam are therefore in most cases an underestimate of the true
mass.  We estimated the bias from this effect by using a Gaussian approximation
for the shape of the ALFA beams \citep{beam} and the redshift of each detected
source to compute the total \hi\ mass corresponding to the detected signal as a
function of Declination, $M_{\rm HI}(\delta)$.  We then used the \hi\ mass
function $\Phi(M_{\rm HI})$ by \cite{zwaan2003} to derive the probability
$P(\delta)$ that the detected source is at that position,

\begin{equation}
        P(\delta) =  {\Phi(M_{\rm HI}(\delta))\over\int_{ -\pi/2}^{\pi/2} \Phi(M_{\rm HI}(\delta)) {\rm d} \delta}.
\end{equation}
Finally, we computed the expected \hi\ mass as 

\begin{equation}
    \MHI^{\rm corr} =  \int M_{\rm HI}(\delta) P(\delta) d \delta
\end{equation}
The corrected \hi\ masses range from 2 to $11\times10^{9}$~\Msun, the detailed
distribution is shown in Fig.~\ref{fig:mhist}.

{For six sources in our sample, a unique optical counterpart has been found
(see Sec.~\ref{sec:sdss}). For those sources, the accuracy of the bias
correction can be independently checked. One of the sources, AUDS 3-1, is at a
position about 2.6 arcmin from the center of the strip. At this position, the
sensitivity of individual beams is only about 10\%, but the effective beam
shape of the drift scans is highly uncertain. Our model predicts that the
probability that the detected galaxy is located that far from the center of the
beam is less than 1\%.  The other five optical counterparts are on average at a
distance of 0.66 arcmin from the center of their strip, and the average flux
correction relative to the center of the strip is a factor of 1.15. Our bias
model predicts an average most likely distance from strip center of $0.65$, and
an average correction factor of $1.20$. We therefore conclude that our flux
correction significantly reduces the biases in the mass estimate based on the
sensitivity at the center of each strip.}

The measured and derived \hi\ parameters are listed in
Tab.~\ref{tab:hiresults}. The first line for each galaxy lists the
measurements, and the second line gives the $1\sigma$ error estimates for each
parameter.  In detail, the individual columns contain the following: (1)
assigned name in the form $n$--$m$, where $n$ is the number of the Declination
strip, and $m$ is the running number assigned to each galaxy in order of
increasing Right Ascension; (2) and (3)  coordinates of the detected source;
(4) measured noise in the unsmoothed spectrum close to the detected line; (5)
peak signal to noise ratio of the unsmoothed line spectrum; (6) a quality code:
1 - best detection; 2 - likely detection but possibly affected by RFI; 3 -
uncertain detection because of low noise or RFI; (7) center frequency of line;
(8) optical heliocentric velocity centroid, i.e. $c\cdot z$, measured at 50\%
of peak flux; (9) redshift z; (10) luminosity distances using $H_o = 72$
km/sec/Mpc, $\Omega_m = 0.26$, and $\Omega_\lambda = $ 0.74.; (11) observed
velocity width of line measured at 20\% of the peak flux; (12) integrated line
flux; (13) raw \hi\ mass estimate; (14) \hi\ mass estimate corrected for
position bias. {For sources 1-2, 2-1, 2-4, 2-5, 2-6, and 3-1, column (14) lists
the mass computed assuming the source position is identical to that of its SDSS
counterpart.}

\subsection{Redshift Distribution and SDSS Identifications}\label{sec:sdss}

\begin{figure}
\plotone{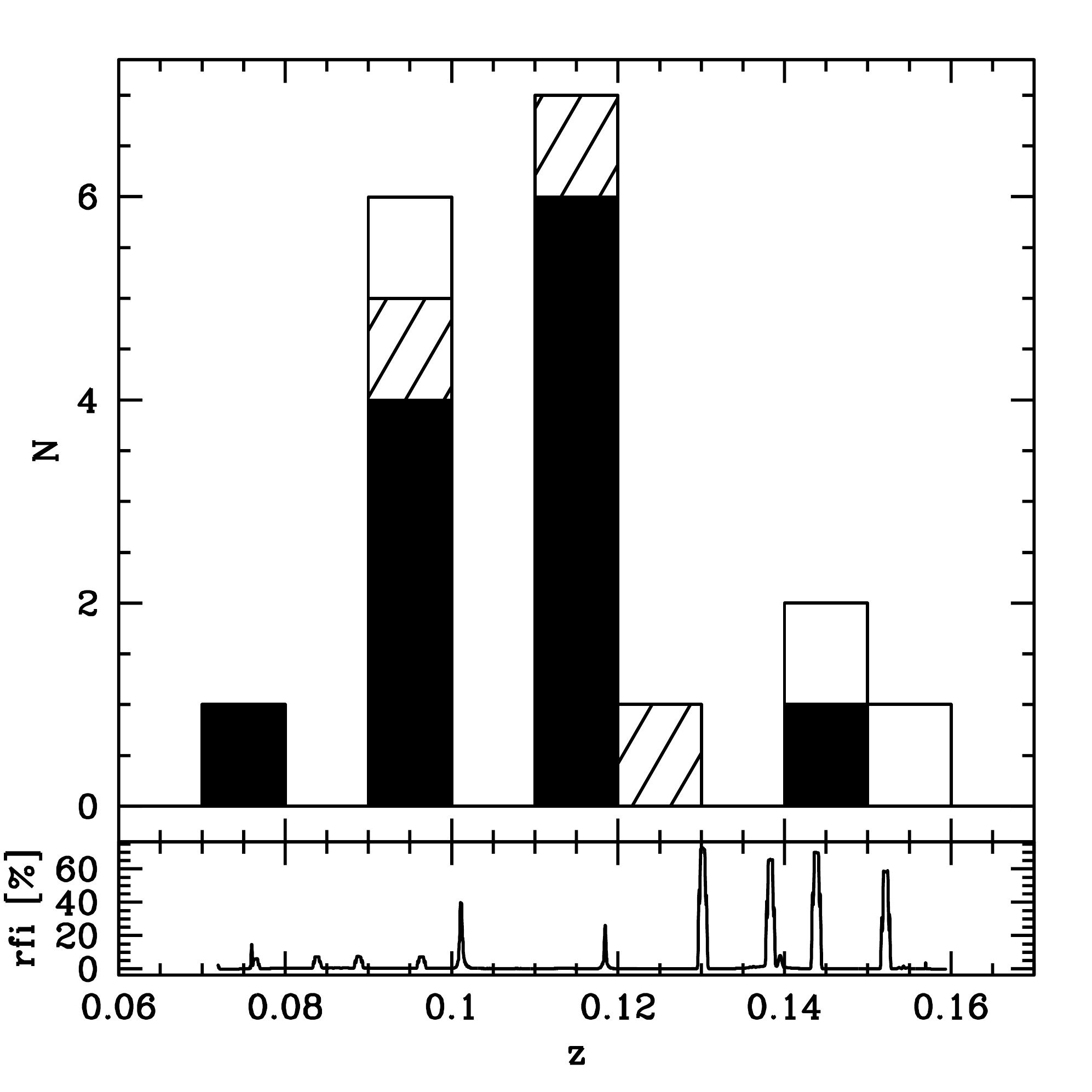}
\caption{Redshift distribution of \hi\ lines detected in the AUDS precursor run.
The filled  parts of the histogram are the highest quality  spectra, the
shaded part is the medium quality, and
open parts are the lowest quality spectra. {For comparison, the RFI
occupancy shown in Fig.~\ref{fig:rfispectrum} is replotted in the lower panel}.
\label{fig:zhist}}
\end{figure}

The redshifts of the detected galaxies range from 0.07 to 0.15, see
Fig.~\ref{fig:zhist} for the redshift distribution.  Nine of the detected
galaxies plus six more candidates are at a redshift above 0.1. This sample of
at least nine \hi\ rich galaxies at z$>0.1$ is one of the largest field samples
of \hi\ sources at that redshift range.  Fig.~\ref{fig:radec} shows the
position and redshifts of the AUDS precursor sources, and also shows all
spectroscopic sources with redshifts in the SDSS DR7 \citep{sloan}. 

{The six \hi\ detections labeled $1-1$, $2-1$, $2-4$, $2-5$ and $2-6$ which
overlap in position, and 3-1 all have SDSS counterparts that match both in}
position and redshift.  In order to search for plausible candidate galaxies for
detections without matching redshifts in the SDSS, we carefully searched the
beam area around each detection for optical galaxies in the SDSS.  The details
of the search are given in Appendix~A, where we  present images obtained with
the SDSS finding chart tool at
\url{http://cas.sdss.org/dr7/en/tools/chart/chart.asp} for each source.
Plausible candidates for the \hi\ emission are marked.  At least one possible
counterpart was found for each of the \hi\ detections.

\begin{figure}
\plotone{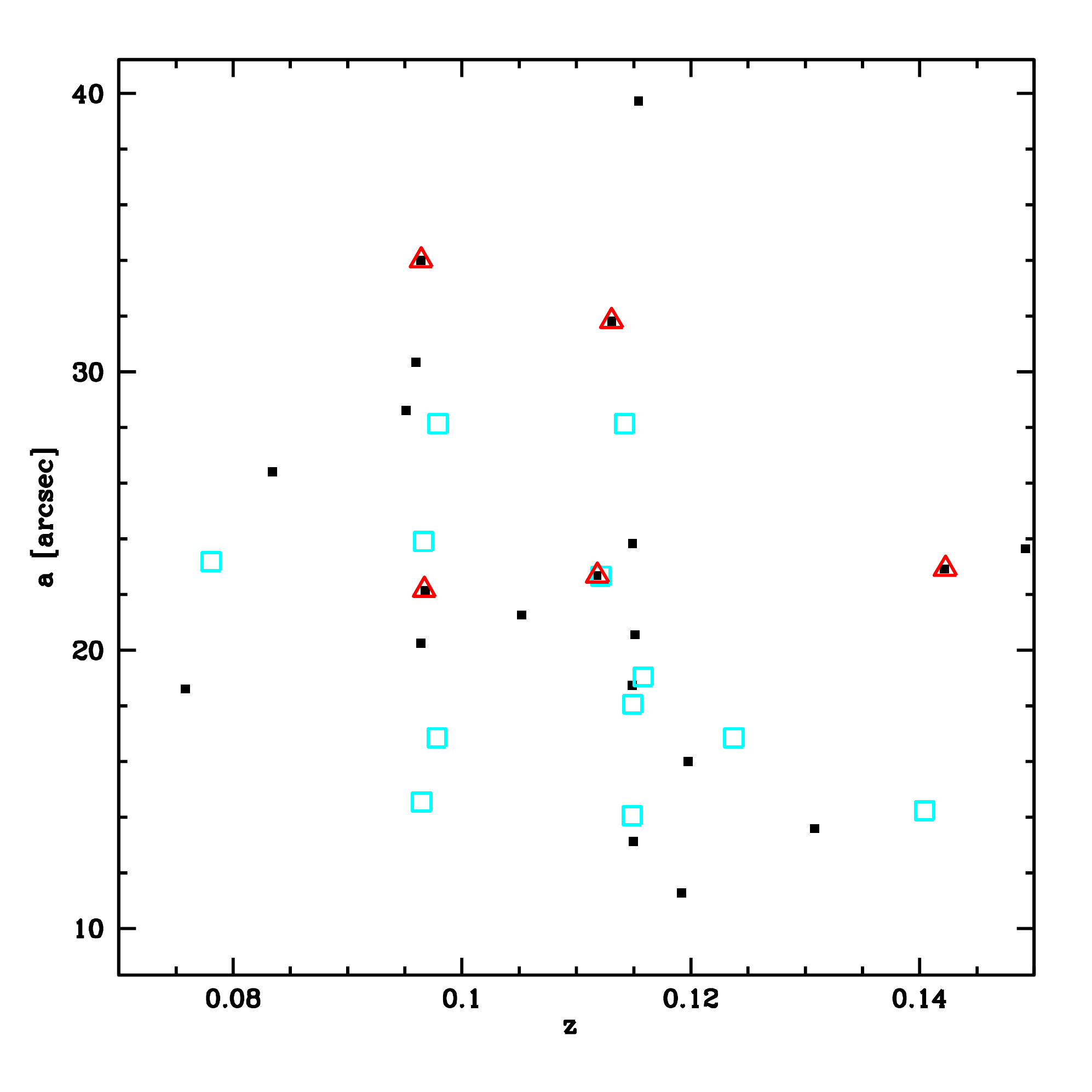}
\caption{Distribution of $r$-band isophotal major axes versus the redshift of
\hi\  and SDSS sources within the target region. Open triangles are the 
\hi\ sources with  SDSS counterparts confirmed by redshift, open squares those
without such confirmation, and solid squares are SDSS galaxies in the AUDS
precursor region.  For sources without counterparts confirmed by redshift, the major
axis of the largest
plausible candidate in the field was used.
\label{fig:z}}
\end{figure}

\begin{figure}
\plotone{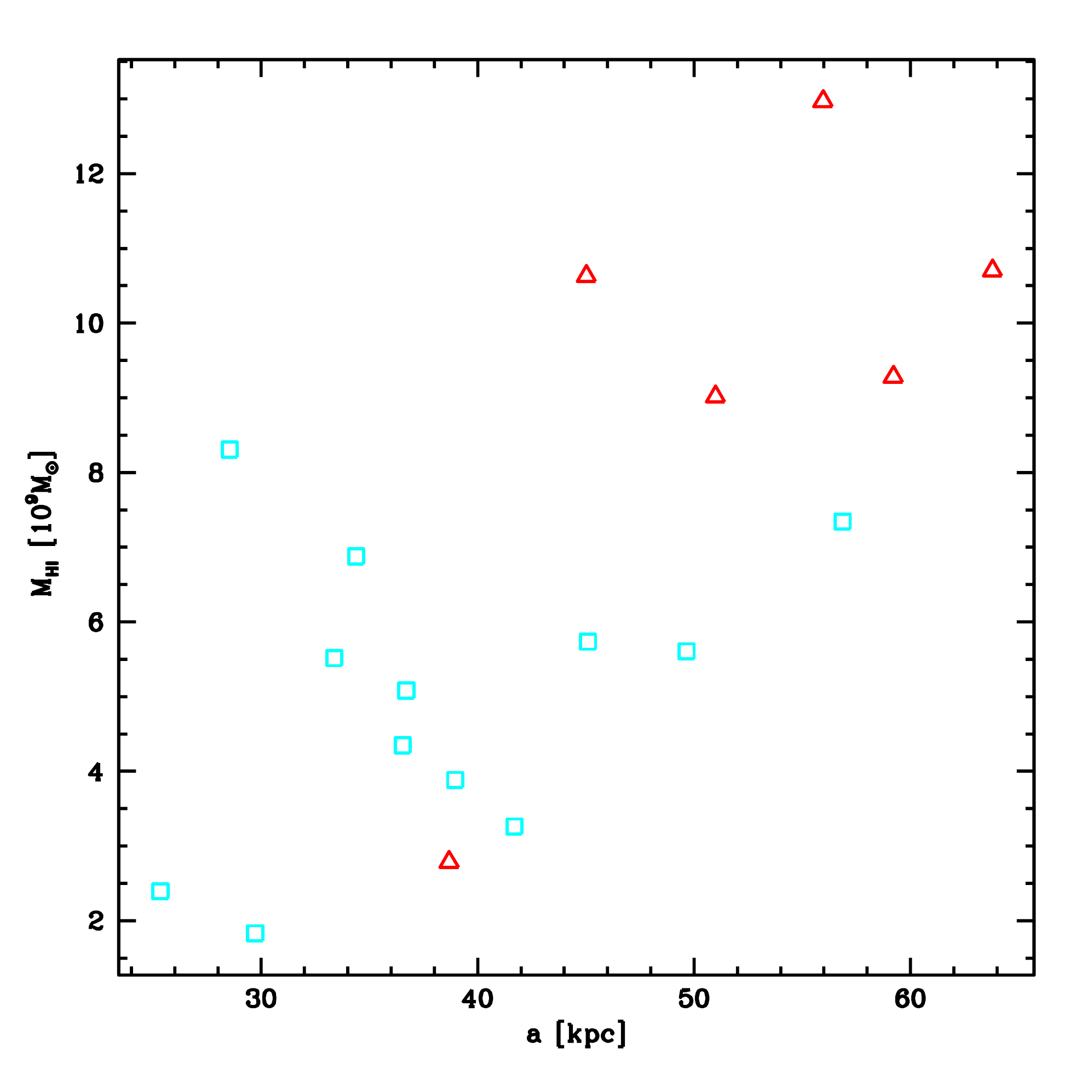}

\caption{Scatter plot of the estimated \hi\ masses versus the $r$-band
isophotal major axes.  Open triangles are the \hi\ sources with SDSS
counterparts confirmed by redshift, and open squares represent those without
such confirmation.  For sources without counterparts confirmed by redshift, the
major axis of the largest plausible candidate in the field was used to compute
the restframe size.\label{fig:masses}}

\end{figure}

To investigate whether the sizes of the identified candidates makes them
plausible sources for the \hi\ emission, we plotted the $r$-band isophotal
major axes as listed in the SDSS as a function of redshift (Fig.~\ref{fig:z})
and estimated \hi\ mass (Fig.~\ref{fig:masses}).  For \hi\ detections without
confirmed SDSS redshift, we used the major axis of the largest of the
candidates.  Fig.~\ref{fig:z} shows that for all \hi\ detections there are
candidates with sizes similar to other SDSS galaxies at that redshift.
Fig.~\ref{fig:masses} shows that most candidates with unknown optical redshift
are smaller than the redshift confirmed SDSS counterpart. These smaller sizes
are plausible given the smaller \hi\ masses of the unconfirmed candidates.  We
therefore conclude that for all our \hi\ detections, plausible counterparts
exist in the SDSS.

\subsection{Detection Rate}

The survey samples a rather small volume. {The total volume within the area
covered within the half-power sensitivity of at least one feed in each scan and
between a redshift of 0.07 and 0.16 excluding regions heavily affected by RFI
is about 2000 Mpc$^3$.} The number of galaxies detected in such a volume will
vary substantially for independent volumes simply due to cosmic variance.  {We
estimated the expected cosmic variance of the number counts in this volume from
the catalog of \cite{millenium}. {The \hi\ mass function recovered from this
catalog resembles the one  of \cite{zwaan2003}.} We extracted one thousand mock
surveys of galaxies with the same sensitivity and volume shape as the real
data.  We found that the 1$\sigma$ variation of the mean number of galaxies in
the mock survey scatter is about 50\%.}

In addition, the selected volume is not a fair sample of that redshift range
because it was not chosen at random.  Nevertheless, it is illustrative to
compare the detection rate of galaxies to that expected from the local \hi\
mass function. 

\begin{figure}
\plotone{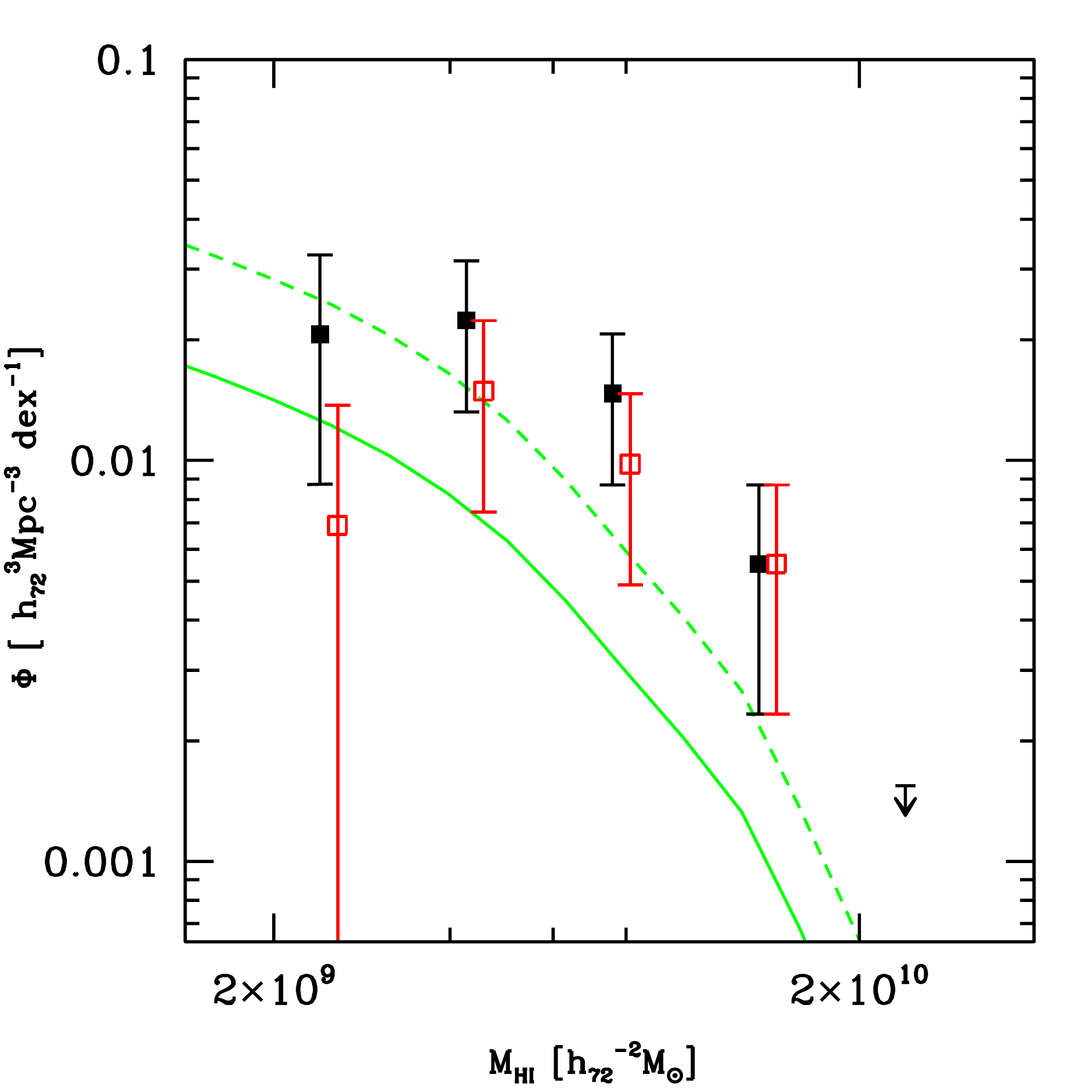}
\caption{\hi\ mass function of detected \hi\ emitters. Filled squares
are the points derived using all detected \hi\ emitters, whereas the open
squares only use the highest quality detection. The solid line
is the local \hi\ mass function by \cite{zwaan_hi}. The dashed line is the local
HI mass function multiplied by the relative overdensity of SDSS
galaxies within the survey region
 (see text). {The arrow marks the resulting density  if a single galaxy
 had been detected in that bin.}}\label{fig:himass}
\end{figure}

For that purpose, we computed the \hi\ mass function from our sample using the
standard 1/$v_{\rm max}$ method \citep{vmax}. {Volumes were computed using the
same Gaussian approximation for the shape of the beam as in
Sec.~\ref{sec:extract}}.  For comparison, we also estimated the \hi\ mass
function assuming a  distribution function for the \hi\ flux and line width.
The results were similar to the ones shown here.  The result with the 1/$v_{\rm
max}$ method  using the full sample is shown as filled squares in
Fig.~\ref{fig:himass}.  The local \hi\ mass function determined from the HIPASS
survey \citep{zwaan_hi} is shown as a solid line. It can be seen that the
{shape of the derived mass function is consistent with the HIPASS \hi\ mass
function.} We used a weighted fit to determine that the relative normalizations
between the two \hi\ mass functions differ by a factor {of 3.2$\pm1.0$.}  We
then recomputed the \hi\ mass function after removing detections with the
lowest quality rating. The normalization of the derived \hi\ mass function from
this reduced sample is still a factor of {2.1$\pm0.7$} higher than the local
\hi\ mass function.  We adopt the mean of the two factors and the mean relative
error, $\delta_{\rm HI} = 2.6\pm0.8$, as our best estimate of the relative HI
overdensity in the AUDS precursor region and its uncertainty.

{For comparison, we estimated the relative density of SDSS galaxies within the
AUDS precursor survey volume. For that purpose, we extracted the SDSS
photometric redshifts for  galaxies with $r$-band isophotal major axes between
5 and 40 arcsec. The selection on the major axes was imposed to eliminate
galaxies with implausible photometric redshifts from the sample.  The limits
were chosen to include all candidate SDSS counterparts, see Sec.~\ref{sec:sdss}
and Fig.~\ref{fig:z}. The results presented below are insensitive to the exact
range of selected isophotal major axes.  There are 55 galaxies selected in that
manner within the AUDS precursor  survey volume.

We tested the success of our selection procedure for the subsample of galaxies
with available spectroscopic redshifts. We found that all galaxies that are
within the AUDS redshift range based on their spectroscopic redshift, are also
correctly placed within that volume by their photometric redshift. There is
only one galaxy which is placed within that volume by its photometric redshift
even though the spectroscopic redshift is outside that range.  The
spectroscopic redshift of that galaxy is 0.1502. We therefore conclude that the
SDSS photometric redshifts are accurate enough to estimate the relative
densities of volumes in the AUDS redshift range.

We then counted the number of galaxies selected in the same manner in 100
close-by regions of the same size and shape. We find that on average, there are
$22.3\pm1.3$ galaxies in regions of that size, where 1.3 is the uncertainty of the
mean based on the scatter in the number counts.  The overdensity of SDSS
galaxies in the AUDS precursor region is therefore about $\delta_{\rm SDSS} =
2.5\pm0.4$, where the error estimate includes the Poisson error on the number
counts within the AUDS volume.  Assuming that the ratio between \hi\ density
and that of SDSS galaxies is independent of the local overdensity, we can
estimate the mean \hi\ density at median redshift of the AUDS precursor survey
in terms of the \hi\ density at zero redshift $\rho_0$ to be
\begin{equation}
\rho_{\rm HI}(z = 0.125) =  {\delta_{\rm HI}\over\delta_{\rm SDSS}}\rho_0 =
( 1.0 \pm 0.3) \cdot \rho_0.
\end{equation}
}

{The error estimate includes only random errors. When we use the SDSS
spectroscopic redshifts to estimate the relative overdensity in the same manner
as above, the result is $\rho_{\rm HI}(z = 0.125) =  ( 1.3 \pm 0.6) \cdot
\rho_0.$ The larger uncertainty  is due to the smaller number of galaxies with
measured spectroscopic redshift.  }

\section{Summary and Conclusion}

We have used about 53 hours of integration time with the Arecibo telescope to
demonstrate the feasibility of high sensitivity \hi\ observations with the
Arecibo L-band Feed Array.  To this depth, the noise in the final spectra still
behaves as $\propto \sqrt{1/t_{\rm exp}}$.  RFI renders a significant fraction
of the spectrum useless. However, regions of the spectrum where RFI is rarely
present are little affected by the strong RFI in neighboring channels. 

In total, we detected \nhitotal\ \hi\ emitters in the survey. For six of them,
there are optical counterparts with  measured redshifts in the SDSS. For all
other detections, there are plausible SDSS counterparts without redshift
confirmation.  The detection rate is consistent with the total \hi\ density in
the local universe.  

The experiment was a pilot study for the ``Arecibo Ultradeep Survey''.  This
survey will be more sensitive, cover more area on the sky with a better
sampling, and probe a  larger redshift range than this pilot study. AUDS will
allow to investigate in detail the \hi\ content in two independent volumes
outside the local Universe.

\acknowledgments

This research has made use of the Sloan Digital Sky Survey archive. Its full
acknowledgment can be found at \url{http://www.sdss.org}.

\begin{deluxetable}{lccccccl}
\tablecolumns{8}
\tablewidth{0pc}
\rotate
\tablecaption{\hi\ Survey at z $>$ 0.1\label{tab:hisurveys}}
\tablehead{
\colhead{Reference} &
\colhead{$a\times b$} &
\colhead{rms} &
\colhead{$\Delta v$} &
\colhead{z$_{\rm min}$ -- z$_{\rm max}$} &
\colhead{$\Omega_s$} &
\colhead{n$_{\rm det}$} &
\colhead{target}  \\
\colhead{} &
\colhead{$\arcsec\times \arcsec$} &
\colhead{mJy} &
\colhead{km\,s$^{-1}$} &
\colhead{} &
\colhead{$\square \deg$} &
\colhead{} &
\colhead{}  \\}
\startdata
\cite{zwaan}        &        18$\times$20    &  0.11    &    39     &      0.17-0.19         &   0.4    &     1 &  A2218    \\
\cite{ver}         &        17$\times$27    &  0.068   &    20     &      0.164--0.224  &   0.81   &    42 &  A963, A2192    \\
\cite{sdsssample}  &        230$\times$270  &  0.11    &   27      &     0.17--0.245    &  \nodata &   $\approx30$  & targeted emission line \\
                   &                        &          &           &                    &          &       & galaxies    \\
                   \cite{lah}         &        2.9$\times$2.9  &  0.13    &    33     &      0.218--0.253  &  \nodata &     0\tablenotemark{a} &  emission line galaxies\\
                   &                 &          &           &      &                    &          &  in field around Quasar\\
                   &                 &          &           &      &                    &          &  SDSSp J104433-012502    \\
                   \cite{lah09}       &        3.3$\times$3.3  &  0.16    &    36     &      0.33--0.40    &  \nodata &     0\tablenotemark{a} &  galaxies in A370 field   \\
AUDS precursor     &        200$\times$200  &  0.08    &    5.2    &      0.07--0.16    &   0.09\tablenotemark{b}   &    18 &  field around 4 emission\\
                   &                        &          &           &                    &          &       &  line galaxies    \\
                   AUDS               &        200$\times$200  &  0.08\tablenotemark{c}    &    5.2    &      0--0.16       &   0.35   &   $>$200\tablenotemark{c}&  random pointings    \\

\enddata
\tablenotetext{a}{\hi\ detected in stacked spectra.}
\tablenotetext{b}{Only a small fraction of the survey area reached the listed rms sensitivity.}
\tablenotetext{c}{Expected numbers based on extrapolation of the precursor data.}
\tablecomments{The columns are: {$a\times b$}: beam size, rms: sensitivity of survey,
$\Delta v$: channel width, z$_{\rm min}$ -- z$_{\rm max}$: redshift range, $\Omega_s$: survey area given unless
individual galaxies were targeted, {n$_{\rm det}$}: number of detected \hi\ lines.}
\end{deluxetable}

\begin{deluxetable}{rrrrrrrrrrrrrrr}
\tablecolumns{14}
\tablewidth{0pc}
\rotate
\tablecaption{Properties of AUDS Galaxies\label{tab:hiresults}}
\tablehead{
\colhead{ID} & \colhead{$\alpha_{2000}$} & \colhead{$\delta_{2000}$} &
\colhead{rms} &\colhead{s/n} & \colhead{q\tablenotemark{a}} & \colhead{$\nu$} & \colhead{$v_{50}$} & \colhead{$z$} & \colhead{$d_{lum}$} &
\colhead{$w_{20}$} & \colhead{$\int{fdv}$} & \colhead{$M_{\rm HI}$} &
 \colhead{$M_{\rm HI, corr}$}     \\
&&&\colhead{$\mu$Jy} &&& \colhead{MHz} & \colhead{km/s} && \colhead{Mpc} & \colhead{km/s} & \colhead{mJy$\cdot$km/s} & \colhead{$10^9$M$_{\rm \odot}$} & \colhead{$10^9$M$_{\rm \odot}$} \\
(1)&(2)\phantom{xx}&(3)\phantom{xx}&(4)&(5)\phantom{x}&(6)&(7)\phantom{xx}&(8)\phantom{x}&(9)\phantom{xx}&(10)\phantom{x}&(11)&(12)\phantom{xx}&(13)\phantom{x}&(14)\phantom{xx}
}
\startdata
1-1&23 59 18.3&15 37 28 &   152 &  3.5 & 2 & 1272.9 & 34728&  0.11584&     525.7&  197  &       46 &  2.66 &  3.89 \\ &&&&&&   0.8 &    22& 0.00007&       0.3&   66&      16& 0.94& 1.37 
\\ 
1-2&23 59 22.4&15 37 28 &   125 &  4.6 & 1 & 1276.1 & 33899&  0.11308&     512.2&  662  &      160 &  8.88 & 10.02\tablenotemark{b} \\ &&&&&&   0.8 &    21& 0.00007&       0.3&   62&      29& 1.58& 1.91 
\\ 
3-1&23 59 39.0&15 47 52 &   100 &  3.9 & 3 & 1295.1 & 28997&  0.09672&     433.2&  346  &       43 &  1.73 &  8.97\tablenotemark{b} \\ &&&&&&   0.6 &    14& 0.00005&       0.2&   43&      20& 0.80& 1.29 
\\ 
2-1&23 59 47.3&15 42 40 &    80 & 10.8 & 1 & 1295.5 & 28910&  0.09643&     431.8&  480  &      189 &  7.58 &  7.59\tablenotemark{b} \\ &&&&&&   0.3 &     7& 0.00002&       0.1&   21&      27& 1.07& 1.30 
\\ 
2-2&23 59 51.5&15 42 40 &    73 &  5.4 & 3 & 1245.5 & 42099&  0.14043&     647.9&  497  &       62 &  5.37 &  6.88 \\ &&&&&&   0.8 &    28& 0.00009&       0.4&   85&      20& 1.70& 2.18 
\\ 
2-3&23 59 59.8&15 42 40 &   167 &  5.9 & 1 & 1317.5 & 23408&  0.07808&     345.1&  277  &      158 &  4.11 &  5.52 \\ &&&&&&   0.3 &     6& 0.00002&       0.1&   18&      24& 0.62& 0.84 
\\ 
1-3&00 00 08.1&15 37 28 &    93 &  4.1 & 2 & 1293.9 & 29322&  0.09781&     438.3&  127  &       24 &  0.99 &  1.83 \\ &&&&&&   0.6 &    14& 0.00005&       0.2&   41&      11& 0.46& 0.86 
\\ 
3-2&00 00 08.1&15 47 52 &   105 &  5.1 & 1 & 1274.0 & 34452&  0.11492&     521.2&  263  &       65 &  3.72 &  5.08 \\ &&&&&&   0.5 &    14& 0.00005&       0.2&   41&      16& 0.91& 1.24 
\\ 
1-4&00 00 12.3&15 37 28 &    83 &  6.8 & 2 & 1264.0 & 37100&  0.12375&     564.6&  189  &       46 &  3.06 &  4.35 \\ &&&&&&   0.3 &     9& 0.00003&       0.1&   26&      11& 0.74& 1.06 
\\ 
2-4&00 00 12.3&15 42 40 &    80 &  3.5 & 3 & 1235.0 & 45011&  0.15014&     697.2&  448  &       74 &  7.34 &  7.71\tablenotemark{b} \\ &&&&&&   0.5 &    17& 0.00006&       0.3&   52&      22& 2.21& 2.71 
\\ 
\hline\tablebreak
2-5&00 00 16.4&15 42 40 &    82 & 10.4 & 1 & 1277.5 & 33533&  0.11185&     506.2&  313  &      162 &  8.82 & 11.12\tablenotemark{b} \\ &&&&&&   0.2 &     6& 0.00002&       0.1&   18&      19& 1.04& 1.25 
\\ 
2-6&00 00 16.4&15 42 40 &    76 &  6.2 & 1 & 1243.5 & 42645&  0.14225&     657.1&  446  &      123 & 10.93 & 14.18\tablenotemark{b} \\ &&&&&&   0.3 &    10& 0.00003&       0.2&   30&      23& 2.05& 2.43 
\\ 
1-5&00 00 20.6&15 37 28 &    89 &  3.8 & 1 & 1295.4 & 28922&  0.09647&     432.0&  218  &       35 &  1.42 &  2.39 \\ &&&&&&   0.7 &    15& 0.00005&       0.2&   45&      14& 0.55& 0.93 
\\ 
2-7&00 00 28.9&15 42 40 &    79 &  5.9 & 1 & 1295.2 & 28971&  0.09664&     432.7&  205  &       53 &  2.13 &  3.26 \\ &&&&&&   0.5 &    10& 0.00003&       0.2&   31&      13& 0.51& 0.79 
\\ 
3-3&00 00 33.0&15 47 52 &   111 &  5.2 & 1 & 1277.2 & 33626&  0.11216&     507.7&  194  &       79 &  4.31 &  5.74 \\ &&&&&&   0.3 &     7& 0.00002&       0.1&   21&      15& 0.81& 1.08 
\\ 
3-4&00 00 41.4&15 47 52 &   128 &  5.3 & 1 & 1274.8 & 34235&  0.11420&     517.7&  365  &      102 &  5.79 &  7.34 \\ &&&&&&   0.3 &     7& 0.00002&       0.1&   22&      22& 1.23& 1.56 
\\ 
3-5&00 00 49.7&15 47 52 &   138 &  4.4 & 1 & 1293.7 & 29354&  0.09791&     438.9&  455  &      102 &  4.20 &  5.61 \\ &&&&&&   0.7 &    16& 0.00005&       0.2&   49&      22& 0.89& 1.20 
\\ 
2-8&00 00 58.0&15 42 40 &   127 &  4.6 & 1 & 1274.0 & 34440&  0.11488&     521.0&  593  &      117 &  6.68 &  8.31 \\ &&&&&&   0.3 &     9& 0.00003&       0.1&   27&      25& 1.42& 1.76 
\\ 
\enddata
\tablenotetext{a}{quality code for detection: 1 - likely detection; 2 - possible detection; 3 - doubtful detection}
\tablenotetext{b}{based on position of SDSS source}
\end{deluxetable}

\clearpage

\appendix
\section{Appendix: SDSS Counterparts  of AUDS \hi\ Detections}

The target field for the AUDS precursor survey was chosen to include galaxies
that were likely to be strong \hi\ emitters as judged by the SDSS spectrum.
Those galaxies were intentionally placed close to the center of the central
strip in the most sensitive part of the survey.  The \hi\ counterparts for
these galaxies can easily be identified by redshift.  The final data cube of
the survey was searched for \hi\ detections without knowledge of the location
or redshift of SDSS galaxies in this volume.  Most of the \hi\ detections are
lower in \hi\ content than the preselected targets, and most of them are
without matching redshift listed in the SDSS.  Identifying the optical
counterparts for those sources is difficult because of the large beam size.

{Because of the geometry of the survey, the Declination of a source is very
uncertain.  The sensitivity of the telescope beam between the strips drops to
about 10\% of the sensitivity in the center of the strips. A strong source can
therefore be detected at any position within the survey region, including the
region between the strips.  If a source is located at a large angular distance
from the center of the strip, the measurement error in  Right Ascension becomes
dominated by the lack of our knowledge of the exact beam shape.  The ALFA beam
shape at larger distance form the center varies strongly with the pointing of
the telescope \citep{beam}, and the effective beam shape of our repeated drifts
is highly  uncertain.  An  accurate estimate of the positional uncertainties as
a function of signal-to-noise ratio is therefore difficult. For simplicity, we
adopt 20\% of the FWHM beam as our estimated uncertainty in Right
Ascension for all of our sources.}

We searched for optical counterparts  within 3.2 arcminutes of the
\hi\ position in Right Ascension and 6 arcminutes in Declination. This
corresponds to a sensitivity of 50\% and 10\% of that at the center of the beam.
Within that region, we selected  possible optical counterparts from the SDSS
that were classified as galaxies and had a major axis of 10 arcseconds or more
in the SDSS $r$-band.  We manually cleaned the extracted source list by
inspecting their images and removing misidentification such as double stars or
image defects. If the SDSS redshift of any of the candidates was within the
detected \hi\ line, we adopted this source as certain detection of a unique
optical counterpart.

In Fig.~A1, we show a region of 6$\times$6 arcmin$^2$ around each AUDS source
obtained from the SDSS finding chart generator. For better clarity, the images
are shown with inverted color scale, i.e. bluer colors represent redder
galaxies.  {The sensitivity at the edges of the images is approximately 10\%
of the sensitivity in the center. In each image, the approximate size of the beam at 50\% and
25\% of the beam center is marked as circles}. The outer part of the shown
region corresponds to about 10\% the sensitivity of the center of the beam.  For
sources with a unique optical counterpart, the counterpart galaxy is circled.
For all other sources, all possible counterparts are marked with circles. The
diameter of each circle is the size of the major axis of each source as listed
in the SDSS. For each source, we selected up to four candidates we judged to be
the most likely optical counterparts based on the following criteria: 1)
counterparts are more likely to be located close to the center of the field, 2)
\hi\ line widths larger than 400  km\,s$^{-1}$ are likely to correspond to
galaxies with highly inclined disks, whereas those smaller than 200
km\,s$^{-1}$ should correspond to almost face-on disks, and  3) larger sources
are more likely to contain large amounts of \hi.  The most likely candidate
counterpart galaxies are also shown separately in cutouts 20 $\times$ 20
arcseconds in size, and their major axis $a$ is given.  For easy comparison, we
also reproduce the \hi\ spectrum for each source. {In this spectrum, the measure
\hi\ redshift is marked by an arrow at the top, and if available, the SDSS
redshift  with error estimated is marked by a bar below the \hi\ line.}

\newcommand{\galcomment}[1]{
\ifthenelse{\equal{#1}{1-1}}{The narrow line width of 197  km\,s$^{-1}$ suggests a relatively face-on
galaxy. Because of the low signal-to-noise ratio, we cannot exclude a substantially larger 
line width. The \hi\ line is also visible in the spectrum of source 1-2, which is consistent with
the position of SDSS 235914.9+153457. Source 1-2 is within the beam, and its line is visible 
in the spectrum between 1275 and 1279 MHz.}{}
\ifthenelse{\equal{#1}{1-2}}{The redshift of SDSS 255918.8+153641 is marked in the \hi\ spectrum.
Source 1-1 is also in the beam and its line is visible at around 1272 MHz.}{}
\ifthenelse{\equal{#1}{1-3}}{The narrow line width suggest a relatively face-on
galaxy, whereas the SDSS counterparts are fairly inclined disks. Because of the low 
signal-to-noise ratio, the line width could be underestimated.  The beam overlaps with 
the one of source 1-4, and they share two of the candidate counterparts.}{}
\ifthenelse{\equal{#1}{1-4}}{The narrow line width suggest a relatively face-on
galaxy, whereas the SDSS counterparts are fairly inclined disks.The beam overlaps with 
the one of source 1-3.}{}
\ifthenelse{\equal{#1}{1-5}}{Because of its high signal-to-noise
ratio, the positional uncertainty in right ascension should be small. Another potential
counterpart is partially hidden by the bright star close to 000021.8+153909. It is not 
marked because the SDSS measurement of the major axis is not reliable.}{}
\ifthenelse{\equal{#1}{2-1}}{SDSS 235948.4+154243 was one of the sources used to select
the target region. Its SDSS redshift is marked in the spectrum. }{}
\ifthenelse{\equal{#1}{2-2}}{\hi\ emission is between regions with frequent RFI. Both 
candidate SDSS counterparts are low surface brightness galaxies.}{}
\ifthenelse{\equal{#1}{2-3}}{The line width of 277  km\,s$^{-1}$ suggests a highly inclined
galaxy. This makes SDSS 235960.0+154235 the most likely optical counterpart.}{}
\ifthenelse{\equal{#1}{2-4}}{Because of the low signal-to-noise ratio and the closeness
of a region with frequent RFI, the quality of the spectrum was judged to be low. However,
the exact match in redshift confirms the reality of the detected line. The redshift of
SDSS 000011.1+154212 is marked in the spectrum.}{}
\ifthenelse{\equal{#1}{2-5}}{SDSS 000016.8+154140 was one of the sources used to select
the target region. Its SDSS redshift is marked in the spectrum. The measured position of 
source 2-5 is identical to source 2-6.}{}
\ifthenelse{\equal{#1}{2-6}}{SDSS 000014.9+154342 was one of the sources used
to select the target region. Its SDSS redshift is marked in the spectrum. The measured 
position of source 2-6 is identical to source 2-5. The line is close to a region in the 
spectrum with  well known RFI at Arecibo. We used this fact to test our ability to detect 
lines close to RFI. }{}
\ifthenelse{\equal{#1}{2-7}}{The edge-on disk galaxy in the center of
this beam is at a redshift of 0.115 and was not detected. }{}
\ifthenelse{\equal{#1}{2-8}}{The high line width of 593  km\,s$^{-1}$ suggests a highly inclined
galaxy. For that reason, 000059.3+154034 is the most likely optical counterpart.}{}
\ifthenelse{\equal{#1}{3-1}}{The redshift of SDSS 235936.4+155029 is marked in the spectrum .
It is located close to the edge of the image cutout, where the sensitivity is about 10\% of the sensitivity at 
the center of the beam. The relatively low \hi\ mass estimate of $1.7\times10^9{\rm M}_{\rm \odot}$ is consistent with such an off-center position.}{}
\ifthenelse{\equal{#1}{3-2}}{A cluster at z$\approx$ 0.152 to the north of 
the beam was not detected. Only one of the cluster members has a measured
optical redshift. It is therefore possible that one of the apparent cluster members is
the optical counterpart of the \hi\ emission.}{}
\ifthenelse{\equal{#1}{3-3}}{SDSS 000039.0+154551 is probably too far off center to 
be the optical counterpart given the high signal-to-noise ratio of the detected \hi\ line.
{The peaks of the  double-horned profile of source 3-4 are visible in the
spectrum at $\approx$1274 and 1275.5 MHz.} }{}
\ifthenelse{\equal{#1}{3-4}}{ The beam overlaps with the ones around sources
3-5 and 3-3. The second line in the spectrum centered at $\approx$1277 MHz originates from
source 3-3. }{}
\ifthenelse{\equal{#1}{3-5}}{.}{}
}

\begin{tabular}{ccl}
\multicolumn{3}{l}{Fig. A1. -- SDSS counterparts.} \\
\\
\epsscale{0.3}

\epsscale{0.3}\plotone{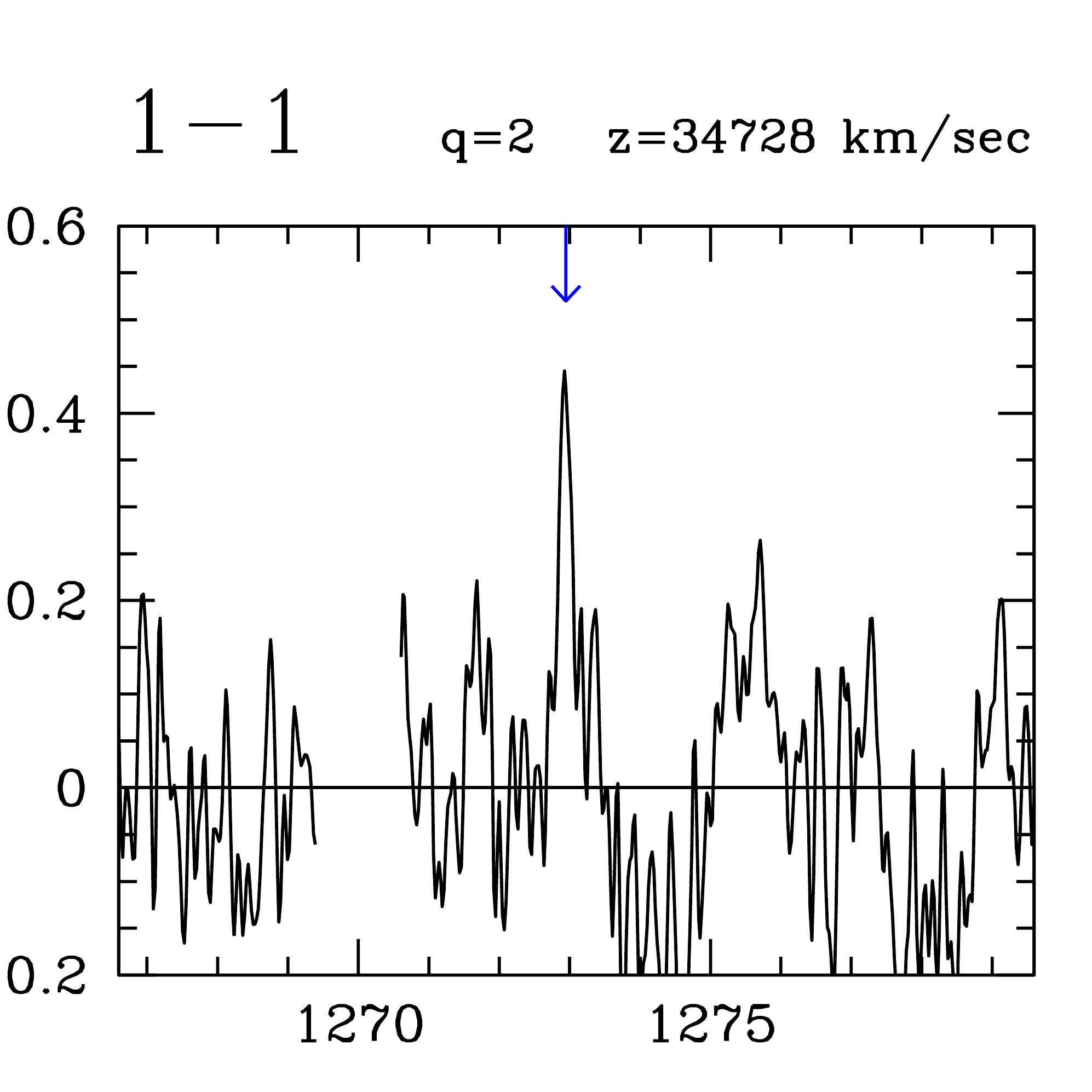} & \plotone{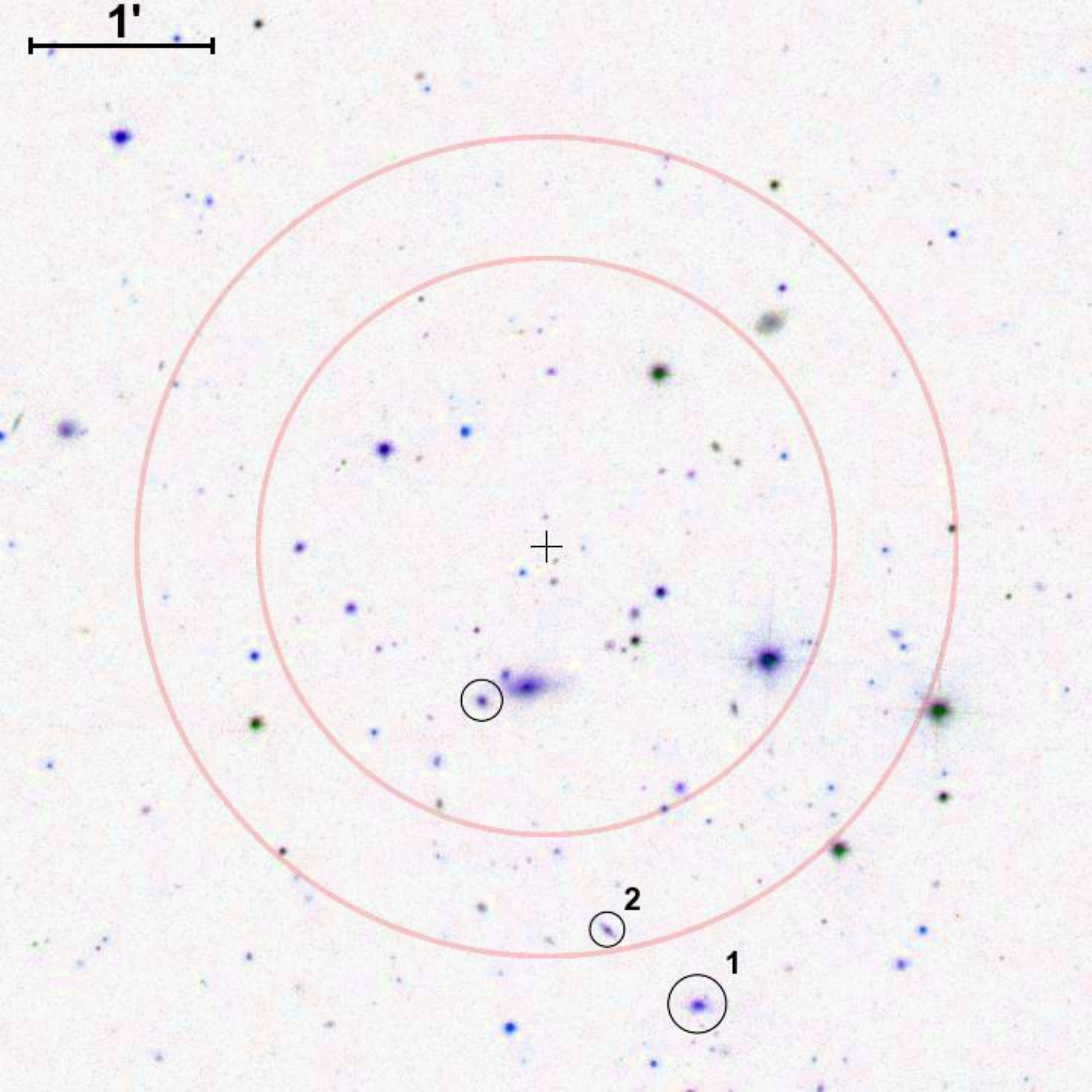} & 
\parbox[b]{4truecm}{\epsscale{0.07}
\plotone{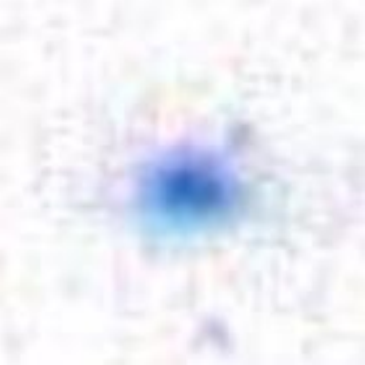}\parbox[b]{4.0truecm}{1: 235914.9+153457 \newline \phantom{1:} a=19" \vspace{1mm}}\\
\plotone{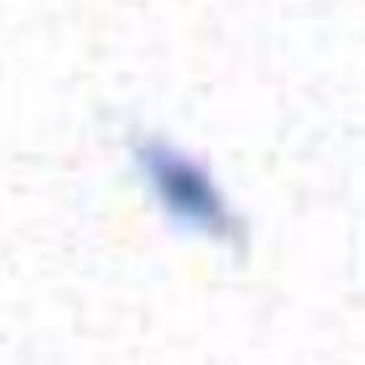}\parbox[b]{4.0truecm}{2: 235916.9+153521 \newline \phantom{2:} a=11" \vspace{1mm}}\\
\epsscale{0.3} } \\
\multicolumn{3}{l}{\parbox[t]{16truecm}{
\galcomment{1-1}
\newline}}\\\epsscale{0.3}\plotone{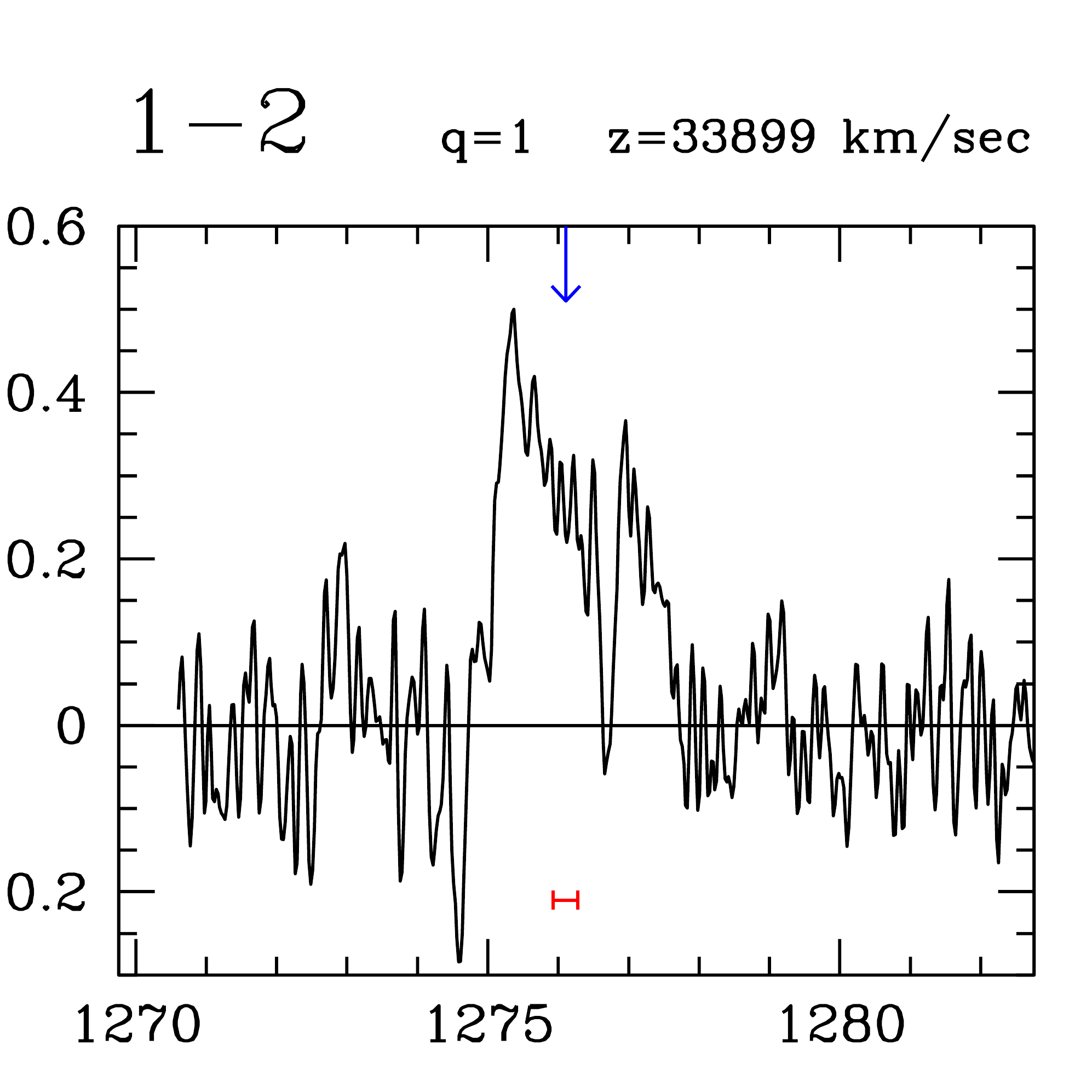} & \plotone{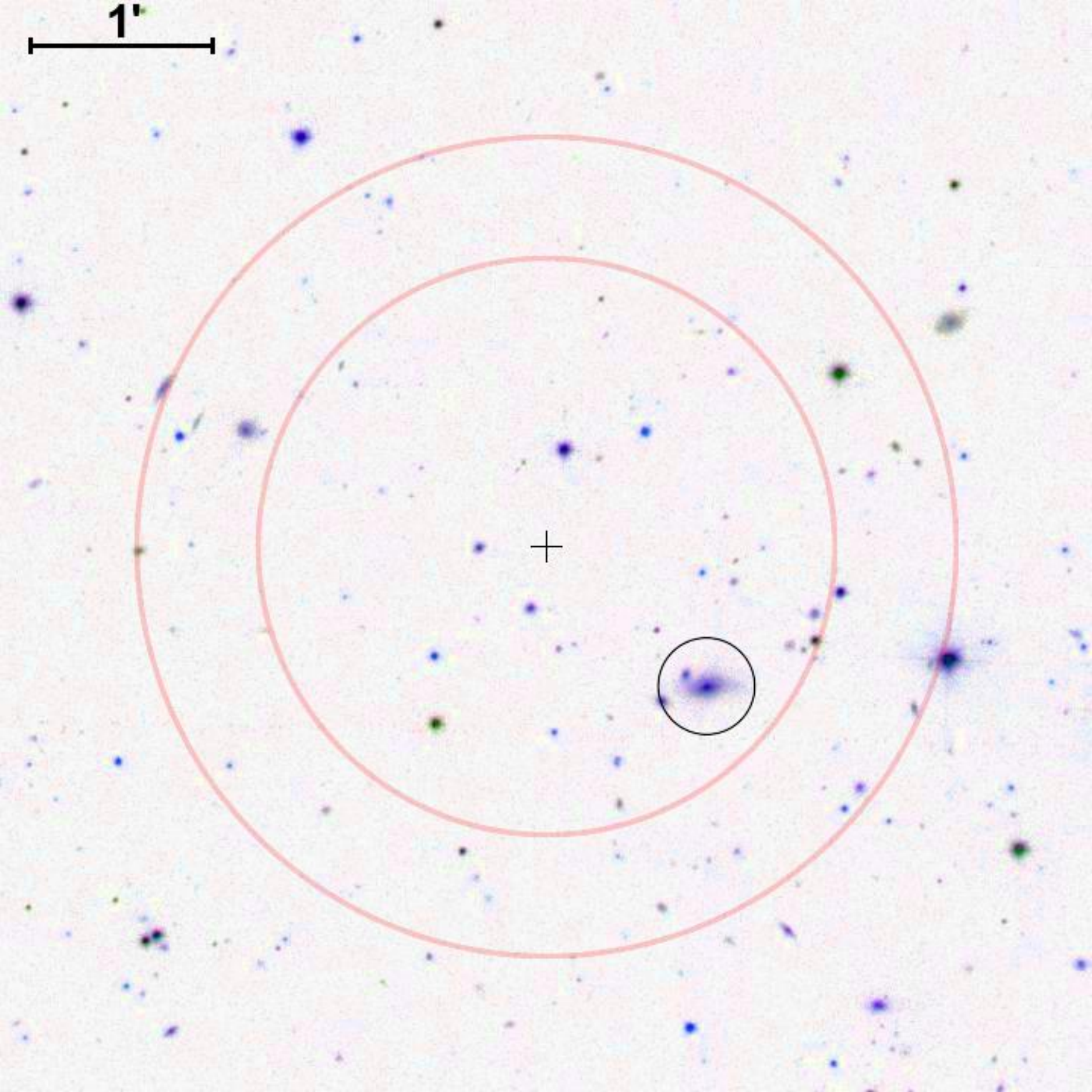} & 
\parbox[b]{4truecm}{\epsscale{0.07}
\plotone{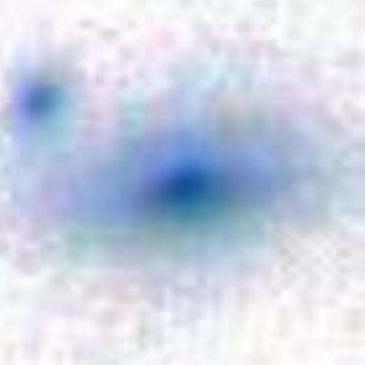}\parbox[b]{4.0truecm}{ 235918.8+153641 \newline  a=31"\newline z= 0.1131 }\\
\epsscale{0.3} } \\
\multicolumn{3}{l}{\parbox[t]{16truecm}{
\galcomment{1-2}
\newline}}\\\epsscale{0.3}\plotone{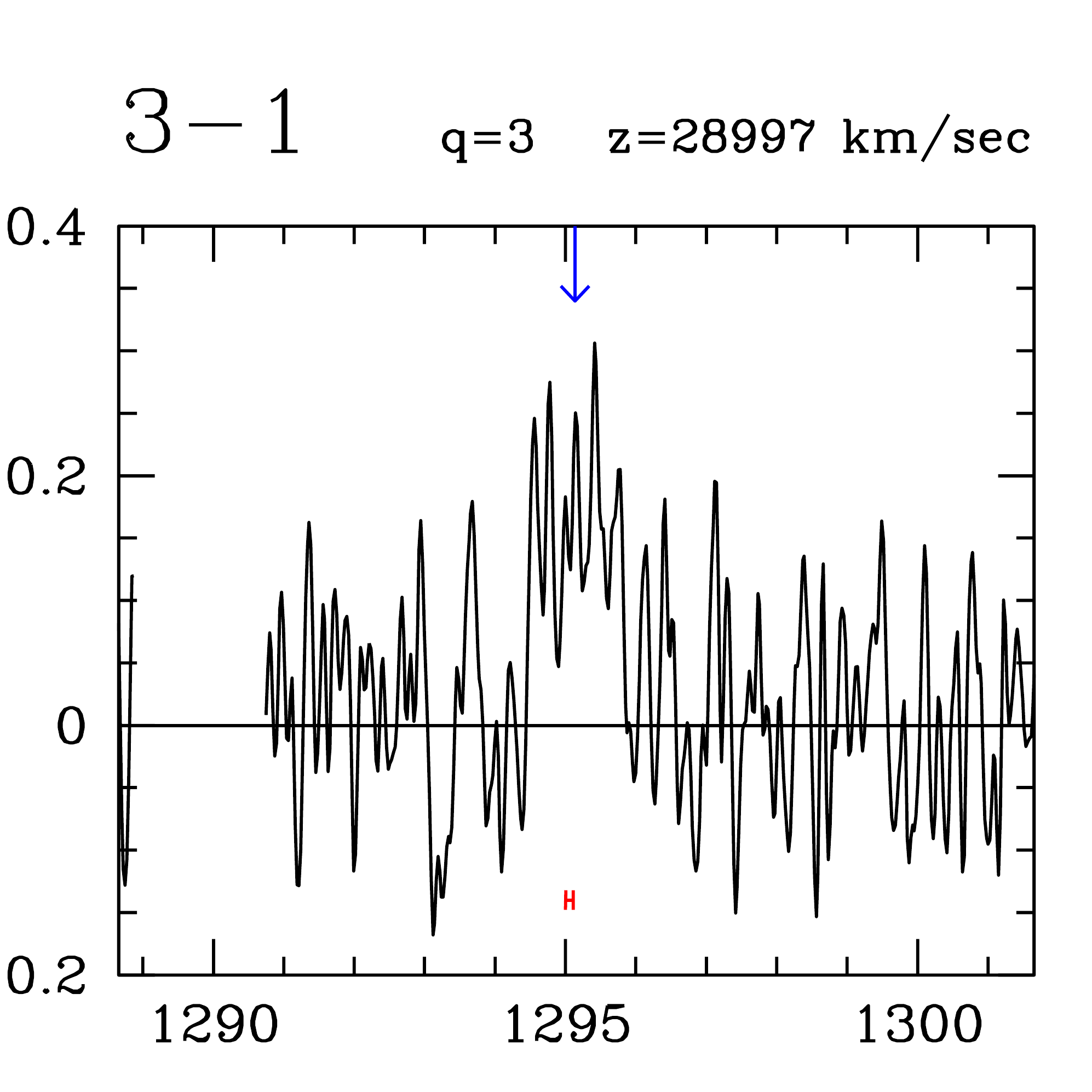} & \plotone{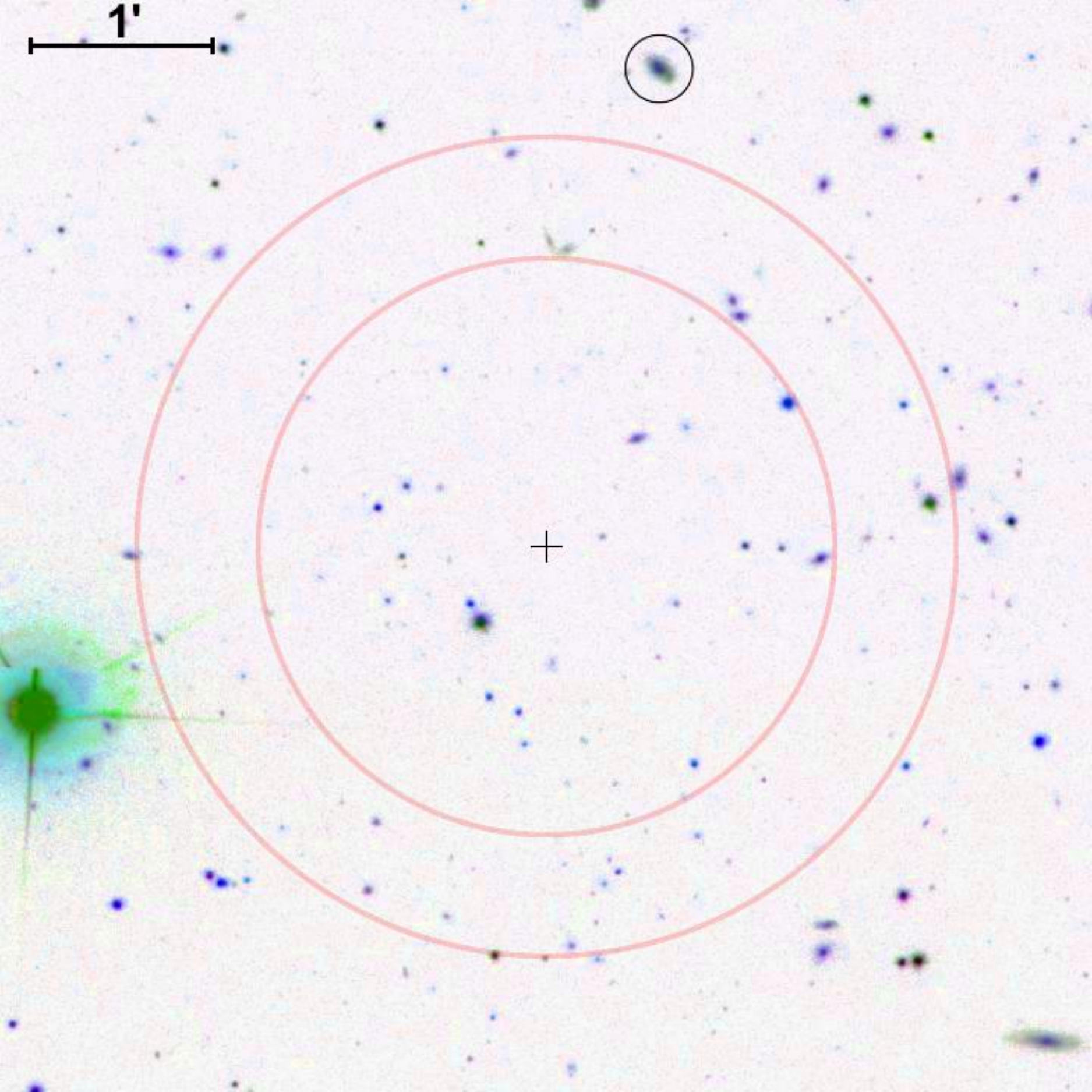} & 
\parbox[b]{4truecm}{\epsscale{0.07}
\plotone{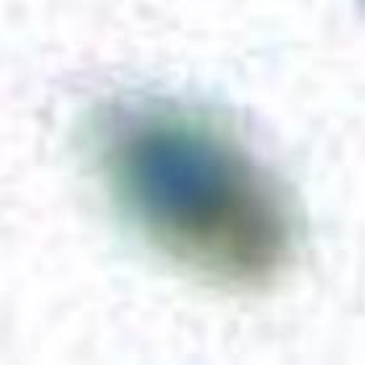}\parbox[b]{4.0truecm}{ 235936.4+155029 \newline  a=22"\newline z= 0.0968 }\\
\epsscale{0.3} } \\
\multicolumn{3}{l}{\parbox[t]{16truecm}{
\galcomment{3-1}
\newline}}\\\end{tabular}

\begin{tabular}{ccl}
\multicolumn{3}{l}{Fig. A1. -- continued} \\
\\
\epsscale{0.3}\plotone{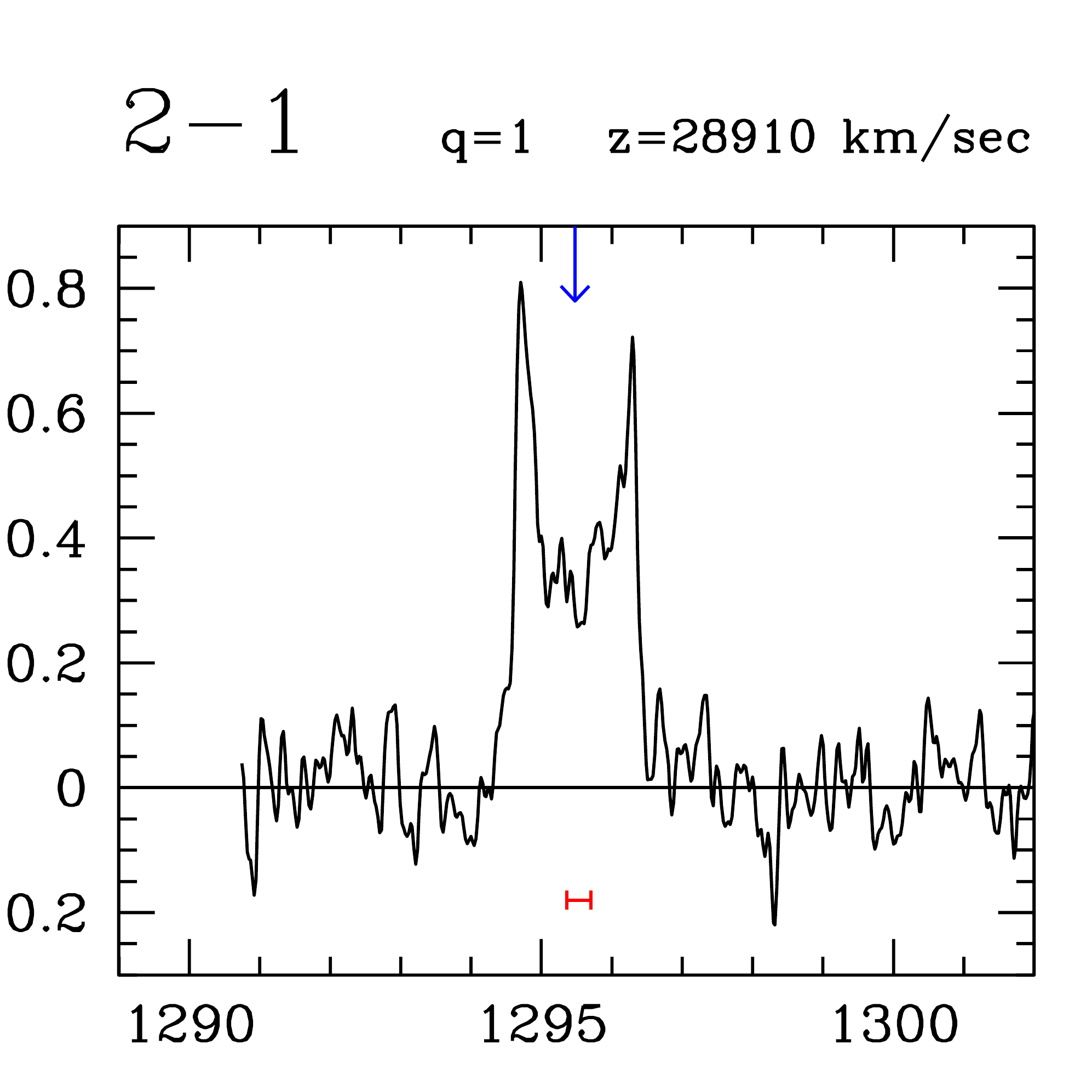} & \plotone{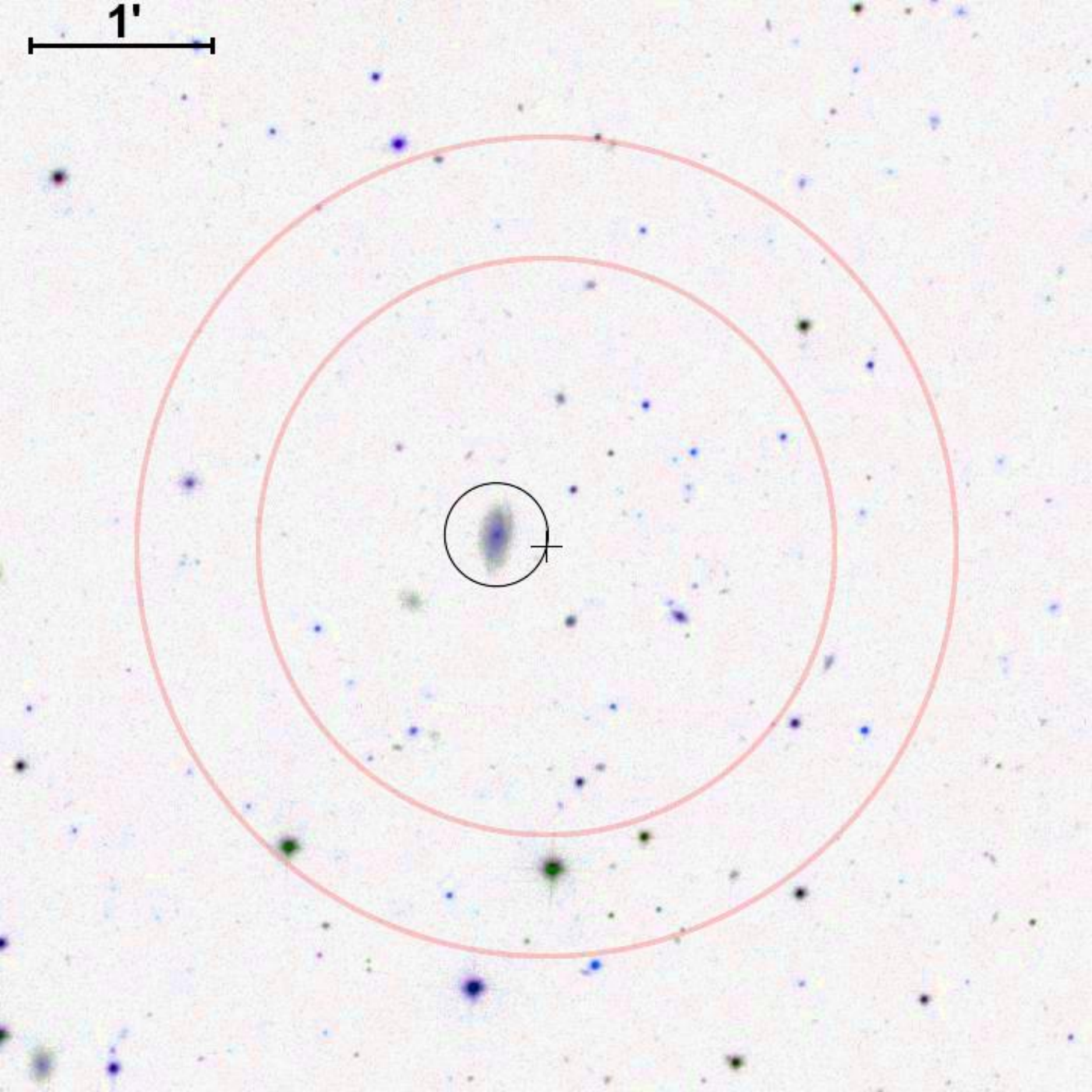} & 
\parbox[b]{4truecm}{\epsscale{0.07}
\plotone{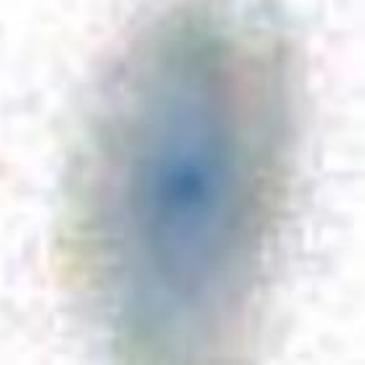}\parbox[b]{4.0truecm}{ 235948.4+154243 \newline  a=34"\newline z= 0.0964 }\\
\epsscale{0.3} } \\
\multicolumn{3}{l}{\parbox[t]{16truecm}{
\galcomment{2-1}
\newline}}\\\epsscale{0.3}\plotone{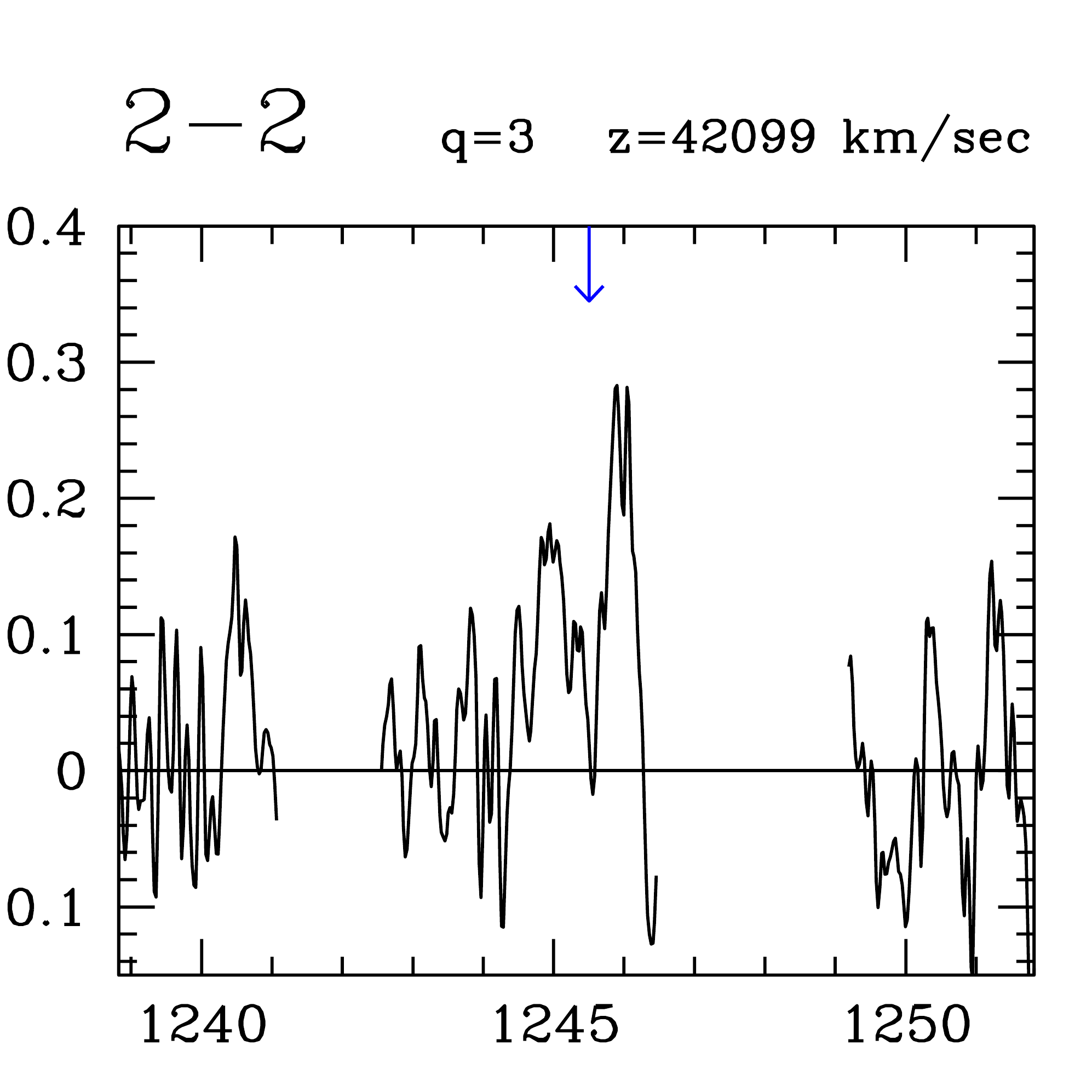} & \plotone{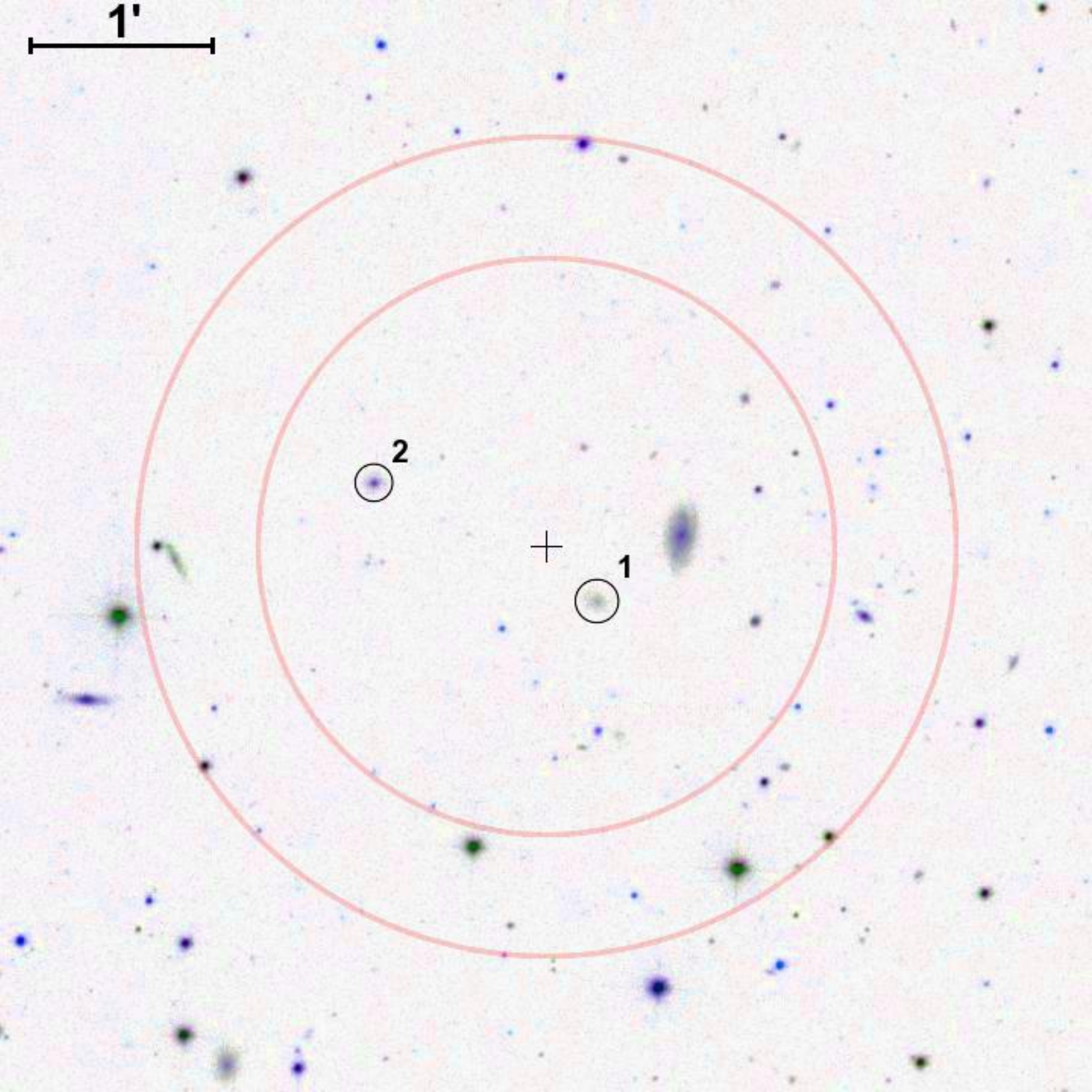} & 
\parbox[b]{4truecm}{\epsscale{0.07}
\plotone{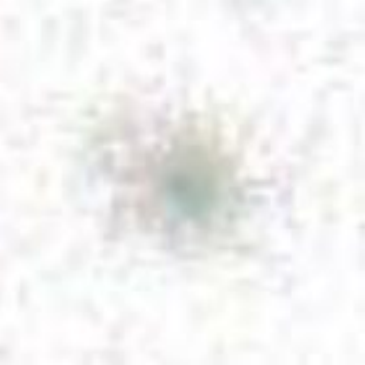}\parbox[b]{4.0truecm}{1: 235950.4+154222 \newline \phantom{1:} a=14" \vspace{1mm}}\\
\plotone{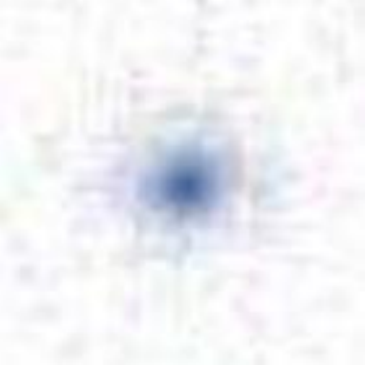}\parbox[b]{4.0truecm}{2: 235955.4+154301 \newline \phantom{2:} a=12" \vspace{1mm}}\\
\epsscale{0.3} } \\
\multicolumn{3}{l}{\parbox[t]{16truecm}{
\galcomment{2-2}
\newline}}\\\epsscale{0.3}\plotone{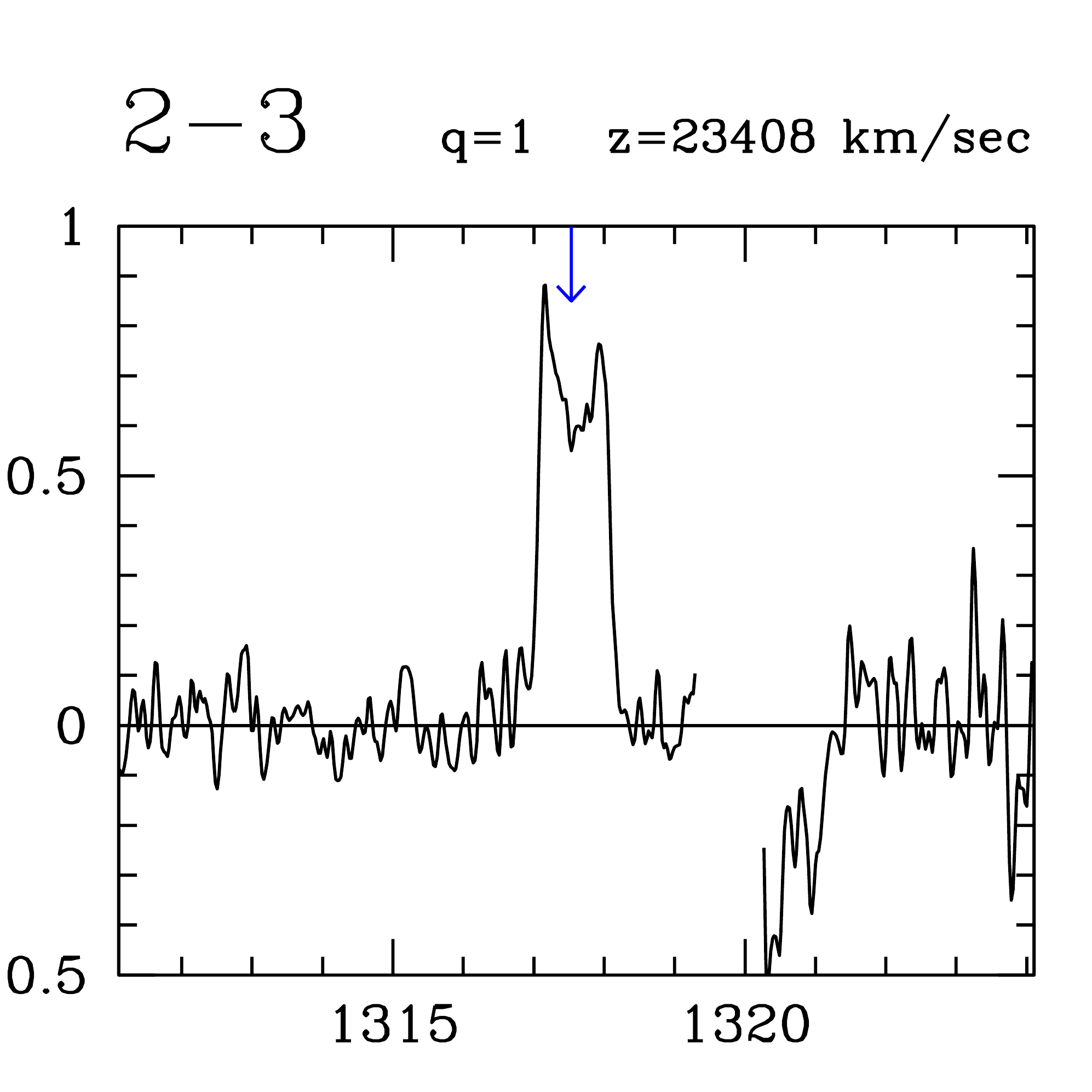} & \plotone{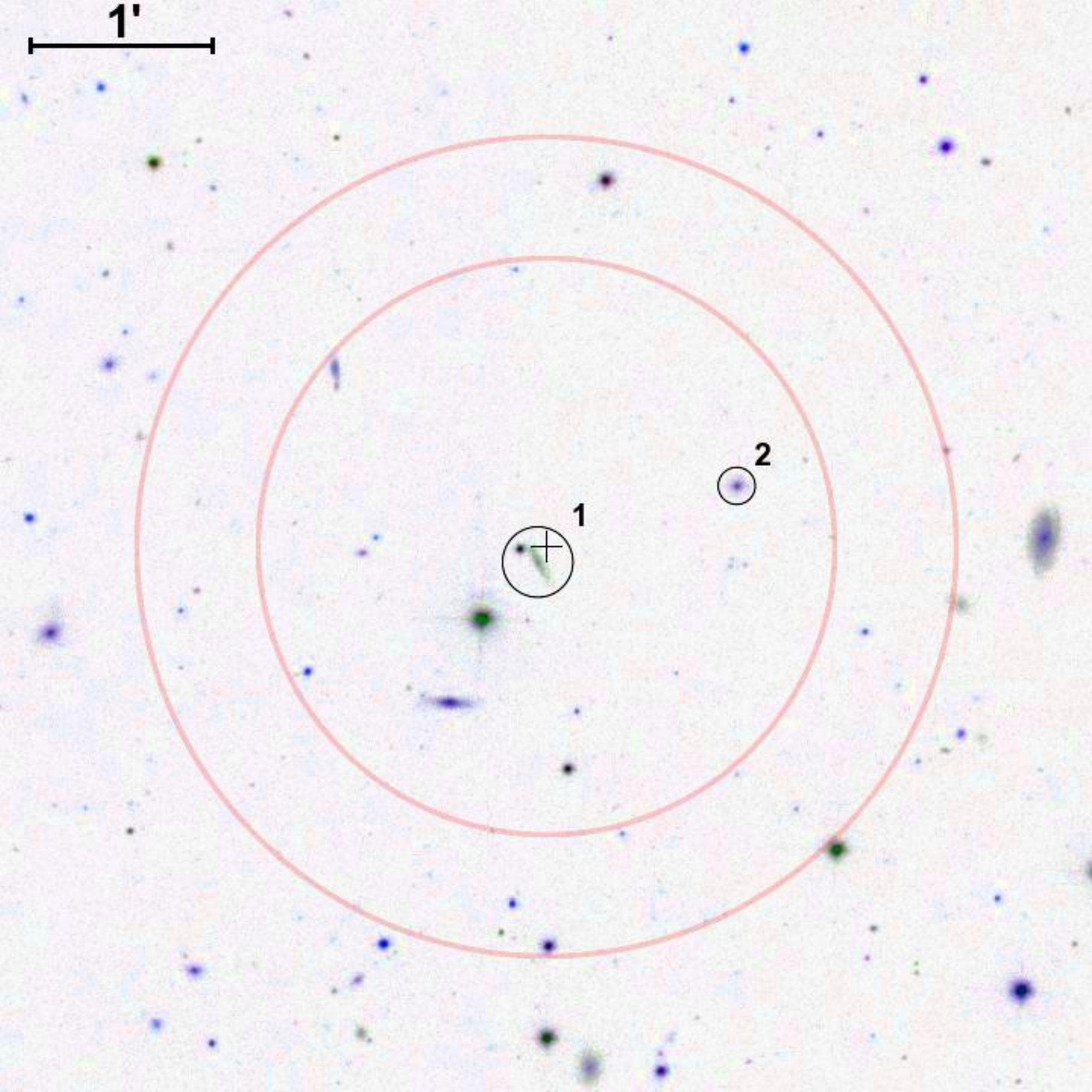} & 
\parbox[b]{4truecm}{\epsscale{0.07}
\plotone{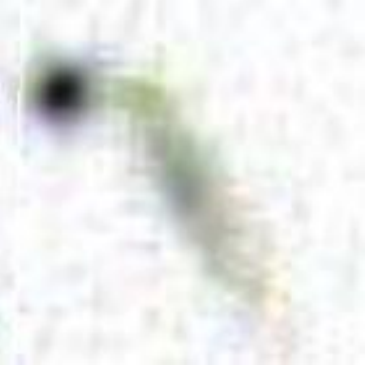}\parbox[b]{4.0truecm}{1: 235960.0+154235 \newline \phantom{1:} a=23" \vspace{1mm}}\\
\plotone{sdss235954+154301.pdf}\parbox[b]{4.0truecm}{2: 235955.4+154301 \newline \phantom{2:} a=12" \vspace{1mm}}\\
\epsscale{0.3} } \\
\multicolumn{3}{l}{\parbox[t]{16truecm}{
\galcomment{2-3}
\newline}}\\\end{tabular}

\begin{tabular}{ccl}
\multicolumn{3}{l}{Fig. A1. -- continued} \\
\\
\epsscale{0.3}\plotone{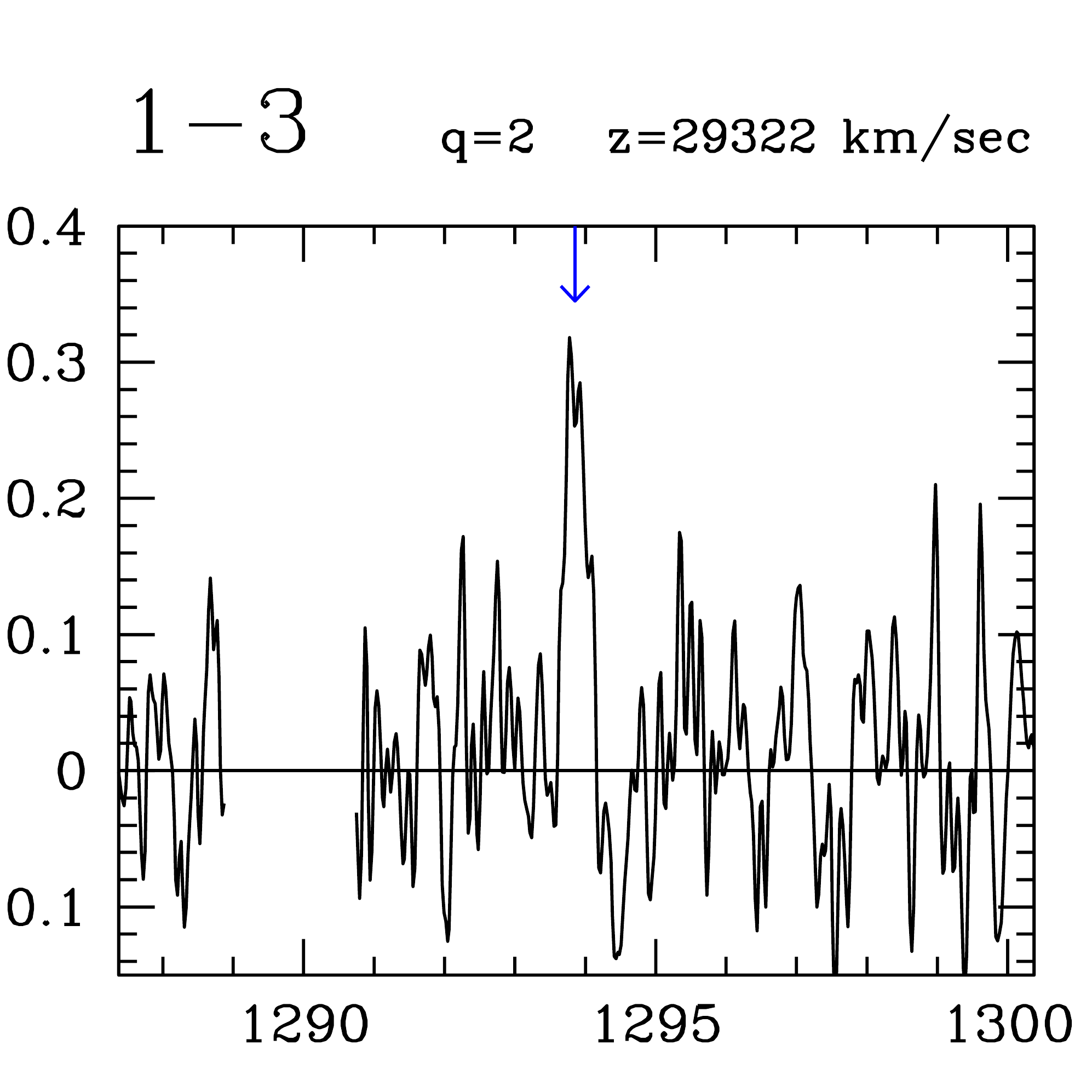} & \plotone{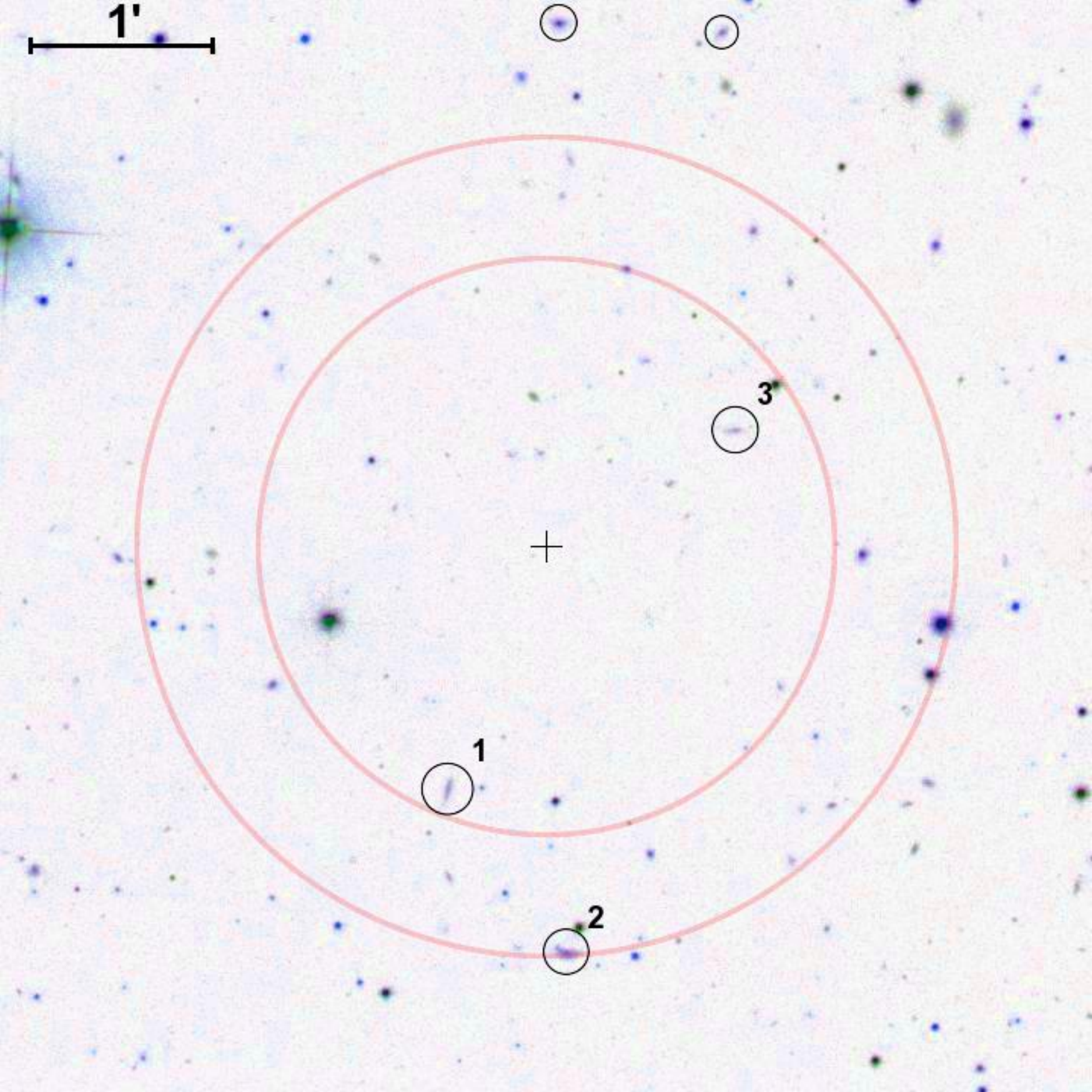} & 
\parbox[b]{4truecm}{\epsscale{0.07}
\plotone{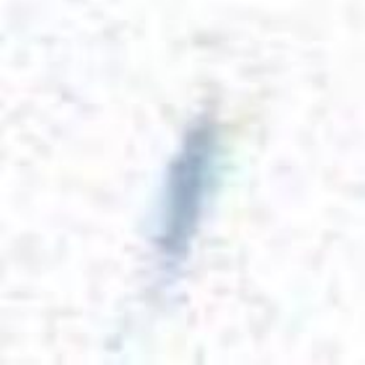}\parbox[b]{4.0truecm}{1: 000010.4+153609 \newline \phantom{1:} a=16" \vspace{1mm}}\\
\plotone{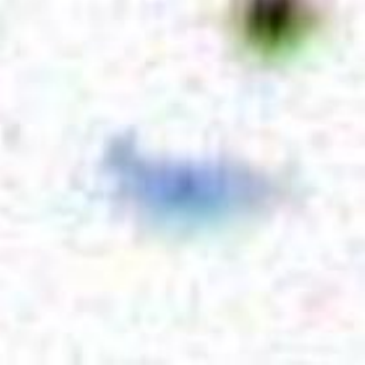}\parbox[b]{4.0truecm}{2: 00007.7+153515 \newline \phantom{2:} a=14" \vspace{1mm}}\\
\plotone{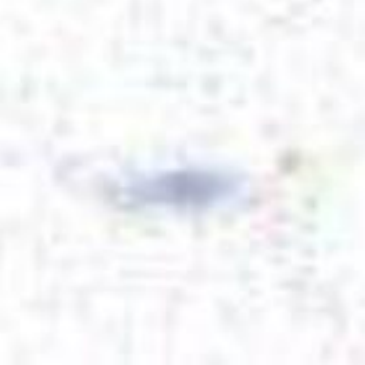}\parbox[b]{4.0truecm}{3: 00003.8+153807 \newline \phantom{3:} a=15" \vspace{1mm}}\\
\epsscale{0.3} } \\
\multicolumn{3}{l}{\parbox[t]{16truecm}{
\galcomment{1-3}
\newline}}\\\epsscale{0.3}\plotone{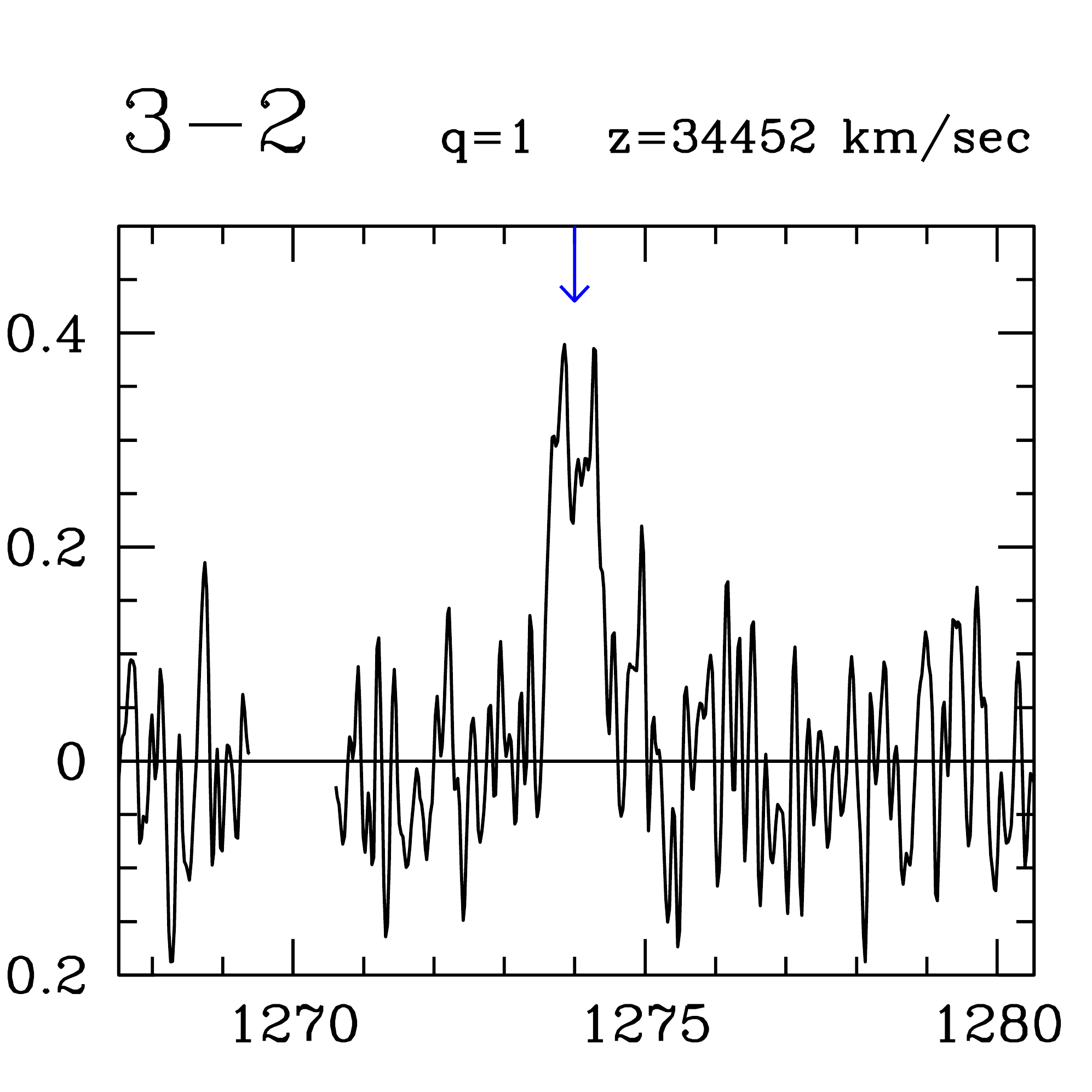} & \plotone{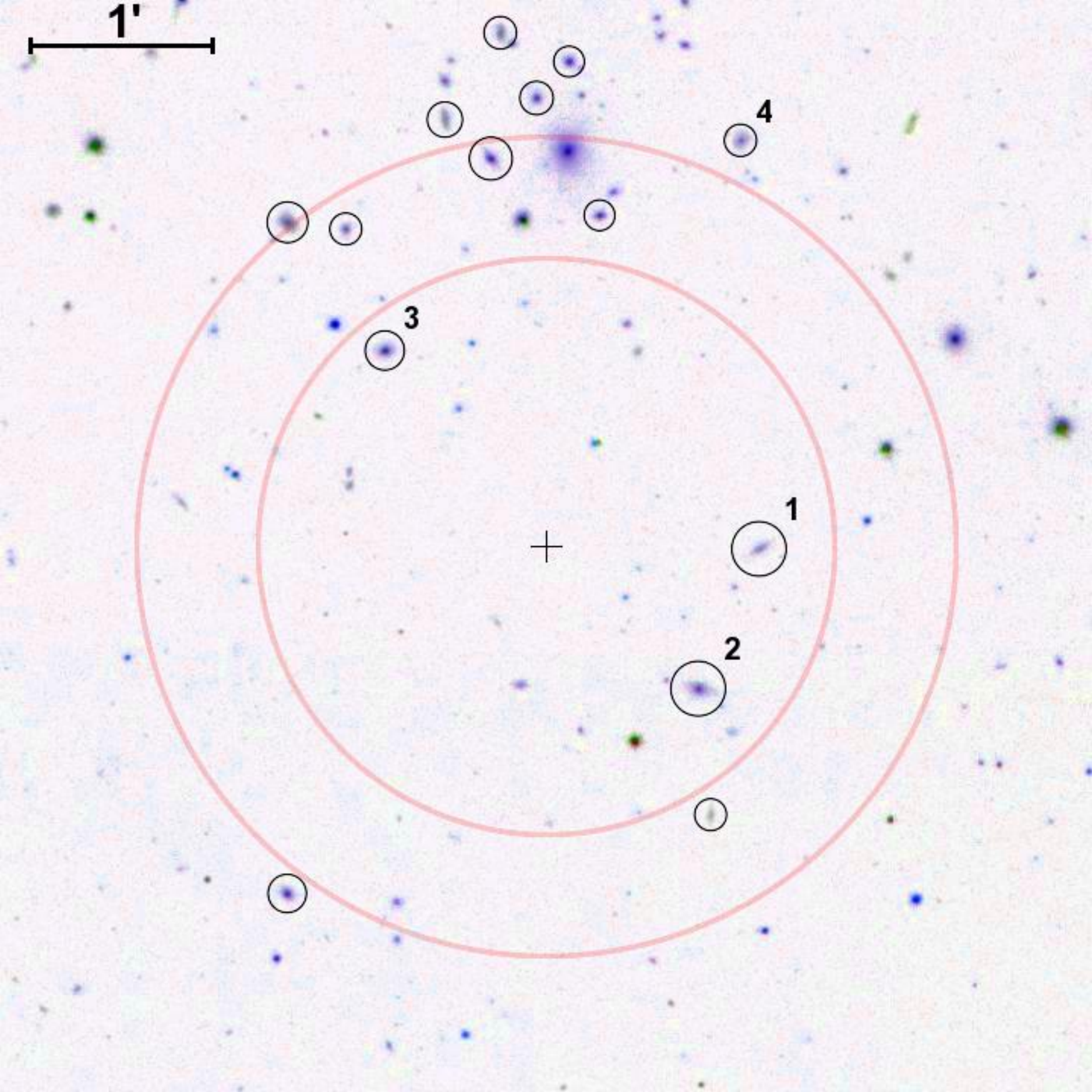} & 
\parbox[b]{4truecm}{\epsscale{0.07}
\plotone{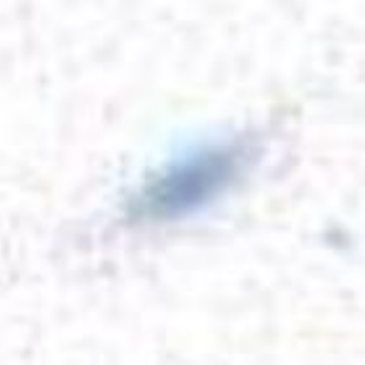}\parbox[b]{4.0truecm}{1: 00003.2+154752 \newline \phantom{1:} a=17" \vspace{1mm}}\\
\plotone{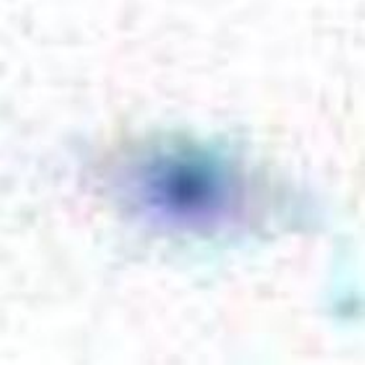}\parbox[b]{4.0truecm}{2: 00004.6+154706 \newline \phantom{2:} a=18" \vspace{1mm}}\\
\plotone{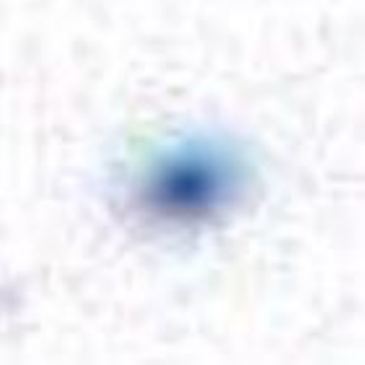}\parbox[b]{4.0truecm}{3: 000011.8+154857 \newline \phantom{3:} a=12" \vspace{1mm}}\\
\plotone{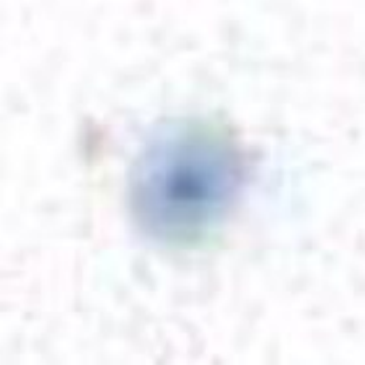}\parbox[b]{4.0truecm}{4: 00003.7+155006 \newline \phantom{4:} a=10" \vspace{1mm}}\\
\epsscale{0.3} } \\
\multicolumn{3}{l}{\parbox[t]{16truecm}{
\galcomment{3-2}
\newline}}\\\epsscale{0.3}\plotone{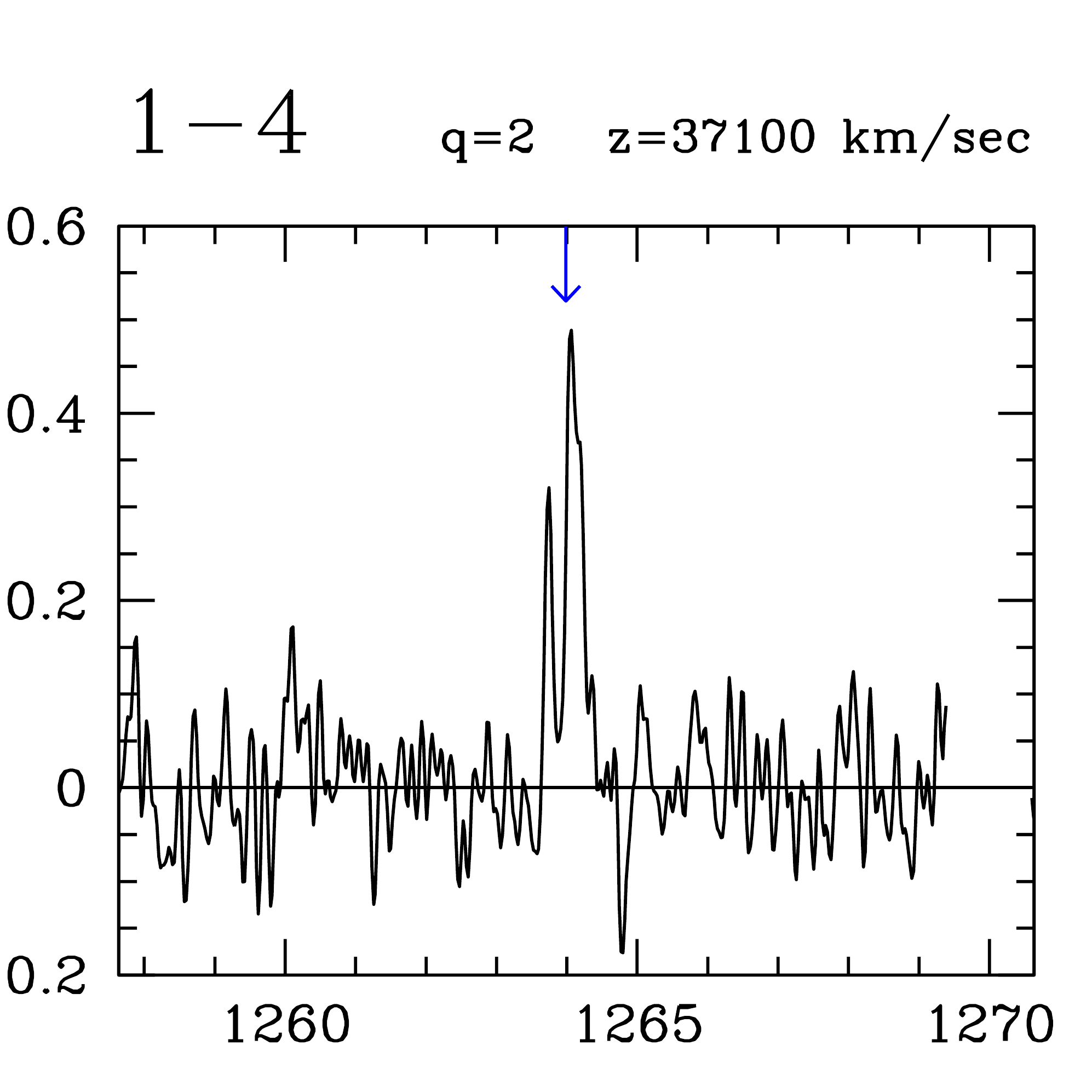} & \plotone{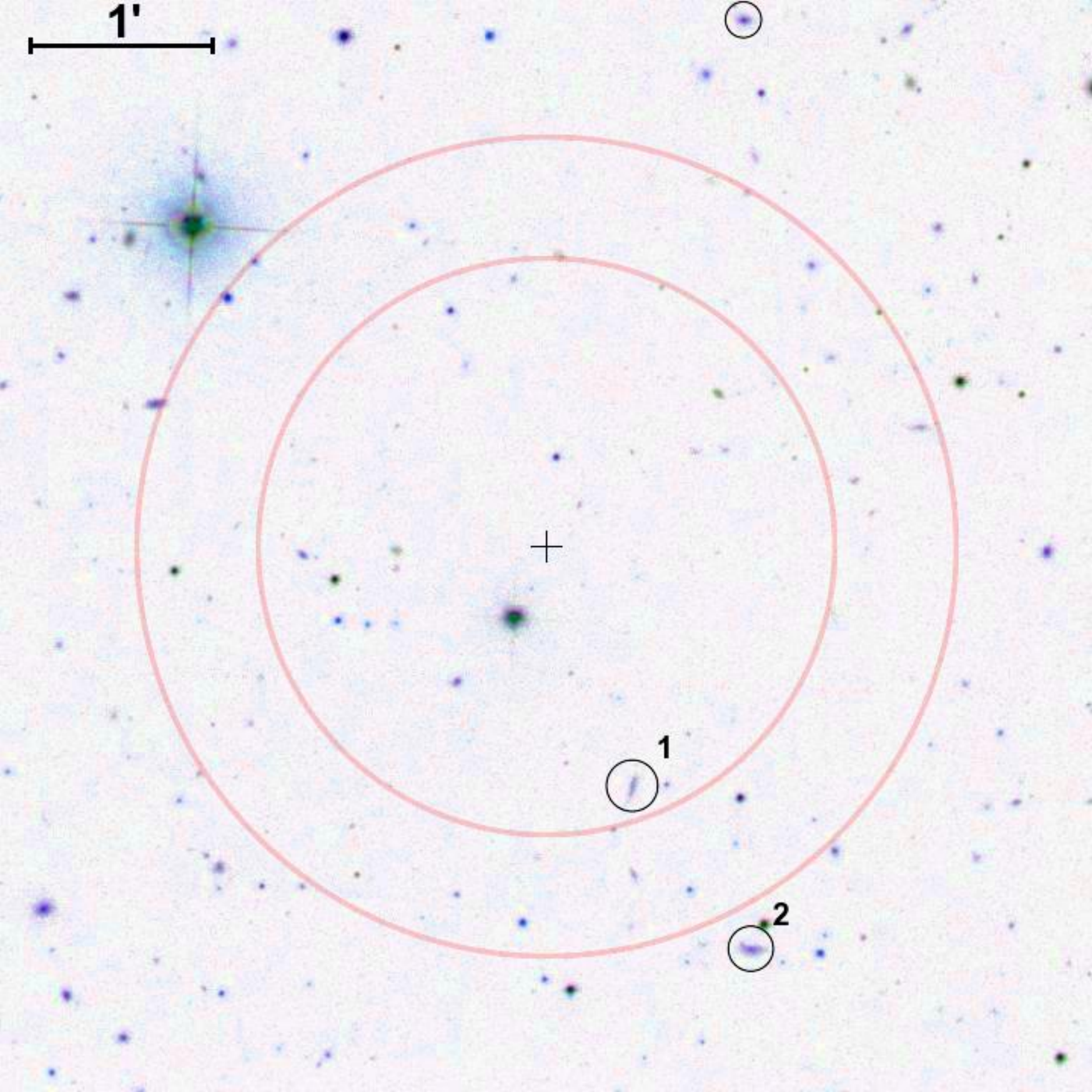} & 
\parbox[b]{4truecm}{\epsscale{0.07}
\plotone{sdss000014+153609.pdf}\parbox[b]{4.0truecm}{1: 000010.4+153609 \newline \phantom{1:} a=16" \vspace{1mm}}\\
\plotone{sdss00007+153515.pdf}\parbox[b]{4.0truecm}{2: 00007.7+153515 \newline \phantom{2:} a=14" \vspace{1mm}}\\
\epsscale{0.3} } \\
\multicolumn{3}{l}{\parbox[t]{16truecm}{
\galcomment{1-4}
\newline}}\\\end{tabular}

\begin{tabular}{ccl}
\multicolumn{3}{l}{Fig. A1. -- continued} \\
\\
\epsscale{0.3}\plotone{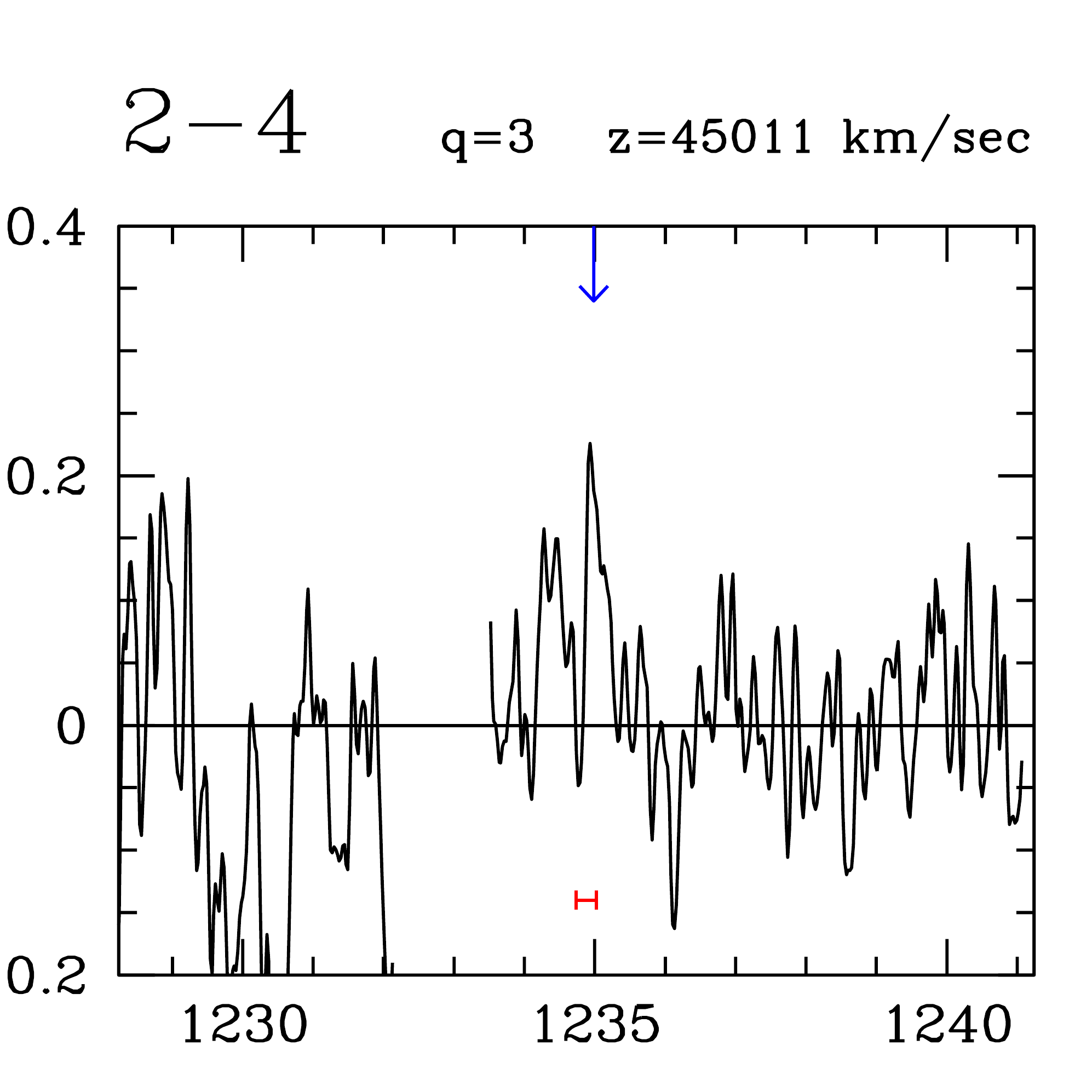} & \plotone{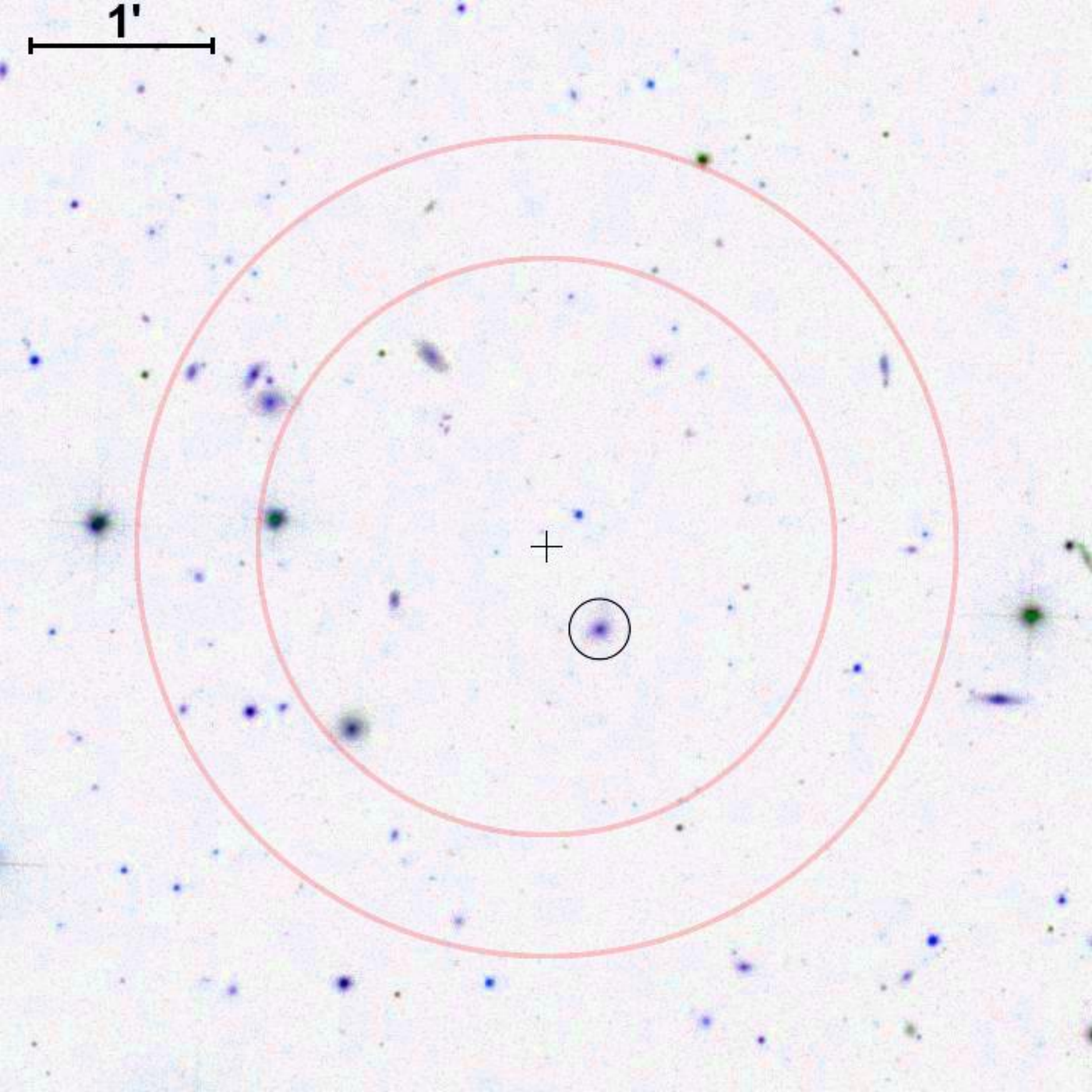} & 
\parbox[b]{4truecm}{\epsscale{0.07}
\plotone{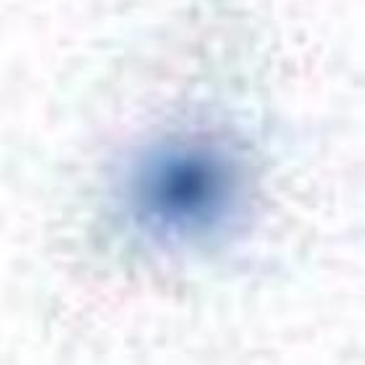}\parbox[b]{4.0truecm}{ 000011.1+154212 \newline  a=19"\newline z= 0.1502 }\\
\epsscale{0.3} } \\
\multicolumn{3}{l}{\parbox[t]{16truecm}{
\galcomment{2-4}
\newline}}\\\epsscale{0.3}\plotone{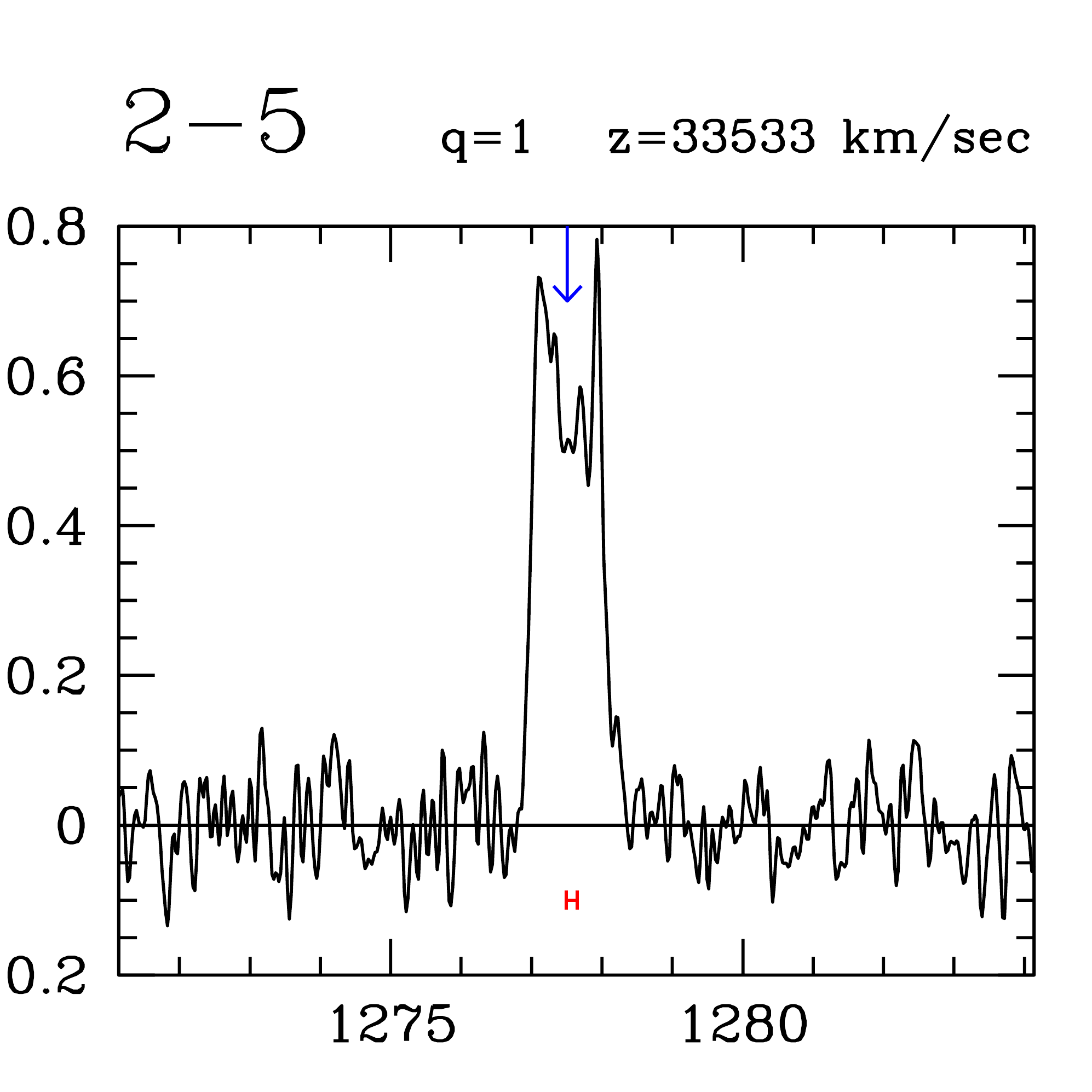} & \plotone{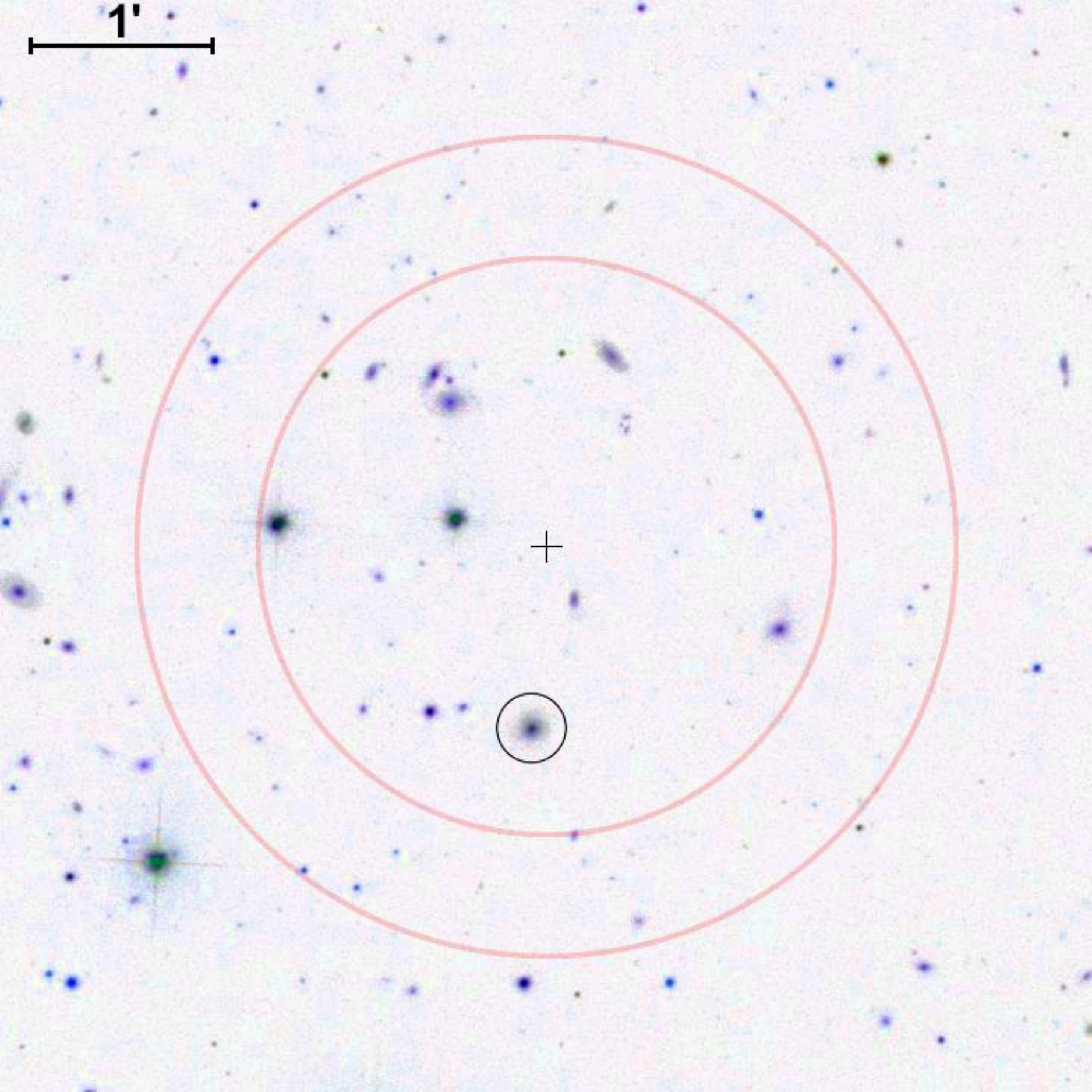} & 
\parbox[b]{4truecm}{\epsscale{0.07}
\plotone{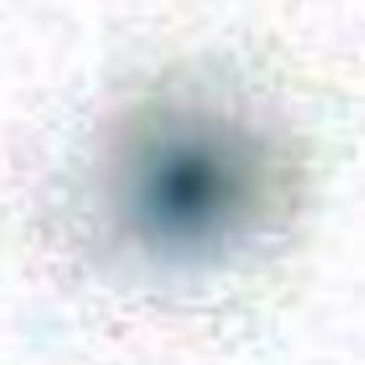}\parbox[b]{4.0truecm}{ 000016.8+154140 \newline  a=22"\newline z= 0.1118 }\\
\epsscale{0.3} } \\
\multicolumn{3}{l}{\parbox[t]{16truecm}{
\galcomment{2-5}
\newline}}\\\epsscale{0.3}\plotone{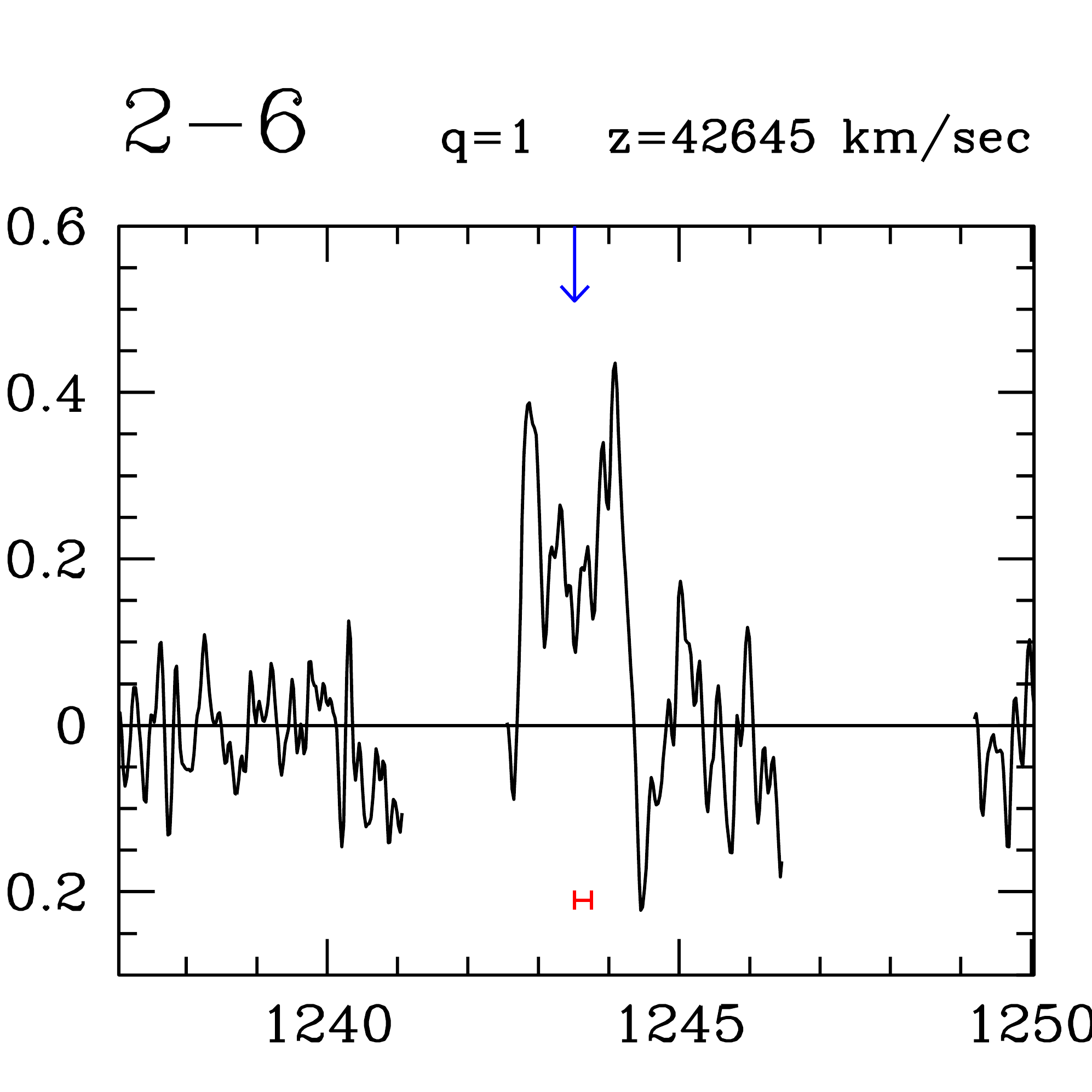} & \plotone{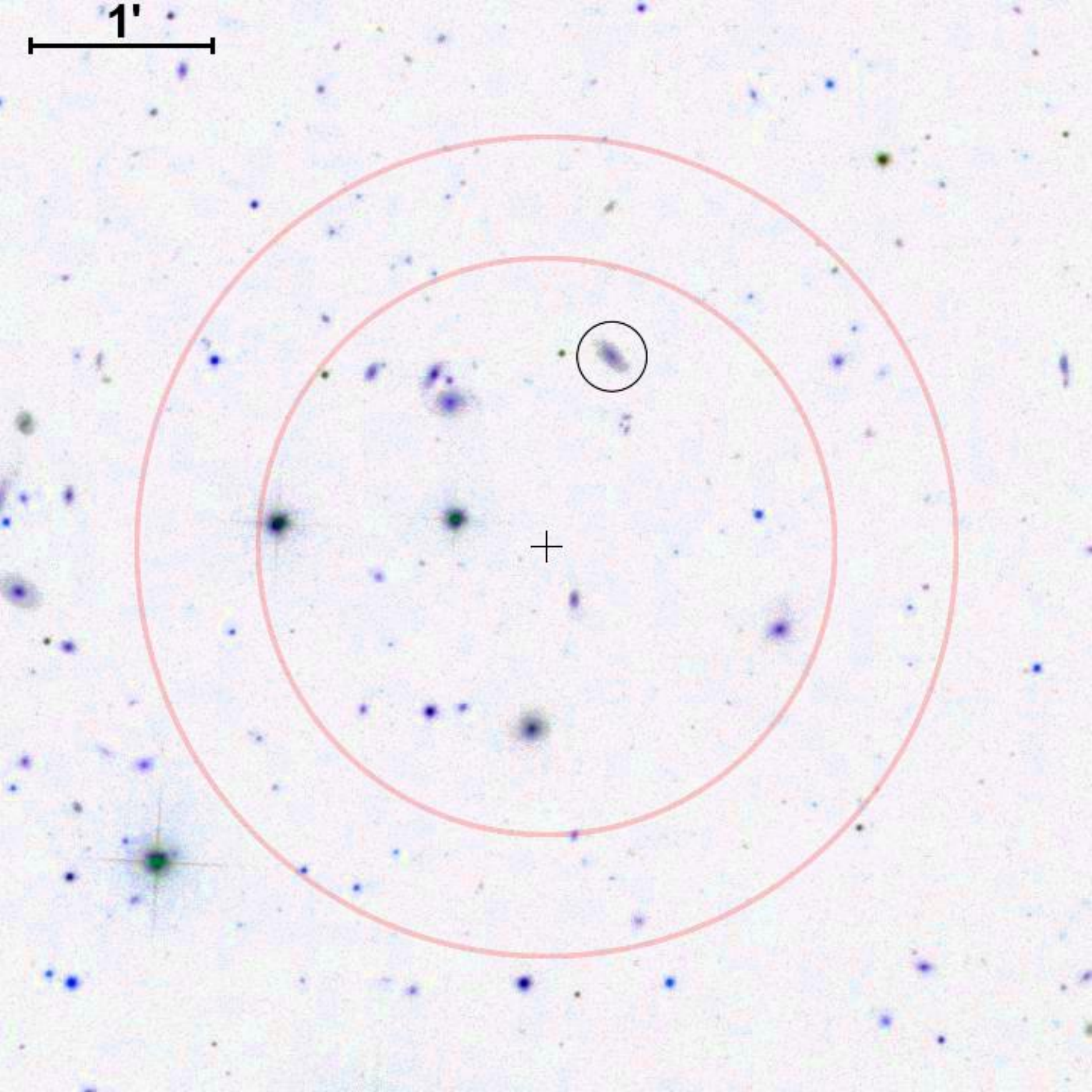} & 
\parbox[b]{4truecm}{\epsscale{0.07}
\plotone{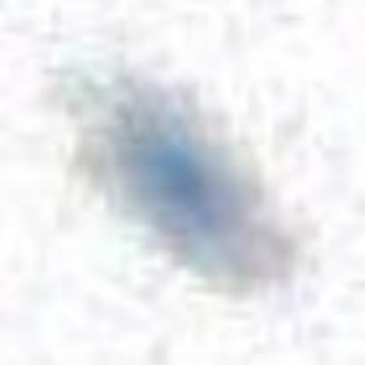}\parbox[b]{4.0truecm}{ 000014.9+154342 \newline  a=22"\newline z= 0.1421 }\\
\epsscale{0.3} } \\
\multicolumn{3}{l}{\parbox[t]{16truecm}{
\galcomment{2-6}
\newline}}\\\end{tabular}

\begin{tabular}{ccl}
\multicolumn{3}{l}{Fig. A1. -- continued} \\
\\
\epsscale{0.3}\plotone{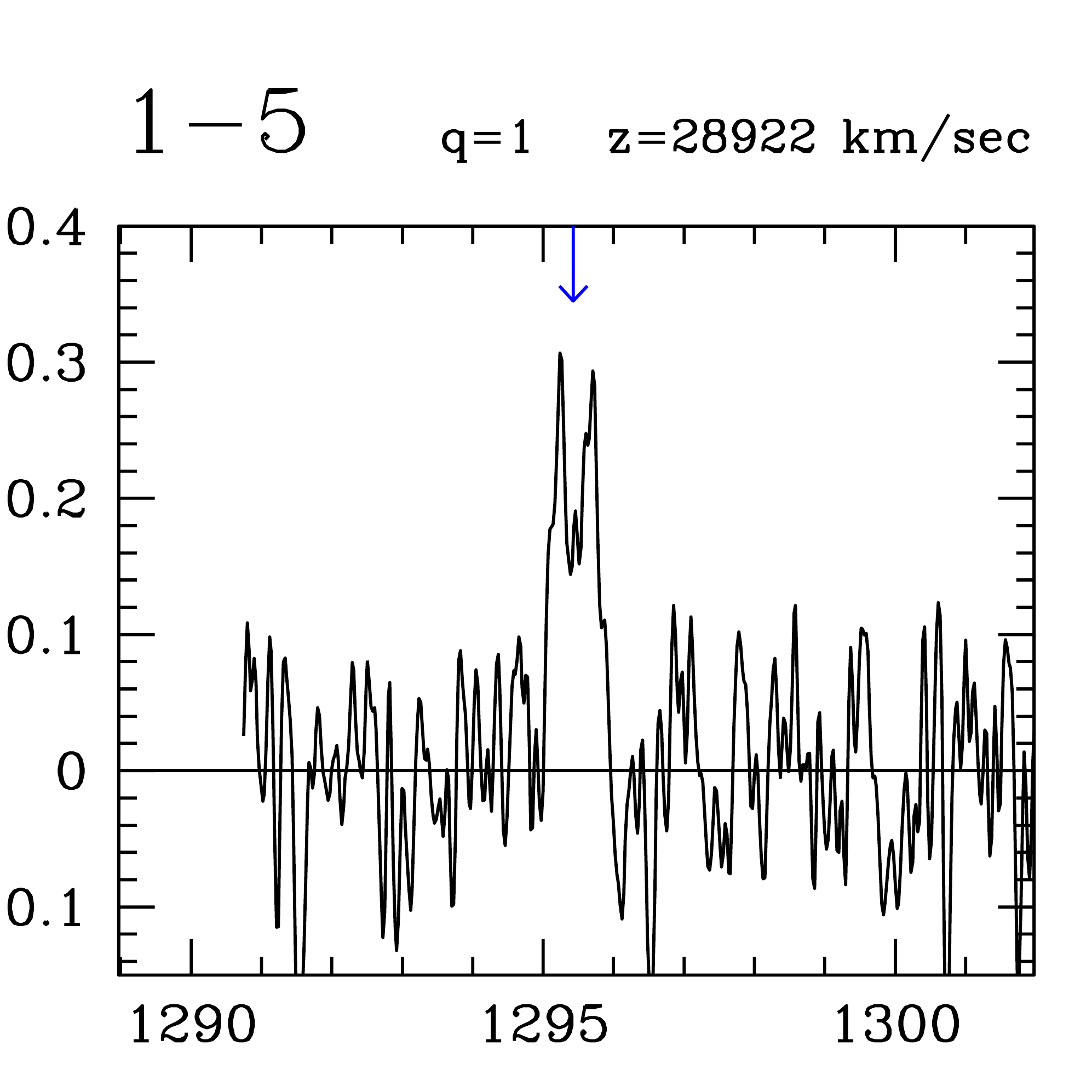} & \plotone{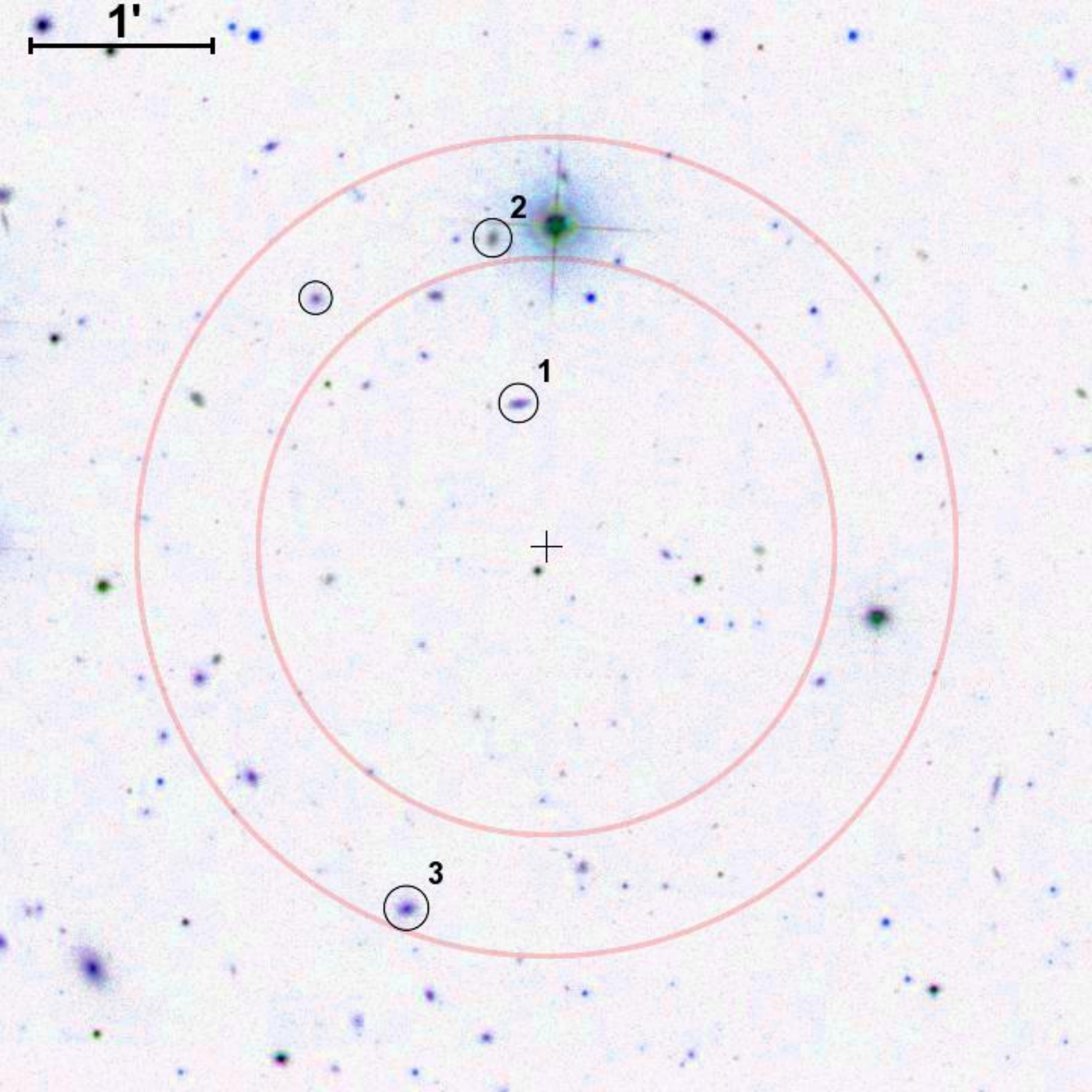} & 
\parbox[b]{4truecm}{\epsscale{0.07}
\plotone{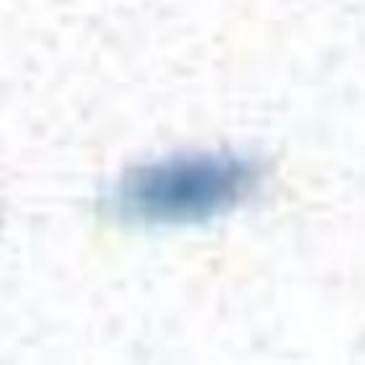}\parbox[b]{4.0truecm}{1: 000021.2+153815 \newline \phantom{1:} a=12" \vspace{1mm}}\\
\plotone{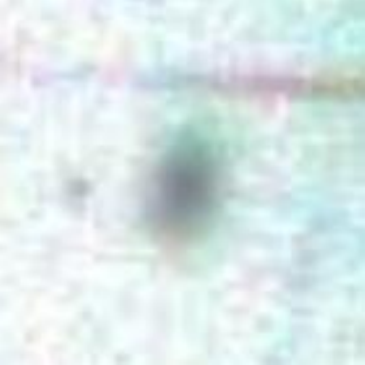}\parbox[b]{4.0truecm}{2: 000021.8+153909 \newline \phantom{2:} a=12" \vspace{1mm}}\\
\plotone{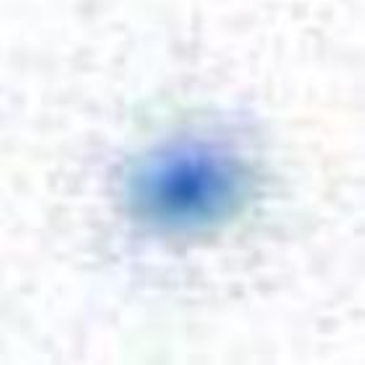}\parbox[b]{4.0truecm}{3: 000023.8+153528 \newline \phantom{3:} a=14" \vspace{1mm}}\\
\epsscale{0.3} } \\
\multicolumn{3}{l}{\parbox[t]{16truecm}{
\galcomment{1-5}
\newline}}\\\epsscale{0.3}\plotone{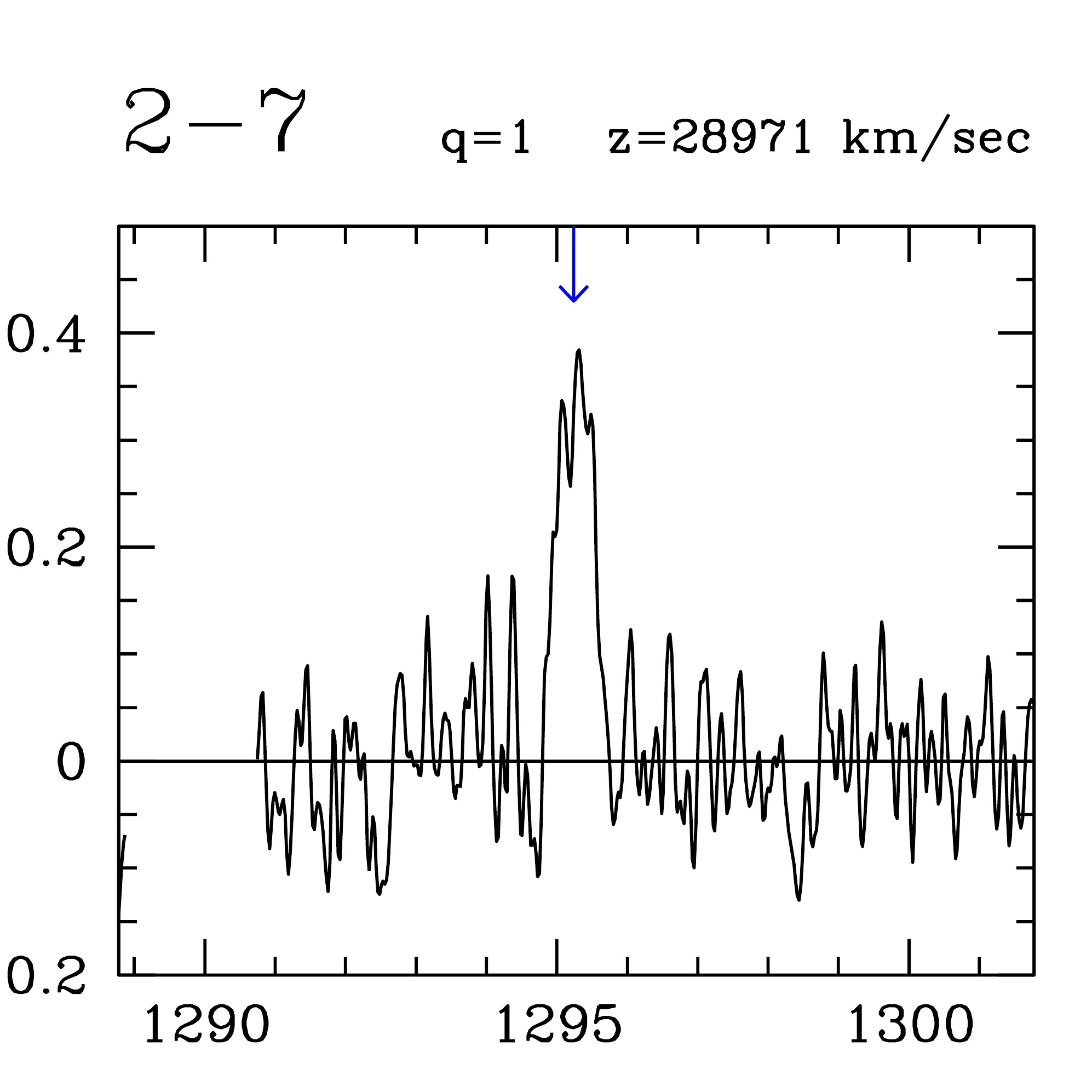} & \plotone{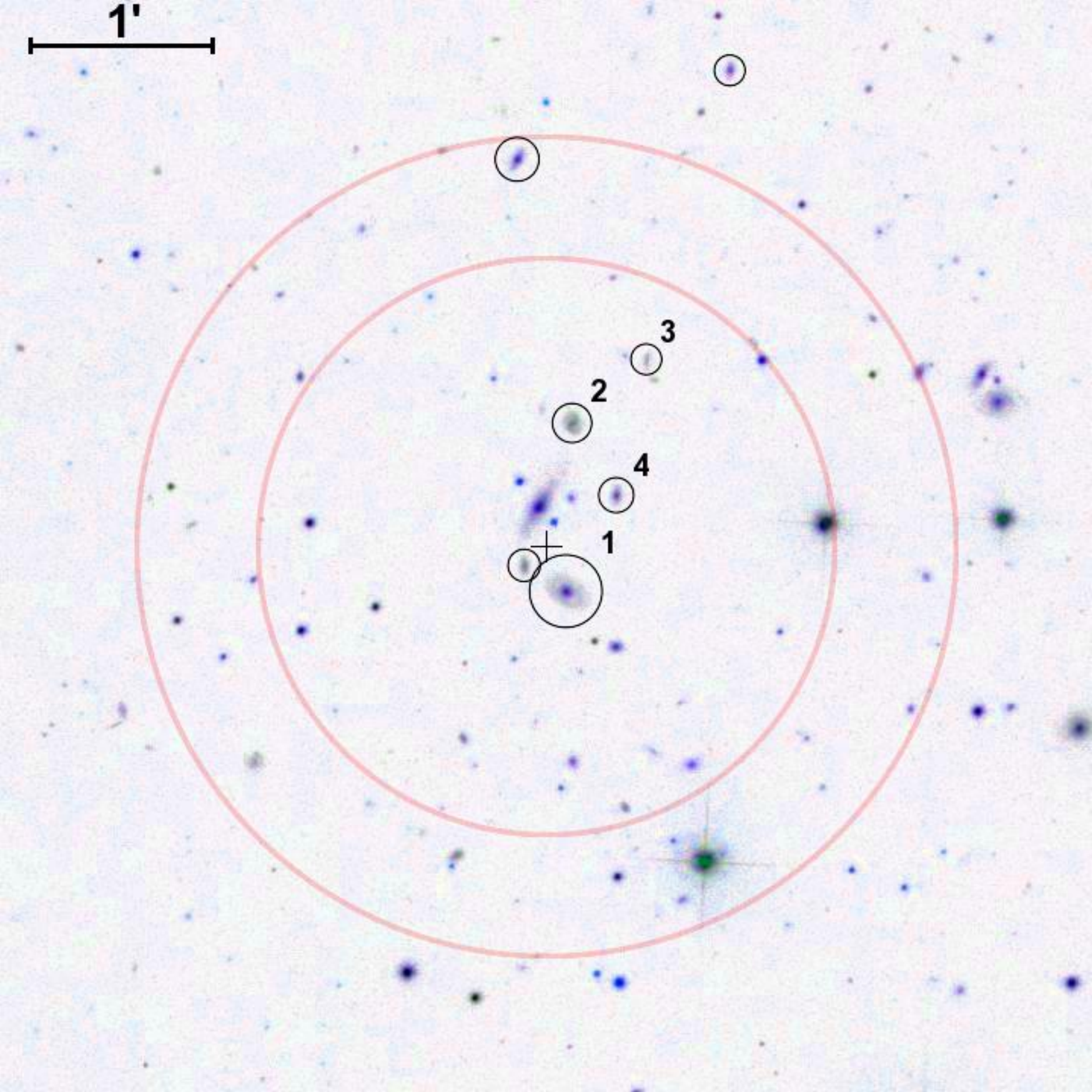} & 
\parbox[b]{4truecm}{\epsscale{0.07}
\plotone{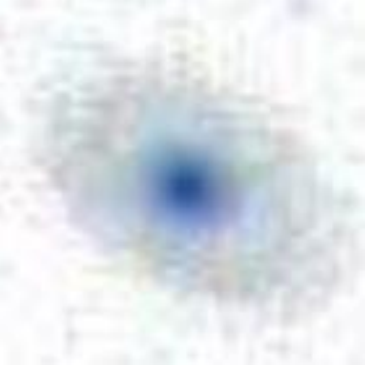}\parbox[b]{4.0truecm}{1: 000028.5+154225 \newline \phantom{1:} a=23" \vspace{1mm}}\\
\plotone{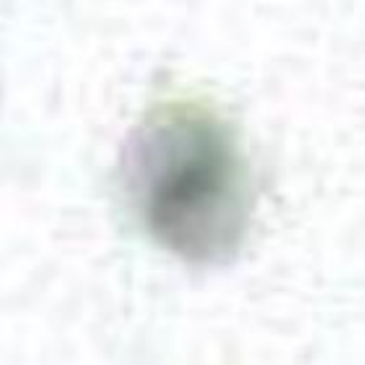}\parbox[b]{4.0truecm}{2: 000028.3+154320 \newline \phantom{2:} a=12" \vspace{1mm}}\\
\plotone{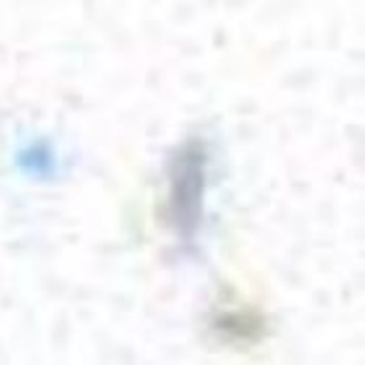}\parbox[b]{4.0truecm}{3: 000026.6+154341 \newline \phantom{3:} a=10" \vspace{1mm}}\\
\plotone{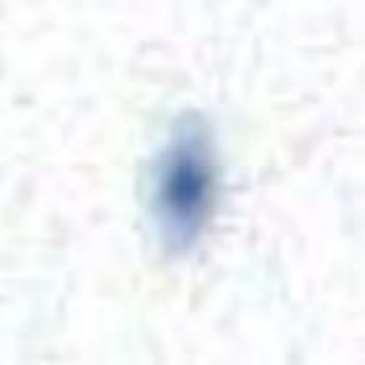}\parbox[b]{4.0truecm}{4: 000027.3+154256 \newline \phantom{4:} a=11" \vspace{1mm}}\\
\epsscale{0.3} } \\
\multicolumn{3}{l}{\parbox[t]{16truecm}{
\galcomment{2-7}
\newline}}\\\epsscale{0.3}\plotone{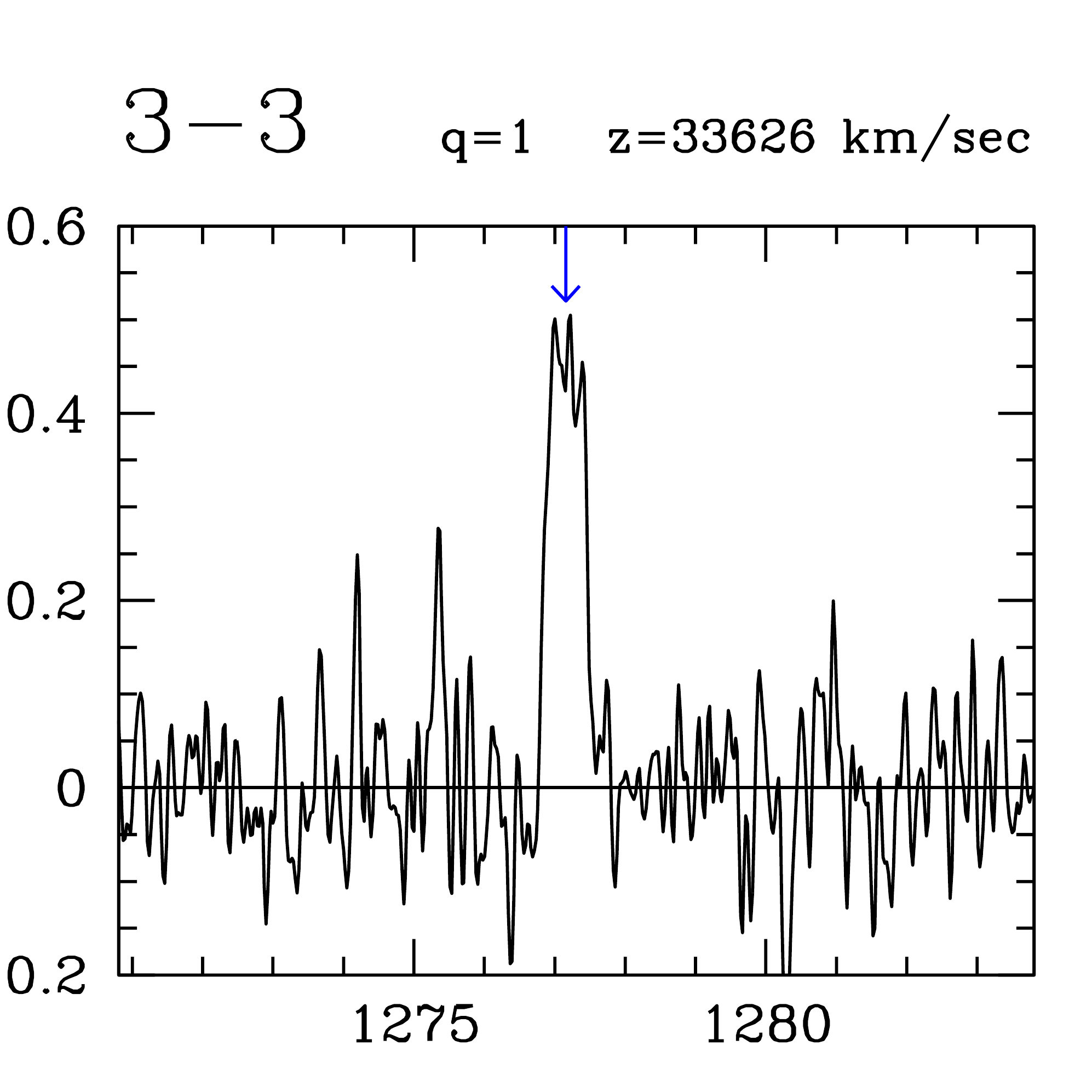} & \plotone{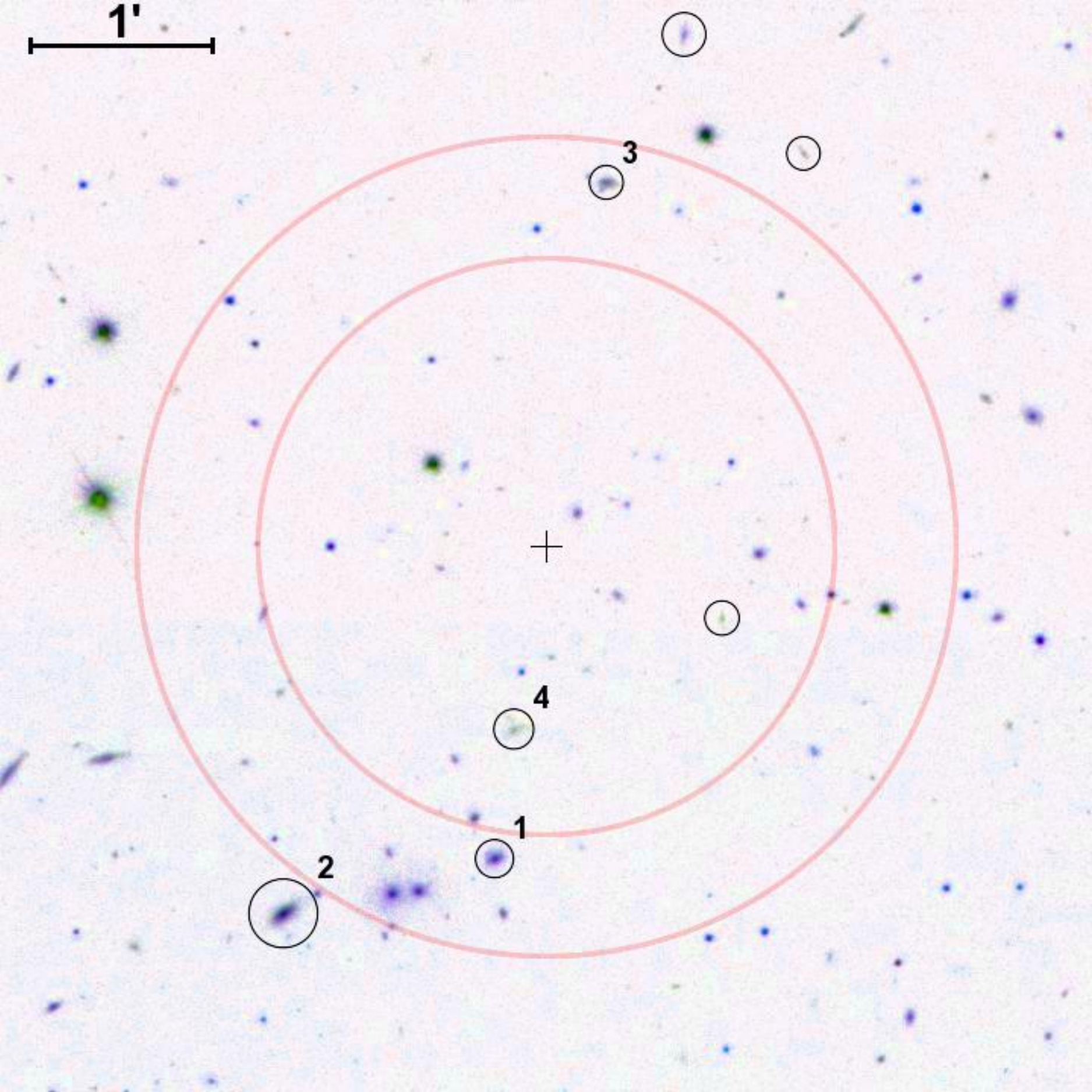} & 
\parbox[b]{4truecm}{\epsscale{0.07}
\plotone{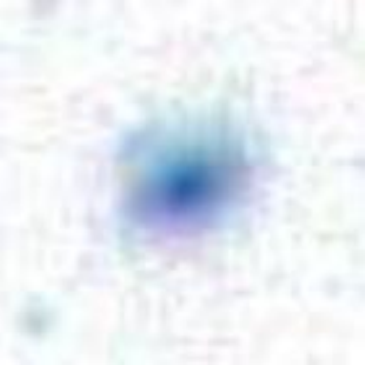}\parbox[b]{4.0truecm}{1: 000034.2+154609 \newline \phantom{1:} a=12" \vspace{1mm}}\\
\plotone{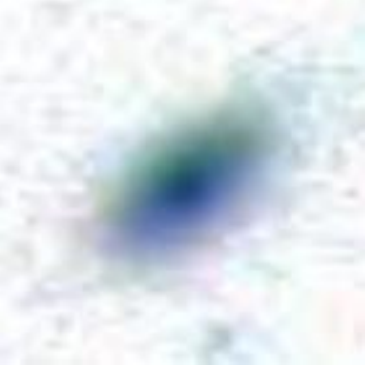}\parbox[b]{4.0truecm}{2: 000039.0+154551 \newline \phantom{2:} a=22" \vspace{1mm}}\\
\plotone{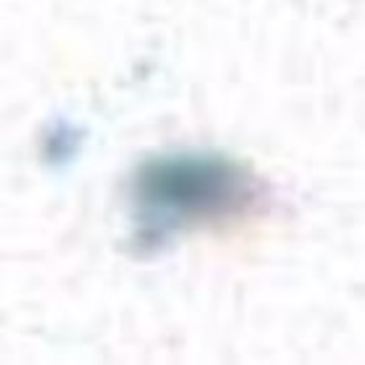}\parbox[b]{4.0truecm}{3: 000031.6+154952 \newline \phantom{3:} a=11" \vspace{1mm}}\\
\plotone{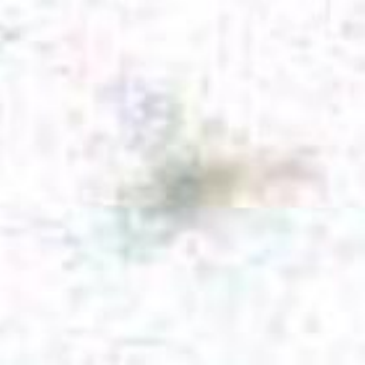}\parbox[b]{4.0truecm}{4: 000033.8+154651 \newline \phantom{4:} a=13" \vspace{1mm}}\\
\epsscale{0.3} } \\
\multicolumn{3}{l}{\parbox[t]{16truecm}{
\galcomment{3-3}
\newline}}\\\end{tabular}

\begin{tabular}{ccl}
\multicolumn{3}{l}{Fig. A1. -- continued} \\
\\
\epsscale{0.3}\plotone{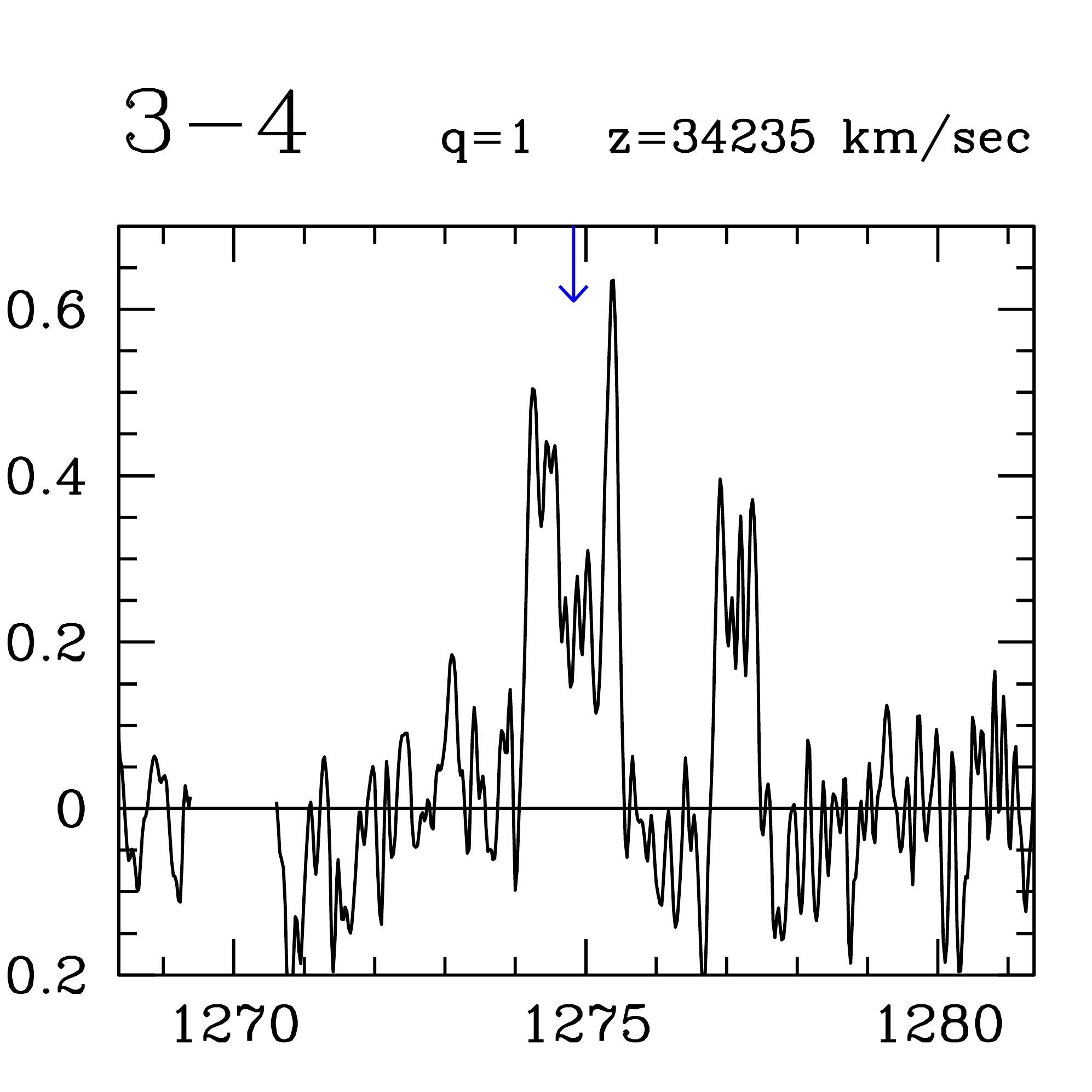} & \plotone{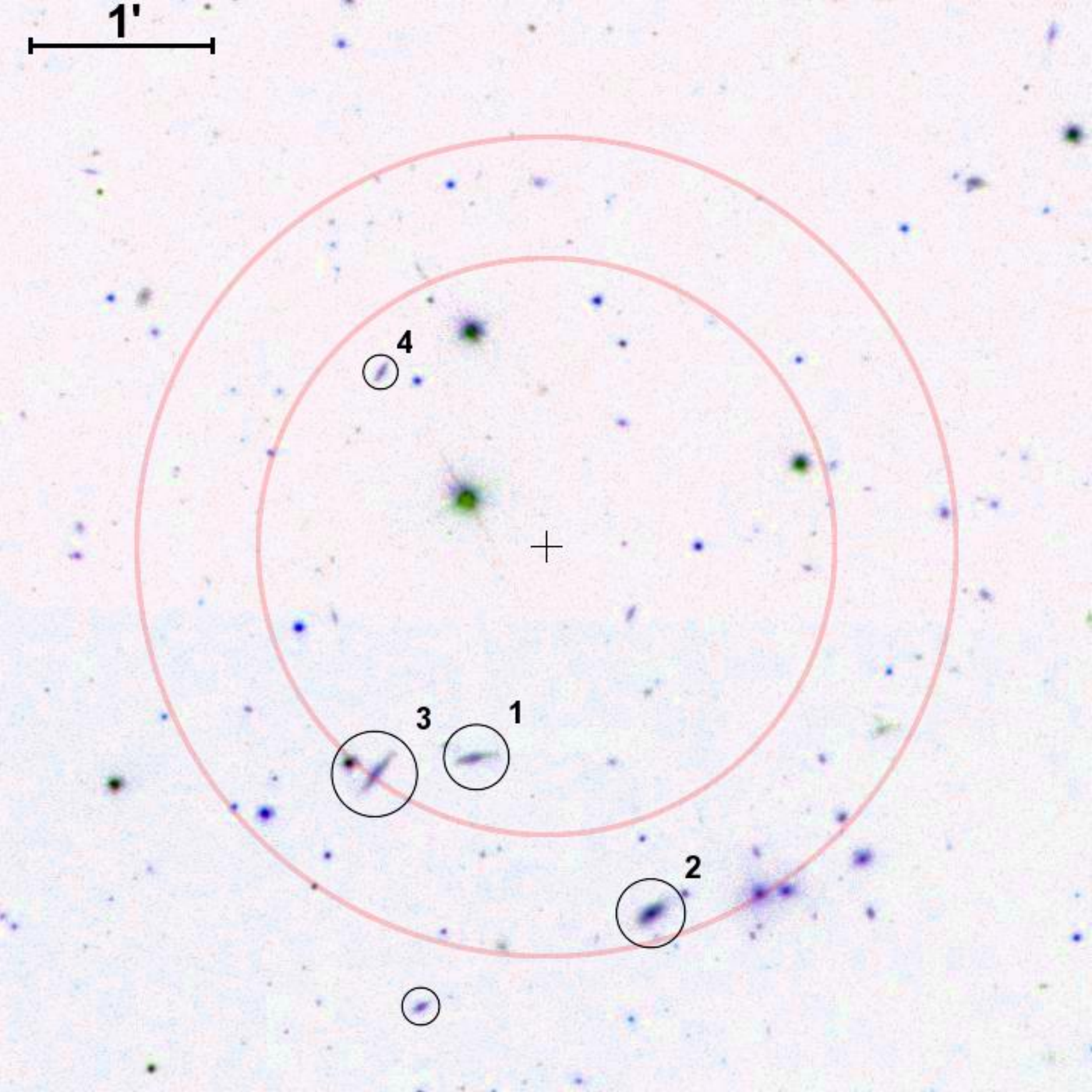} & 
\parbox[b]{4truecm}{\epsscale{0.07}
\plotone{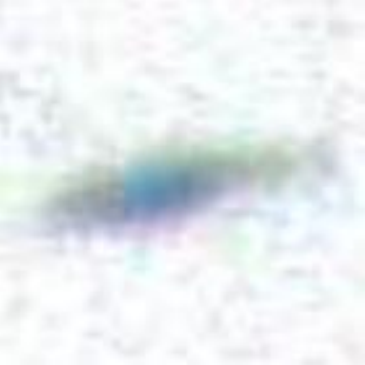}\parbox[b]{4.0truecm}{1: 000043.0+154642 \newline \phantom{1:} a=21" \vspace{1mm}}\\
\plotone{sdss000030+154551.pdf}\parbox[b]{4.0truecm}{2: 000039.0+154551 \newline \phantom{2:} a=22" \vspace{1mm}}\\
\plotone{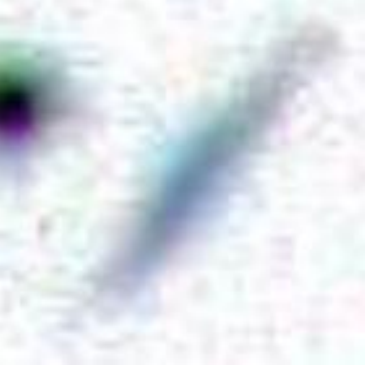}\parbox[b]{4.0truecm}{3: 000045.3+154637 \newline \phantom{3:} a=28" \vspace{1mm}}\\
\plotone{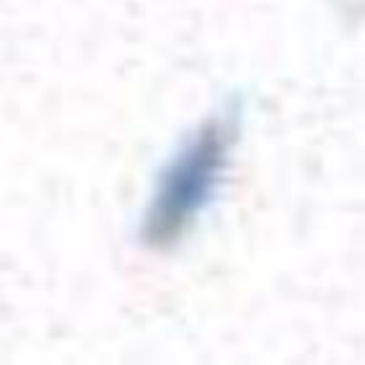}\parbox[b]{4.0truecm}{4: 000045.2+154849 \newline \phantom{4:} a=11" \vspace{1mm}}\\
\epsscale{0.3} } \\
\multicolumn{3}{l}{\parbox[t]{16truecm}{
\galcomment{3-4}
\newline}}\\\epsscale{0.3}\plotone{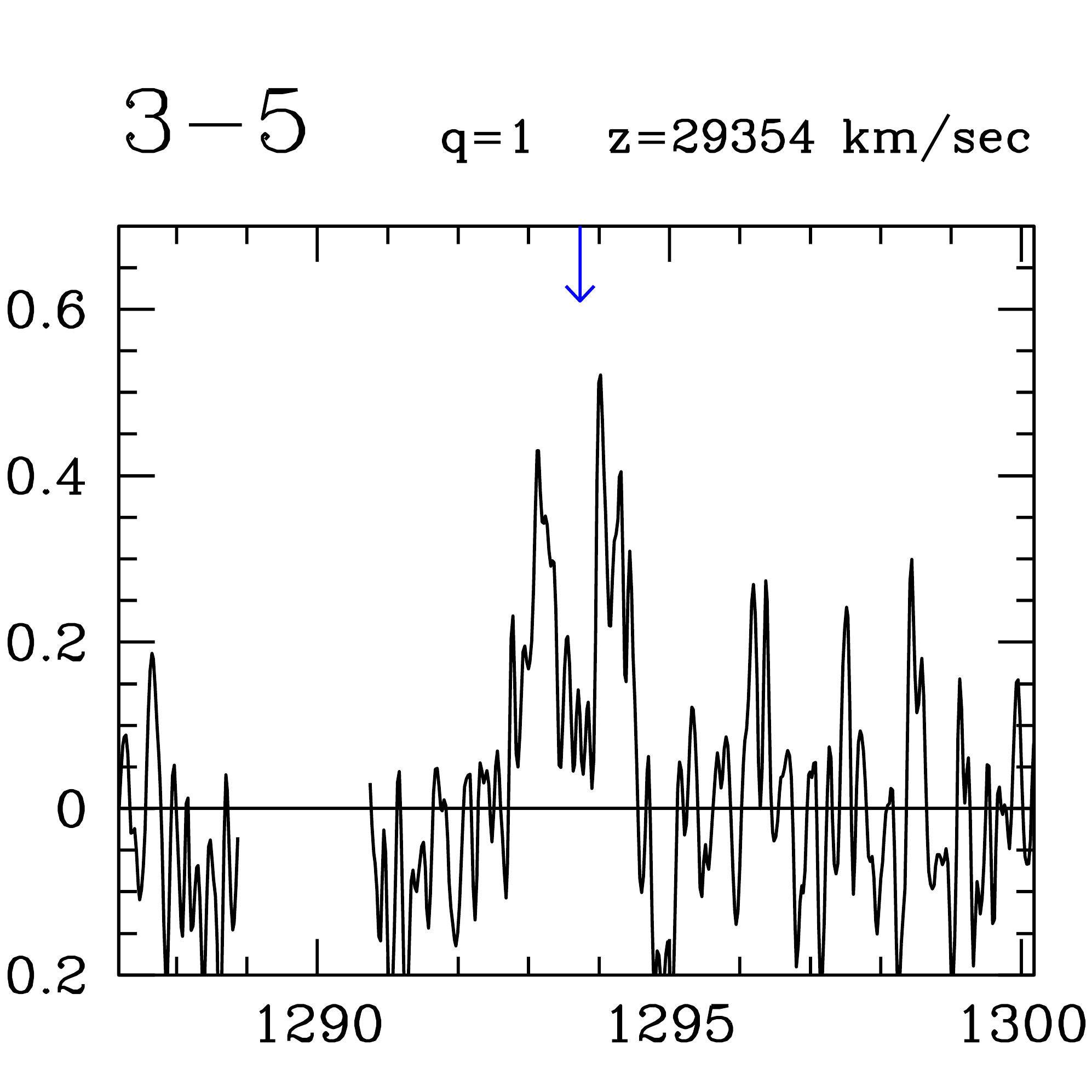} & \plotone{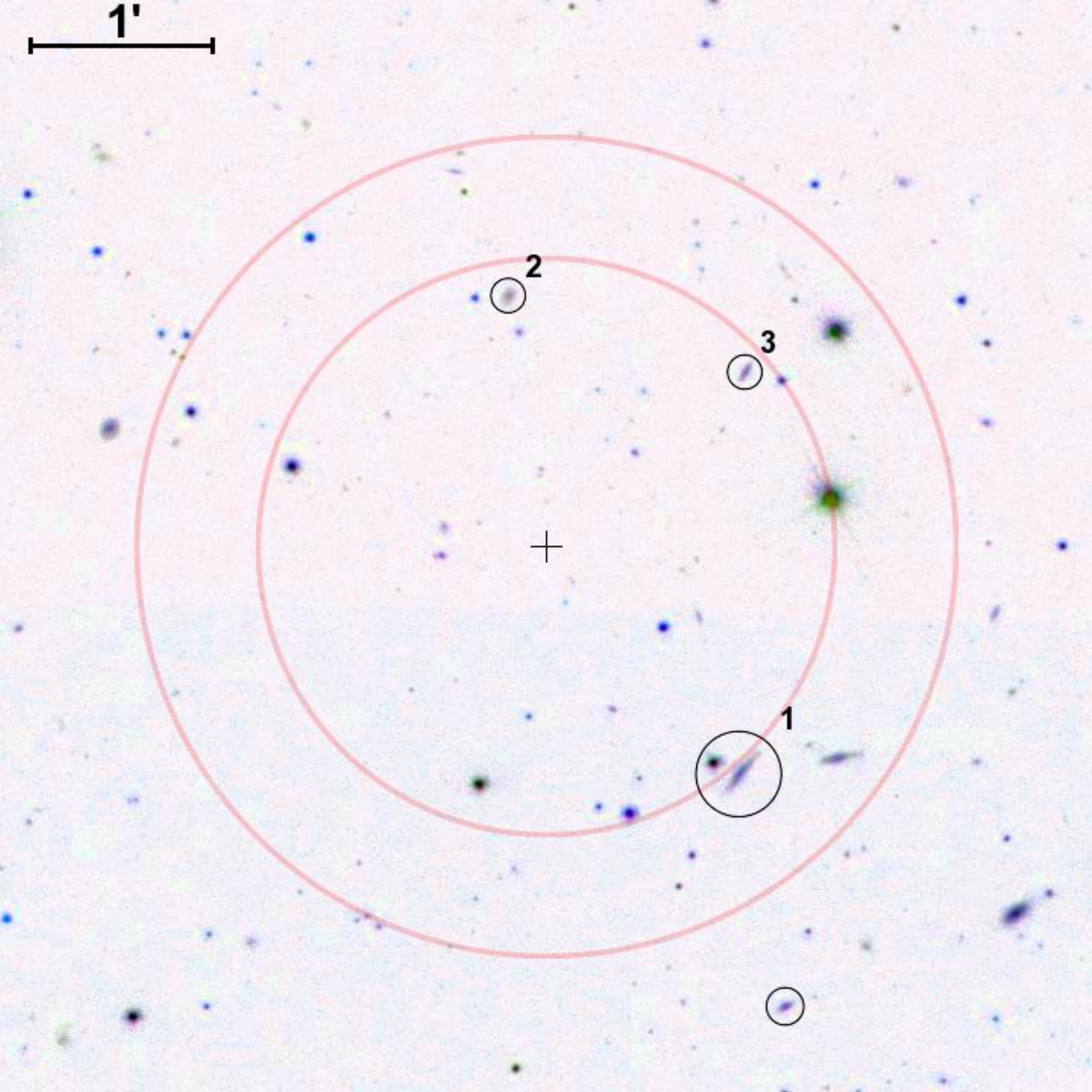} & 
\parbox[b]{4truecm}{\epsscale{0.07}
\plotone{sdss000043+154637.pdf}\parbox[b]{4.0truecm}{1: 000045.3+154637 \newline \phantom{1:} a=28" \vspace{1mm}}\\
\plotone{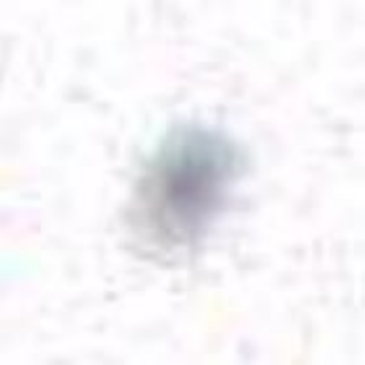}\parbox[b]{4.0truecm}{2: 000050.6+154914 \newline \phantom{2:} a=11" \vspace{1mm}}\\
\plotone{sdss000042+154849.pdf}\parbox[b]{4.0truecm}{3: 000045.2+154849 \newline \phantom{3:} a=11" \vspace{1mm}}\\
\epsscale{0.3} } \\
\multicolumn{3}{l}{\parbox[t]{16truecm}{
\galcomment{3-5}
\newline}}\\\epsscale{0.3}\plotone{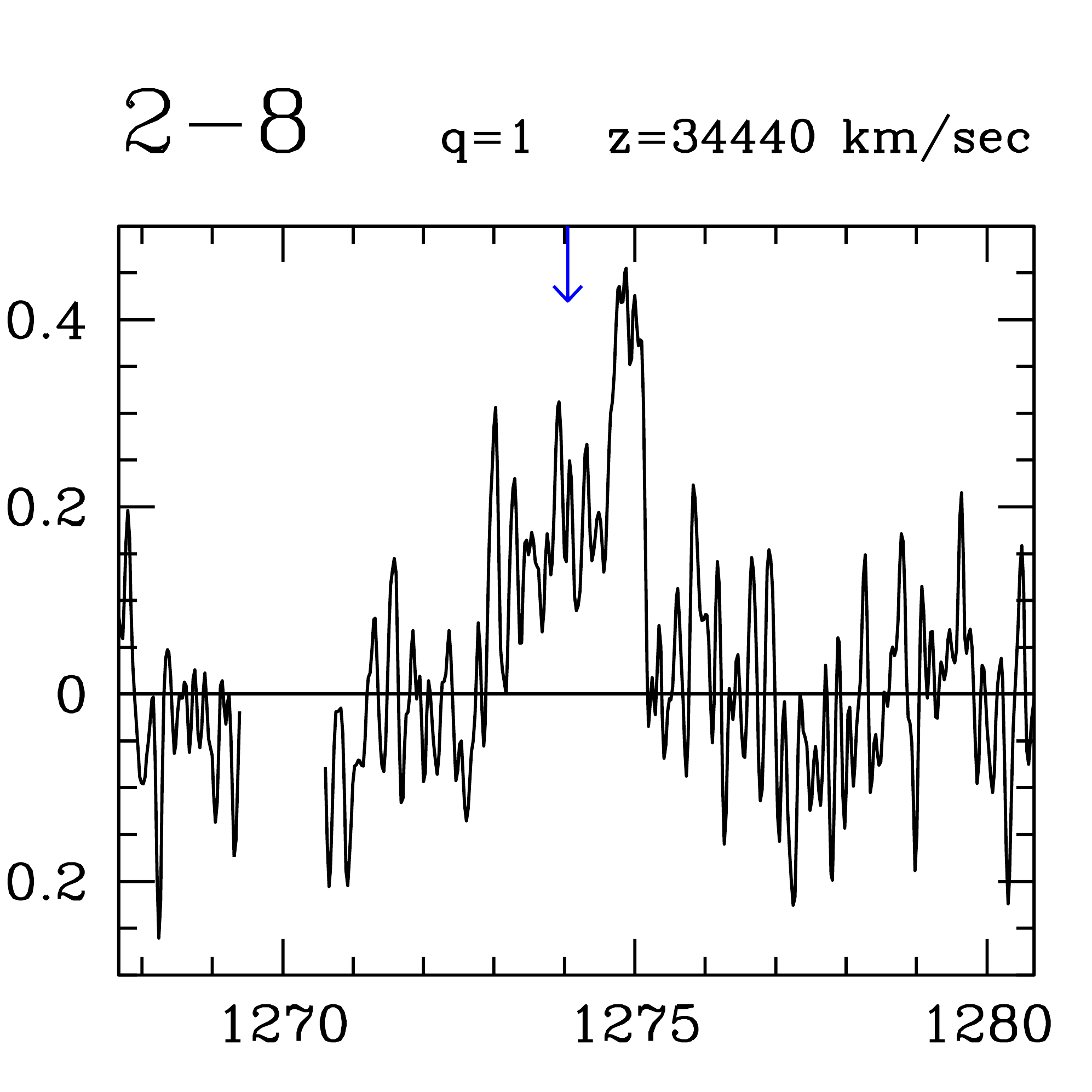} & \plotone{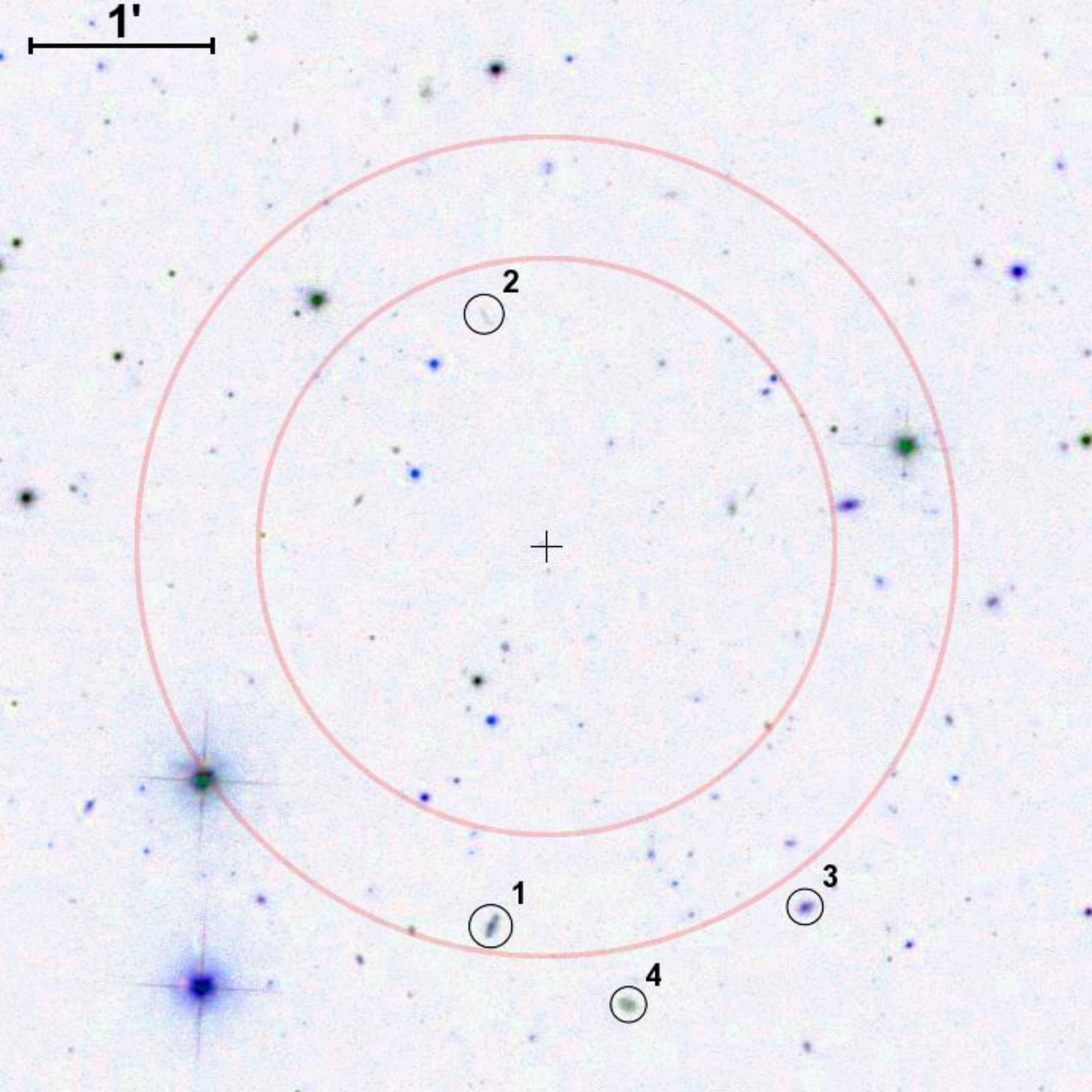} & 
\parbox[b]{4truecm}{\epsscale{0.07}
\plotone{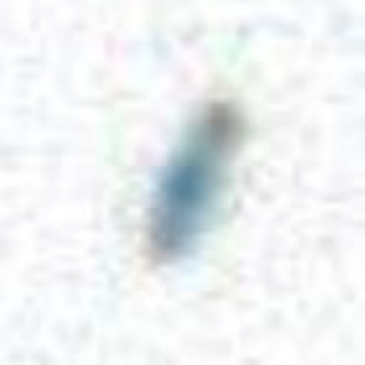}\parbox[b]{4.0truecm}{1: 000059.3+154034 \newline \phantom{1:} a=14" \vspace{1mm}}\\
\plotone{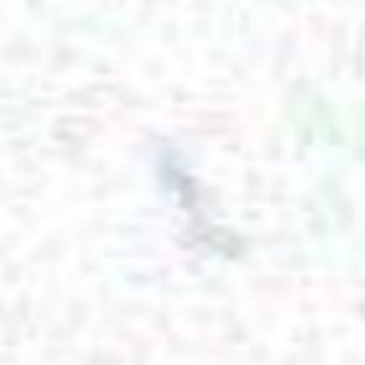}\parbox[b]{4.0truecm}{2: 000059.4+154356 \newline \phantom{2:} a=12" \vspace{1mm}}\\
\plotone{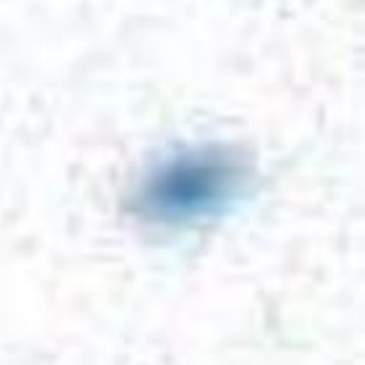}\parbox[b]{4.0truecm}{3: 000052.1+154041 \newline \phantom{3:} a=11" \vspace{1mm}}\\
\plotone{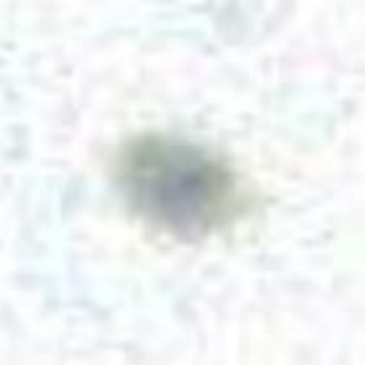}\parbox[b]{4.0truecm}{4: 000056.1+154009 \newline \phantom{4:} a=11" \vspace{1mm}}\\
\epsscale{0.3} } \\
\multicolumn{3}{l}{\parbox[t]{16truecm}{
\galcomment{2-8}
\newline}}\\\end{tabular}


\end{document}